\documentclass[%
 reprint,
 amsmath,amssymb,
 aps, longbibliography,
 pre
]{revtex4-2}
\usepackage[T1]{fontenc}
\usepackage[utf8]{inputenc} 
\usepackage{appendix}
\usepackage{graphicx}
\usepackage{dcolumn}
\usepackage{bm}
\usepackage{hyperref}
\usepackage{csquotes}
\usepackage{booktabs}
\usepackage{xcolor}
\usepackage{comment}

\newcommand{\nl}{\textit{n}$\ell$}
\newcommand{\pinf}{$p_\infty$}
\newcommand{\tm}{\text{m}}

\newcommand{\dc}{d_\mathrm{c}}

\begin{document}
\title{Universality of noise-induced transitions in nonlinear voter models} 
\author{Jaume Llabr\'es}
 \affiliation{Institute for Cross-disciplinary Physics and Complex Systems IFISC (CSIC-UIB), Campus Universitat Illes Balears, 07122 Palma de Mallorca, Spain.}
\author{Maxi San Miguel}
 \affiliation{Institute for Cross-disciplinary Physics and Complex Systems IFISC (CSIC-UIB), Campus Universitat Illes Balears, 07122 Palma de Mallorca, Spain.}
\author{Ra\'ul Toral}
 \affiliation{Institute for Cross-disciplinary Physics and Complex Systems IFISC (CSIC-UIB), Campus Universitat Illes Balears, 07122 Palma de Mallorca, Spain.}

\begin{abstract}
We analyze the universality classes of phase transitions in a variety of nonlinear voter models. By mapping several models with symmetric absorbing states onto a canonical model introduced in previous studies, we confirm that they exhibit a Generalized Voter (GV) transition. We then propose a canonical mean-field model that extends the original formulation by incorporating a noise term that eliminates the absorbing states. This generalization gives rise to a phase diagram featuring two distinct types of phase transitions: a continuous Ising transition and a discontinuous transition we call Modified Generalized Voter (MGV). These two transition lines converge at a tricritical point. We map diverse noisy nonlinear voter models onto this extended canonical form. Using finite-size scaling techniques above and below the upper critical dimension, we show that the continuous transition of these models belongs to the Ising universality class in their respective dimensionality. We also find universal behavior at the tricritical point. Our results provide a unifying framework for classifying phase transitions in stochastic models of opinion dynamics with both nonlinearity and noise.
\end{abstract}

\maketitle

\section{Introduction}
The \textit{voter model}~\cite{vm_clifford,liggett} is a paradigmatic nonequilibrium lattice model~\cite{marroBOOK,rednerBOOK} that has been widely used to study collective phenomena in several contexts, in particular opinion dynamics~\cite{tobias_opinion}, such as electoral processes~\cite{IstheVM} and ecology~\cite{vm_clifford}. This simple model considers a population of $N$ interacting agents, each of them holding a binary state variable, and copying the state of one of its neighbors at random (herding behavior). The model has been studied in regular lattices~\cite{Chate2d} and in complex networks~\cite{Suchecki,EguiluzNJP}. It exhibits two equivalent absorbing states corresponding to collective consensus in each of the two possible states. Still, the model has no phase transition since it has no free parameter. However it has a critical dimension $d=2$, such that there is ordering dynamics (coarsening) for $d\leq 2$, while for $d>2$, and in the thermodynamic limit ($N \rightarrow \infty$), the system remains in a dynamically disordered state. Finite-size fluctuations take the system to one of the absorbing states in an escape time that increases with system size in different forms depending on dimensionality or topology of the complex network of interaction~\cite{redner1, redner2, EguiluzNJP, SucheckiEPL}. 

The universality classes of nonequilibrium phase transitions~\cite{Henkel, Odor2004} for systems with two symmetric absorbing states have been studied in Ref.~\cite{Hammal}. Using a Langevin equation derived from symmetry arguments, the authors demonstrate that these classes include Ising and Directed Percolation (DP) transitions as well as a type of transition named {\it Generalized Voter} (GV), in which the system exhibits an absorbing transition from a disordered state to one of the two absorbing states. The authors confirm what was conjectured first in Ref.~\cite{Droz}, the GV transition is a full transition line which splits into an Ising and a DP transition line. Moreover, the voter model represents the point at which all three lines converge. In Ref.~\cite{Fede_Clopez_AB}, the authors introduce a general framework for deriving the same Langevin equation. By considering a spin model characterized by its flipping probabilities, it is possible to establish a direct connection between the microscopic and the macroscopic dynamics.

The voter model has been generalized in many ways~\cite{rednerCompteRendues}. One class of these extensions is the nonlinear voter models, in which standard pairwise interactions are replaced by group interactions among agents. In these models, absorbing states are preserved, while an agent's rate of state change is determined by a nonlinear function of the fraction of its neighbors holding the opposite state~\cite{Schweitzer,Min,qvoter, Lambiotte_2007, Lambiotte_2008, confident_voters, noise_reduced_voter_model, lucia_NLVM, Sznajd, Sznajd_MF}.
 
A second class of these generalized models consists of noisy models, which incorporate spontaneous changes of state (noise) in the agent's dynamics. As a consequence there are no absorbing states in these models. The paradigmatic example is the noisy voter model (NVM), which has appeared in the literature in different contexts and with different names~\cite{moran, Fichthorn, granovsky, Carro2016, Peralta_sto_2018}. In the socio-economic context it is called the Kirman model~\cite{Kirman1993} where spontaneous changes of state are associated with an idiosyncratic behavior of the agents, independent of the choice or state of the other agents. For finite system sizes, the model exhibits a transition from an ordered to a disordered state at a critical threshold of the parameter ratio quantifying the relative strength of the mechanisms of change of state, namely, copying and noise. However, this transition vanishes in the thermodynamic limit. Another example presenting similar properties is the noisy partisan voter model (NPVM)~\cite{llabres}.

A third class has emerged from the fusion of the previous two classes, the noisy nonlinear voter models. Although a general classification of their possible phase transitions is still lacking, studies have shown that the finite-size noise-induced transitions observed in noisy models become bona-fide phase transitions in the thermodynamic limit ($N\to\infty$) when nonlinearities are taken into account. These phase transitions have been demonstrated to belong to the Ising universality class in their respective dimensionality~\cite{q-voter_noise, Jedrzejewski, Artime_2018, Peralta_2018, Artime_2019, Sznajd_Lama}. The inclusion of contrarian behavior, or anticonformity, has also been explored, yielding effects similar to those induced by random changes~\cite{Sznajd_anti, q-voter_noise}. Additionally, aging, defined as a resistance of the agent to change state the longer is the persistence in that state~\cite{Stark:2008,aging_juan}, has also been shown to represent a form of nonlinearity~\cite{nonmarkovian_to_markovian}.

In this work, we present a unified framework for classifying the universality classes of phase transitions in a broad family of nonlinear voter models, both with and without noise, a summary of these models is given in Table~\ref{tab:models_summary}. We begin by reviewing the canonical model proposed in Ref.~\cite{Hammal} for systems with two symmetric absorbing states, and we use it to classify the transitions of several nonlinear voter models. We then extend this canonical mean-field formulation by incorporating noise, which eliminates the absorbing states. This generalization unveils a phase diagram featuring two distinct transition lines: a continuous Ising transition and a discontinuous transition we term Modified Generalized Voter (MGV), which converge at a tricritical point. This novel model serves as a common framework for analyzing noisy nonlinear voter models.
We further investigate the effect of finite-size fluctuations on the phase transitions observed in the mean-field limit. By applying finite-size scaling techniques on complete graphs, random networks, and regular lattices, we confirm that the continuous transitions in all noisy nonlinear models studied belong to the Ising universality class in their respective dimensionality. We also observe tricritical points exhibiting mean-field behavior. These results unify and extend previous findings, offering a comprehensive characterization of noise-induced transitions in opinion dynamics.

The paper is outlined as follows: In Sec.~\ref{sec:Langevin_eq}, we review a general model that encompasses the universality classes of critical phenomena observed in systems with two absorbing states. In Sec.~\ref{sec:nonlinear_models}, we give a mean-field description of several nonlinear voter models studying the universality class of their transitions from the general model presented in Sec.~\ref{sec:Langevin_eq}. Sec.~\ref{sec:noisynonlinear_sec} is devoted to studying the critical phenomena of nonlinear models without absorbing states due to a noise contribution where absorbing states are destroyed by noise. In Sec.~\ref{sec:noisynonlinear_sec_ABC}, we introduce a canonical model which is an extension of the model studied in Sec.~\ref{sec:Langevin_eq} including noise effects. In Sec.~\ref{sec:noisy_models}, we use this model to study the universality classes of several noisy nonlinear voter models. Finite-size effects in some models are considered in Sec.~\ref{sec:finite}. In Sec.~\ref{sec:Universality_critical}, we verify the predictions made by the canonical model, applying finite-size scaling techniques to some noisy nonlinear voter models in complete graph, 2d-lattice and random networks. In Sec.~\ref{sec:Universality_tricritical} the universality class of the phase transition at the tricritical point is studied in both complete graph and random networks. Finally, some general conclusions are discussed in Sec.~\ref{sec:conclusions}.

\begin{figure*}[t]
 \centering 
 \includegraphics[width=\columnwidth]{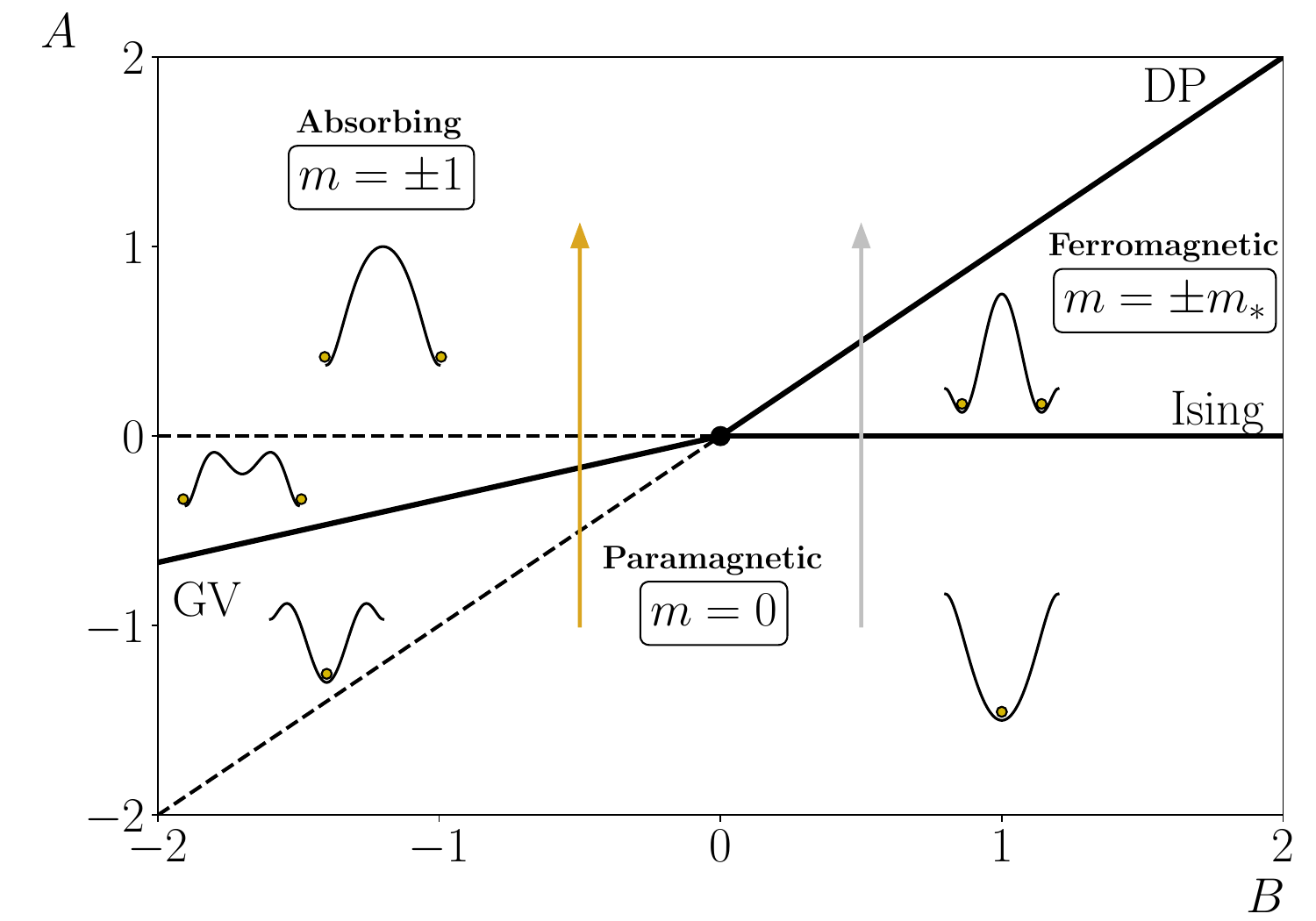}
 \includegraphics[width=\columnwidth]{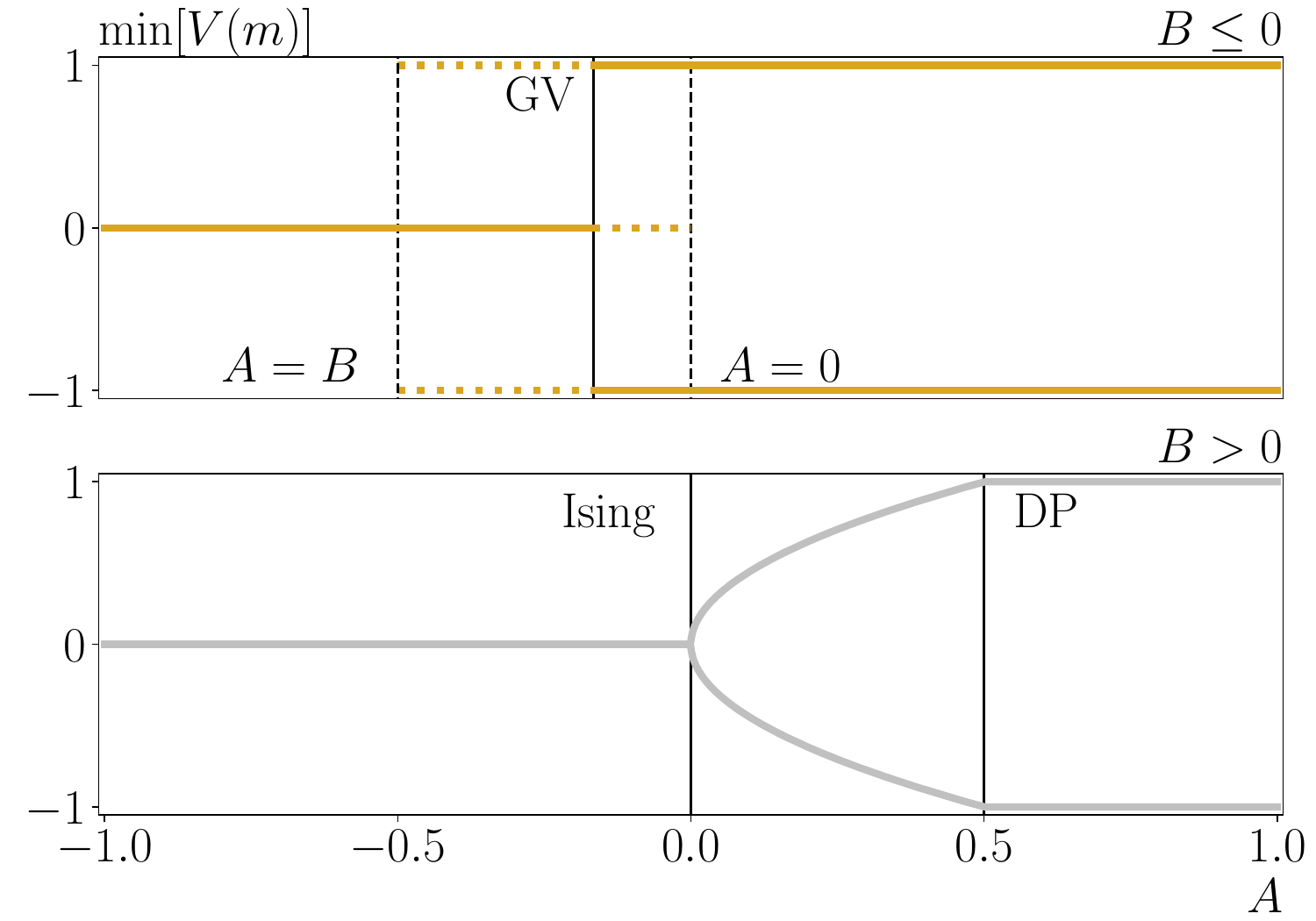}
 \caption{Left panel: Phase diagram in the space $(B,A)$ for the different regimes of the potential $V(m)$ sketched at each region. Transition lines: Ising, DP: Directed Percolation, GV: Generalized Voter. Dashed lines correspond to $A=0$ and $A=B$ delimiting the region for which the potential presents three minima. Right panel: Location of the minimum of the potential $V(m)$ versus the parameter $A$ for $B=-0.5$ (top) and $B=0.5$ (bottom) corresponding, respectively, to the gold and silver vertical arrows of the left panel. For $B\leq0$, solid (resp. dotted) line corresponds to the absolute (resp. relative) minimum of $V(m)$.}
 \label{fig:ab_diagram}
\end{figure*} 

\section{Phase transitions in nonlinear voter models}
In this section, we revisit some paradigmatic nonlinear voter models that exhibit $\mathbb{Z}_2$ symmetry—remaining invariant under a global sign flip of all variables—and admit two symmetric absorbing states. While we introduce these models in their general form, our analysis focuses exclusively on the mean-field limit, defined as the case where agents interact through all-to-all coupling (complete graph) in the thermodynamic limit $N \to \infty$. Within this framework, we establish an exact correspondence between these voter models and the canonical model introduced in Ref.~\cite{Hammal}. This correspondence provides a systematic basis for classifying the nature of the phase transitions exhibited by these systems.

\subsection{General mean-field formulation for systems with two absorbing states}
\label{sec:Langevin_eq}
In the mean-field limit, the canonical model of Ref.~\cite{Hammal} describes the dynamics of a scalar order parameter $m\in[-1,1]$, with the rate equation
\begin{equation}\label{eq:m_general}
 \frac{dm}{dt}=\left(Am-Bm^3\right)(1-m^2)= F(m),
\end{equation}
where $A$ and $B$ are free real parameters. The form of the equation clearly identifies the absorbing states at $m=\pm1$. This equation can be thought of as relaxational dynamics~\cite{SanMiguelToral:2000} in the potential 
\begin{equation}\label{eq:V_general}
 V(m)=-\frac A2 m^2 +\frac{A+B}{4} m^4 - \frac{B}{6}m^6,
\end{equation}
with the so-called drift $F(m)=-\dfrac{dV(m)}{dm}$. The dynamical system given by Eq.~\eqref{eq:m_general} presents up to five fixed points, namely $m=0$ (paramagnetic phase), $m=\pm1$ (absorbing phases), $m=\pm m_*\equiv\pm\sqrt{A/B}$ (ferromagnetic phases). Note that $m_*$ only makes physical sense if $0<A/B<1$, implying $\mathrm{sign}\,A=\mathrm{sign}\, B$. In the context of social dynamics, the paramagnetic phase is interpreted as a state of \textit{coexistence}, as both opinions appear equally in the population, while the absorbing phases are interpreted as \textit{consensus}, as the whole population holds the same opinion. The ferromagnetic phase corresponds to \textit{partial consensus}, an intermediate scenario in which a majority shares the same opinion.

Stable or unstable fixed points correspond, respectively, to minima or maxima of the potential $V(m)$. Equilibrium values of the order parameter correspond to absolute minima of the potential $V(m)$, while relative minima are identified as metastable states. By changing the system parameters $A$ and $B$ it is possible to transit from one equilibrium state to another. As discussed in Refs.~\cite{Hammal,qvoter}, the following scenarios are possible (see Fig.~\ref{fig:ab_diagram} for the corresponding phase diagram):
\begin{itemize}
    \item $B > 0$: When $A < 0$, the system is in the paramagnetic phase, where the equilibrium state $m = 0$ is the sole minimum of $V(m)$. At $A = 0$, the fixed point $m = 0$ becomes a maximum, and two degenerate minima emerge at $\pm m_*$. This indicates a continuous Ising transition, as the system shifts from the paramagnetic state ($m = 0$) to one of the two ferromagnetic states ($\pm m_*$), spontaneously breaking the $\mathbb{Z}_2$ symmetry. As $A$ increases further, the stable states $\pm m_*$ approach continuously the absorbing states $m = \pm 1$ at $A = B$, where the system undergoes a directed-percolation (DP) transition from the ferromagnetic phase to the absorbing phase. 
    \item $B \leq 0$: When $A < B$ the system is in the paramagnetic phase with $m = 0$ being the only minimum of $V(m)$. At $A = B$, two new minima appear at $m = \pm 1$, corresponding to metastable states, while $m = 0$ remains the absolute minimum. At $A = B/3$ the global minimum of $V(m)$ abruptly shifts from $m = 0$ to $m = \pm 1$, making $m = 0$ metastable. This is the generalized voter (GV) transition, where the system moves from the paramagnetic phase to one of the absorbing states ($m = \pm 1$), spontaneously breaking the $\mathbb{Z}_2$ symmetry. For $A > B/3$ no new phases emerge, but at $A = 0$, $m = 0$ becomes a maximum. 

The GV transition can be understood as the superposition of the two competing phenomena present in the voter-like models: the symmetry-breaking Ising transition and the possibility of the system to get trapped in an absorbing state (DP transition). The specific point $A = B = 0$, where the Ising, the GV and DP transition lines converge, corresponds to the classical voter model, in which the order parameter is conserved, $ \displaystyle\frac{dm}{dt}=0$.
\end{itemize}

In the following subsection, we prove the equivalence between several nonlinear voter models and the canonical form given in Eq.~\eqref{eq:m_general}. This equivalence enables a systematic characterization of the phase transitions these models can exhibit.

\subsection{Voter models with nonlinearities} \label{sec:nonlinear_models}
The models we present in this subsection consider a set of $N $ agents connected by links. Each agent $i $ holds a binary state variable $s_i=\pm1 $ and neighboring agents interact with each other taking into account their states. The interaction is such that a particular agent $i=1,\dots,N$, chosen at random from the set of all agents, can switch its value from $s_i=+1 $ to $s_i=-1$, or vice versa, with a probability that depends on the fraction of agents in its neighborhood that hold the opposite state. The different models differ in the functional form of the switching probabilities and may depend on node attributes such as age. Additionally, one needs to define what constitutes an agent's neighborhood, ranging from the set of nearest-neighbors in a regular lattice, to the whole set of agents (all-to-all interactions) in a complete graph, with different network configurations in between. In this subsection, we limit ourselves to the mean-field limit. All models share the feature of having two absorbing configurations, in which all agents are in the same state, either $+1 $ or $-1$, from which the escape probability is zero. The macroscopic state of the systems is characterized by the magnetization, defined as $m = \frac{1}{N} \sum_i s_i$.
\subsubsection{Nonlinear voter model}
As a first example of these models, we review the \textit{nonlinear voter model} (\nl VM)~\cite{lucia_NLVM}. This nonlinear version considers that the probability for an agent to change state is equal to a positive power $\alpha $ of the fraction of neighbor agents in the opposite state. The standard (linear) voter model corresponds to $\alpha=1$. If $\alpha>1$, agents are more resistant than in the linear case to adopt the opinions of their neighbors holding the opposite state. Consequently, a larger proportion of neighbors in the opposite state is required for an agent to change its state compared with the linear voter model counterpart. On the other hand, for $\alpha<1$, agents are more inclined to follow the opinions of their neighbors than in the linear case. As a result, even a small fraction of neighbors in the opposite state is likely to make the agent change its state. Independently of the value of $\alpha$, the consensus states $s_i=+1 $ or $s_i=-1,\,\forall i $ are absorbing.

From the transition probabilities, one can derive in the mean-field limit the following rate equation for the magnetization~\cite{lucia_NLVM},
\begin{equation}\label{eq:rateeq_NLVM}
\frac{dm}{dt}=2^{-\alpha}(1-m^2)\left[(1+m)^{\alpha-1}-(1-m)^{\alpha-1}\right].
\end{equation}
This is also the rate equation of the Abrams-Strogatz model for the dynamics of competing languages with the same prestige~\cite{AS_model}.
A series expansion of the previous equation up to order $O(m^5)$~\cite{Fede_Clopez_AB} allows us to identify the parameters $A,B $ of the canonical model of Eq.~\eqref{eq:m_general} as
\begin{subequations}\label{eq:NLVM} 
\begin{align} 
\label{eq:NLVM_A} A&=2^{1-\alpha}(\alpha-1),\\ 
 \label{eq:NLVM_B} B&=-\frac{2^{-\alpha}}{3}(\alpha-1)(\alpha-2)(\alpha-3).
\end{align} 
\end{subequations}
The standard voter model is recovered for the linear functional dependence $\alpha=1$, for which $A=B=0$. Note that the same null $A,\,B $ values are obtained for $\alpha\to\infty$, a limit in which the switching probability is zero unless all neighbors are in the opposite state and, hence, the system does not evolve dynamically and remains frozen in the initial condition.

\begin{figure}[t] \centering
 \includegraphics[width=\columnwidth]{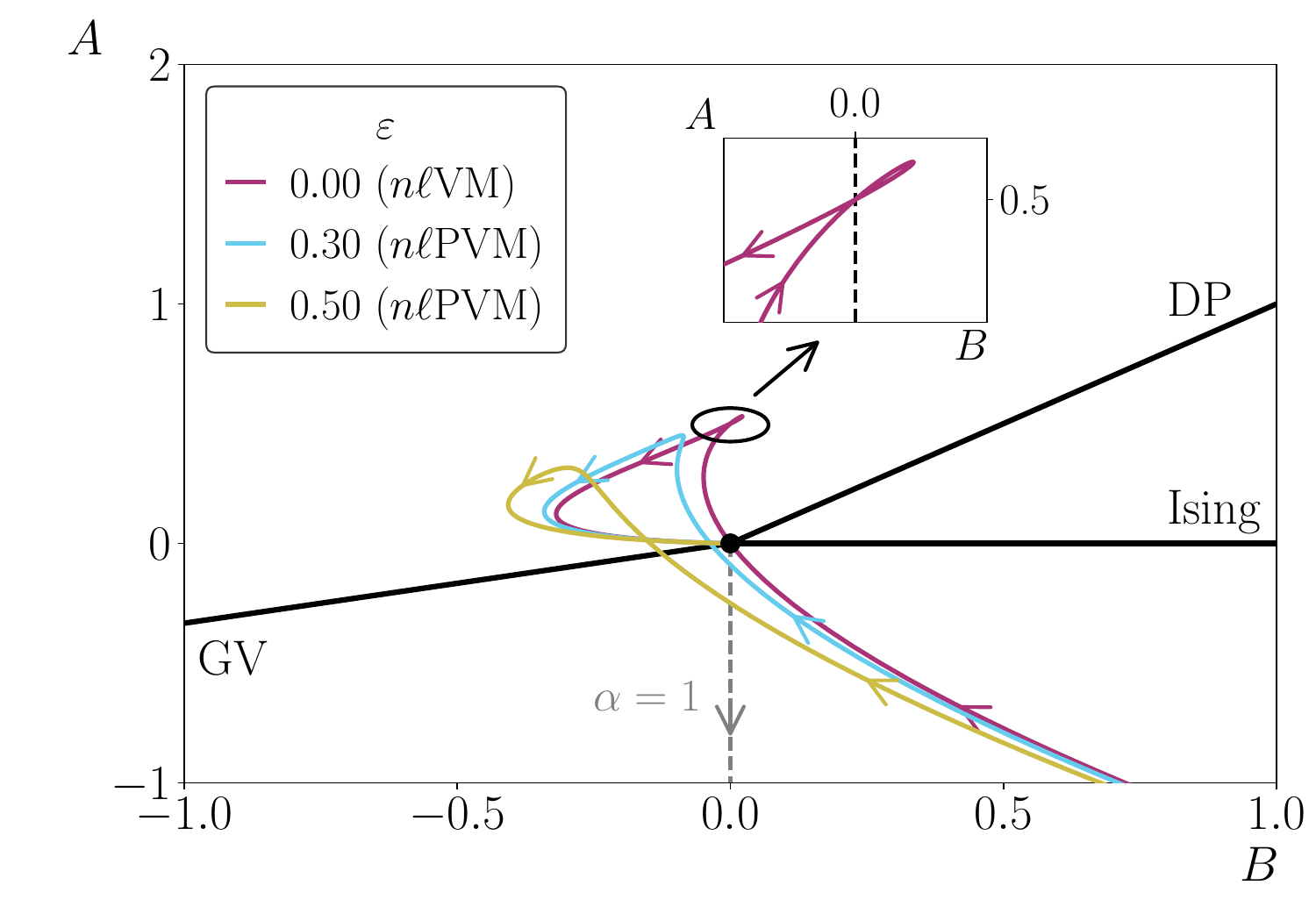}
 \caption{Paths in the parameter space $(B,A) $ for the nonlinear voter model (\nl VM, $\varepsilon=0$) and the nonlinear partisan voter model (\nl PVM, $\varepsilon>0$). Paths are obtained with Eqs.~(\ref{eq:NLPVM}), which reduce to Eqs.~(\ref{eq:NLVM}) for $\varepsilon=0$, varying $\alpha $ for several values of the preference $\varepsilon$, as indicated in the legend. The vertical grey dashed line is generated fixing $\alpha=1 $ and increasing $\varepsilon$, namely $A=-\varepsilon,\,B=0$. Inset: Zoom of the path for $\varepsilon=0$. Intersections of the path with the vertical axis $B = 0$, occurring at $A=1/2, B=0$.}
 \label{fig:ab_diagram_NLPVM}
\end{figure}

In Fig.~\ref{fig:ab_diagram_NLPVM}, we display the parametric curve (path) generated in the parameter space $(B,A)$ as given by Eqs.~(\ref{eq:NLVM_A},\ref{eq:NLVM_B}), as a result of a variation of $\alpha$. Starting from the paramagnetic region, $B=-A=2$ for $\alpha=0$, the system moves towards the absorbing region, crossing the VM point ($A=B=0$) at the critical value $\alpha_\text{c}=1$. It then traces a loop in parameter space before returning to the VM point in the limit $\alpha\to\infty$. For $\alpha<1$, the stable solution is $m=0 $ while for $\alpha>1$, the stable solutions are $m=\pm1$. The transition belongs to the generalized-voter class since the equilibrium value of $m$ changes abruptly from $m=0 $ to $m=\pm1$. Additionally, we observe that both $\alpha=2,3$ correspond to the same location in the parameter space $(B=0,A=1/2)$, resulting in the same rate equation for the magnetization, see inset.

\subsubsection{Nonlinear partisan voter model} \label{sec:nlPVM}
As another variation of the voter model, we introduce the \textit{nonlinear partisan voter model} (\nl PVM), the nonlinear version of the \textit{partisan voter model} (PVM)~\cite{Masuda_2010, Masuda_2011,llabres}. The PVM is an extension of the voter model in which each agent has an innate and fixed preference for one of the two states, as measured by the parameter $\varepsilon \in[0,1]$. Throughout this work, we focus on the case in which exactly half of the population prefer state $+1 $ and the other half prefer state $-1$. 

\begin{figure}[t]
 \centering 
 \includegraphics[width=\columnwidth]{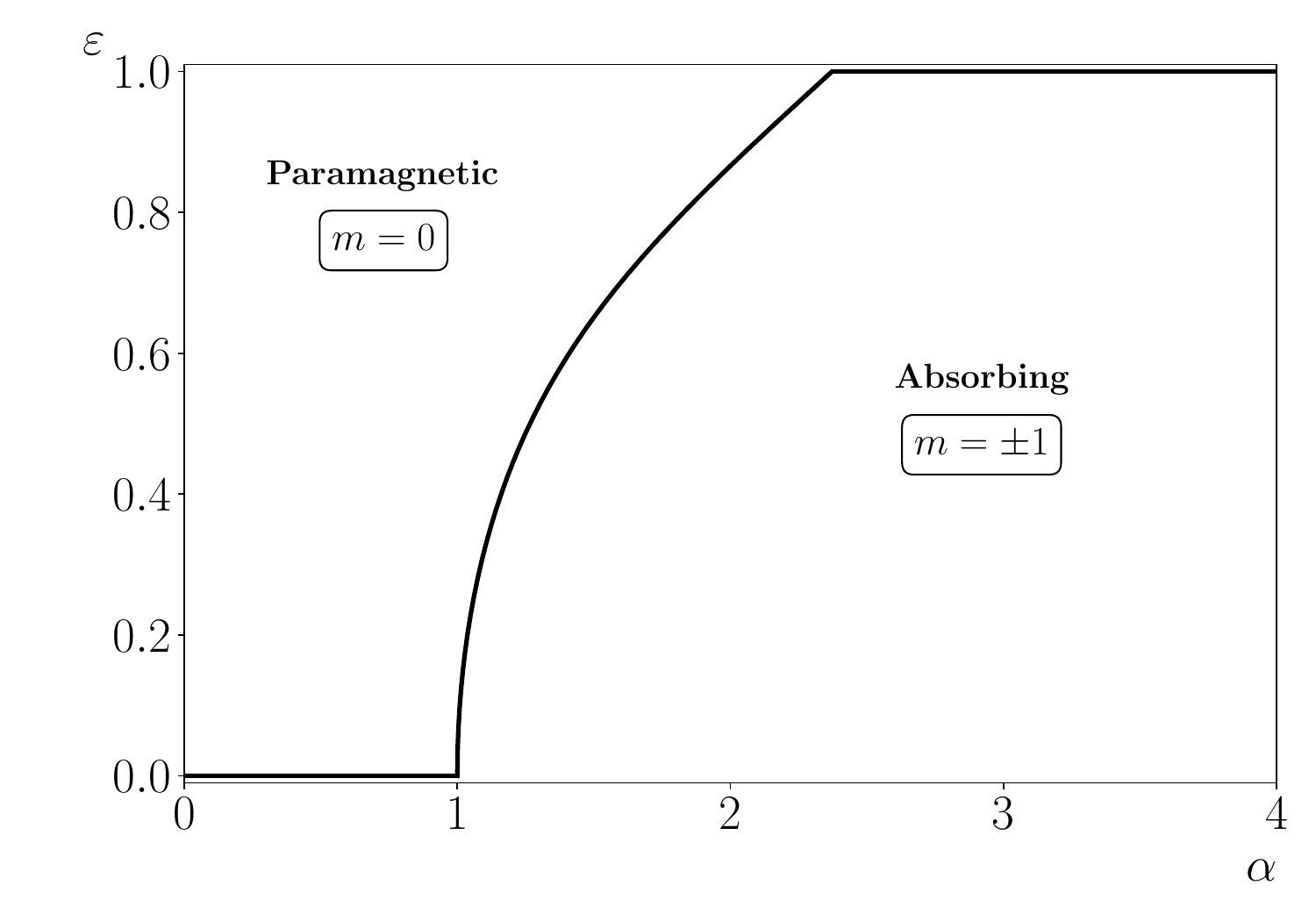}
 \caption{Phase diagram in the parameter space $(\alpha,\varepsilon)$ for the nonlinear partisan voter model. Transition line correspond to the GV transition given by Eq.~\eqref{eq:NLPVM_epsilonc}.}
 \label{fig:NLPVM_ae_diagram}
\end{figure}

The updating rule of the \nl PVM is that a randomly selected agent switches its state with a probability equal to a power $\alpha $ of the fraction of neighbor agents in the opposite state, multiplied by a factor $\frac{1+\varepsilon}{2} $ if the final state is preferred by that agent, or a factor $\frac{1-\varepsilon}{2} $ otherwise. As detailed in Appendix~\ref{app:sec:NLNPVM}, a complete description of the global state of the system requires two macroscopic variables. For the case $\alpha=1$, it has been rigorously shown~\cite{llabres} that an adiabatic elimination provides an exact reduction to an effective one-variable description. As a further approximation, we extend this procedure
to $\alpha\neq 1$, which allows us to obtain the following rate equation for the
magnetization in the mean-field limit
 \begin{align} \label{eq:NLPVM_eqm}
 &\frac{dm}{dt}=2^{-\alpha}(1-m^2) \biggl[ \left[(1+m)^{\alpha-1}-(1-m)^{\alpha-1}\right] \\ 
 & + \frac{ \left[(1-m)^{\alpha }-(1+m)^{\alpha}\right] \left[(1+m)^{\alpha-1 }+ (1-m)^{\alpha-1}\right]}{(1+m)^{\alpha }+(1-m)^{\alpha }}\,\varepsilon ^2\biggr], \nonumber
\end{align}
where we have rescaled $t\to t/2$. This rescaling arises from the fact the rules of the PVM for $\varepsilon=0$ yield a probability $1/2$ of activating the herding mechanism. In this way, for vanishing preference $\varepsilon=0$, we recover the rate equation of the \nl VM, Eq.~\eqref{eq:rateeq_NLVM}. Nevertheless, this overall factor is irrelevant for the study of the fixed points. Eq.~\eqref{eq:NLPVM_eqm} can be expanded up to order $O(m^5) $ in the form of Eq.~\eqref{eq:m_general} yielding
\begin{subequations}\label{eq:NLPVM}
\begin{align} \label{eq:NLPVM_A}
 A&=2^{1-\alpha}\left[\alpha-1-\varepsilon^2\alpha\right],\\ \label{eq:NLPVM_B}
 B&=-\frac{2^{-\alpha}}{3}(\alpha-1)\left[(\alpha-2)(\alpha-3)-\varepsilon^2\alpha(\alpha-8)\right].
\end{align}
\end{subequations}
Paths in parameter space $(B,A)$ varying $\alpha $ for different values of $\varepsilon $ are plotted in Fig.~\ref{fig:ab_diagram_NLPVM}. For fixed $\varepsilon>0$, a GV transition occurs varying $\alpha $ at the critical value $\alpha_\text{c}(\varepsilon)$, such that for $\alpha<\alpha_\text{c}(\varepsilon)$, $m=0 $ is the stable solution while $m=\pm1 $ is stable if $\alpha>\alpha_\text{c}(\varepsilon)$. The relation between $\alpha_\text{c}(\varepsilon)$ and $\varepsilon$ is obtained from the general condition $A=B/3$ as \begin{equation} \label{eq:NLPVM_epsilonc}
\varepsilon(\alpha_\text{c})=\sqrt{\frac{\alpha_\text{c} -1}{\alpha_\text{c}} \cdot\frac{\alpha_\text{c}(\alpha_\text{c} -5) +24}{\alpha_\text{c}(\alpha_\text{c} -9) +26}}.
\end{equation}
This expression defines the transition line that separates the paramagnetic and absorbing phases in the parameter space $(\alpha,\varepsilon)$, see Fig.~\ref{fig:NLPVM_ae_diagram}. The inversion of Eq.~\eqref{eq:NLPVM_epsilonc} to obtain an explicit expression for $\alpha_\text{c}(\varepsilon)$ results in an expression that is too cumbersome to be of practical use, but we can expand it in a power-series of $\varepsilon $ around $\varepsilon=0$, namely
\begin{equation} \label{eq:alpha_c}
 \alpha_\text{c}(\varepsilon)=1+\frac{9}{10} \varepsilon^2+\frac{1233}{2000} \varepsilon^4+O(\varepsilon^6),
\end{equation}
which is accurate to less than $0.5\% $ if $\varepsilon<0.8$. However, this transition line is based on an adiabatic elimination technique, whose validity may not extend to all values of $\alpha$ and $\varepsilon$. This is also evidenced by the fact that Eq.~\eqref{eq:NLPVM_epsilonc} predicts $\varepsilon>1$ for $\alpha>\frac{1}{2}(\sqrt{33}-1)\approx 2.372$. 

Furthermore, the solution $m = 0$ becomes unstable as the system crosses the line $A = 0$, which defines the transition line
\begin{equation} \label{eq:NLPVM_epsilonc_complete}
\alpha_\text{c}(\varepsilon) = \frac{1}{1 - \varepsilon^2},
\end{equation}
which coincides with the exact transition value obtained from the linear stability analysis of the fixed points of the complete two-variable model, given by Eqs.~(\ref{app:eq:nonlinearNPVM_rateeq_D},\ref{app:eq:nonlinearNPVM_rateeq_S}).

In the linear case $\alpha=1 $, corresponding to the standard PVM, the parameters take the values $B=0$ and $A=-\varepsilon^2$, and the system does not exhibit a transition for any value of $\varepsilon>0$. The stable fixed point is always $m=0$, as illustrated by the gray dashed line plotted in Fig.~\ref{fig:ab_diagram_NLPVM}. 

\subsubsection{Nonlinear $q$-voter model}\label{sec:q-voter}
The next example we consider is the \textit{nonlinear $q$-voter model} (\nl$q$VM)~\cite{qvoter}, a variation of the voter model in which an agent selects $q $ neighbors at random (repetitions are allowed). If all $q $ neighbors are in the same state, the agent adopts that common state. Otherwise, if the $q$ neighbors do not coincide, the agent switches state with probability $\epsilon$, a new parameter of the model, not to be confused with the preference parameter $\varepsilon$ of the \nl PVM. The rules are such that the consensus states $m=+1 $ or $m=-1 $ are absorbing. 

\begin{figure}[t]
 \centering
 \includegraphics[width=\columnwidth]{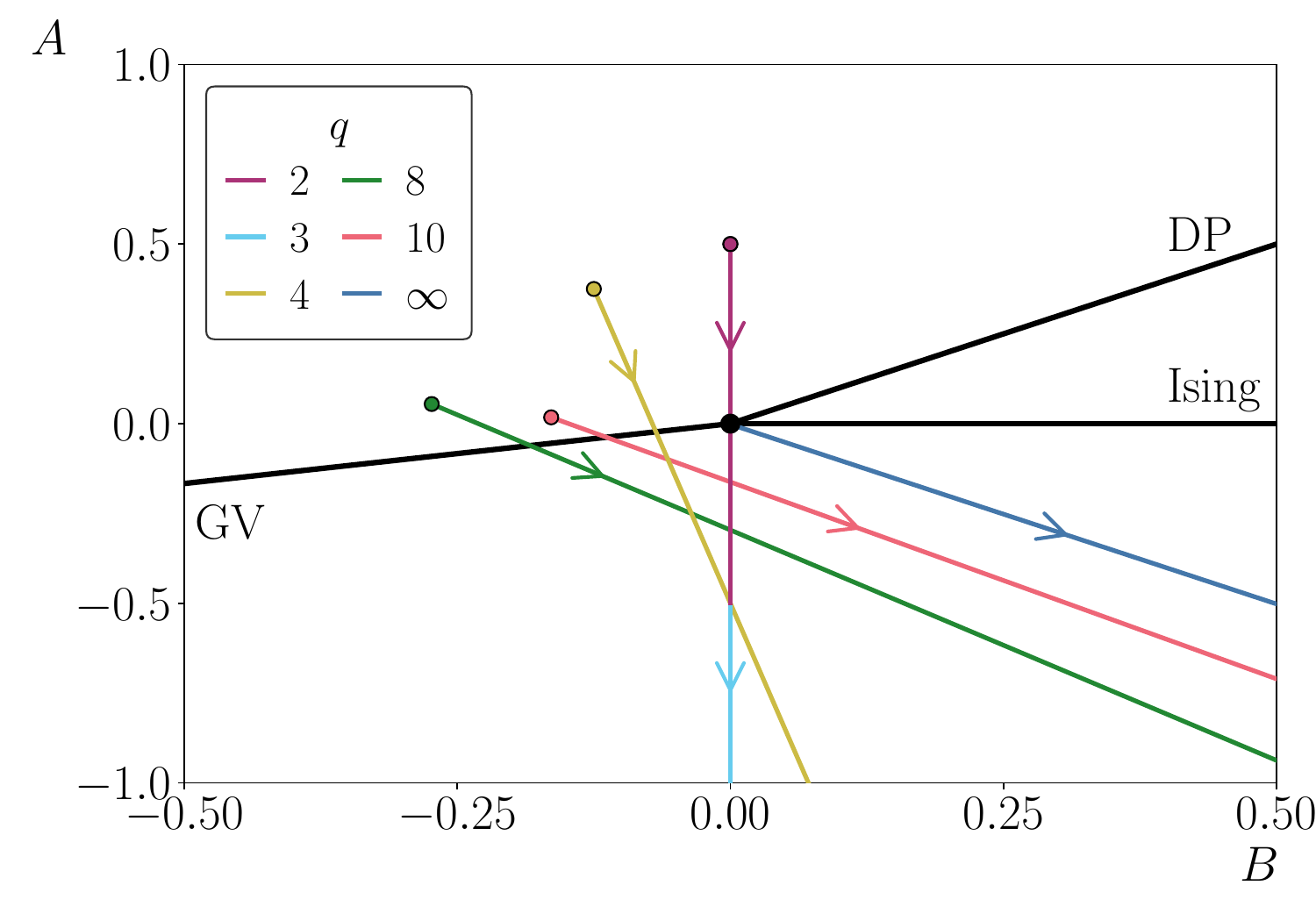}
 \caption{Paths in the parameter space $(B,A) $ for the $q$-voter model, generated from Eqs.~(\ref{eq:q-voter_A},\ref{eq:q-voter_B}) increasing the switching probability $\epsilon $, as indicated by the arrows, for several values of $q$. For $q=2,3 $ the paths are vertical at $B=0$, namely $A=\frac12-\epsilon $ for $q=2 $ and $A=\frac12-\frac{3}{2}\epsilon $ for $q=3$, and cross the VM point $A=B=0$.}
 \label{fig:ab_diagram_qvoter}
\end{figure}

In the mean-field limit, the rate equation for the magnetization in this model is~\cite{qvoter}
\begin{equation}
 \frac{dm}{dt}=\frac{1-m}{2}\,f\left(\frac{1+m}{2}\right)-\frac{1+m}{2}\,f\left(\frac{1-m}{2}\right)
\end{equation}
with $f(x)=x^q+\epsilon \left(1-x^q-(1-x)^q\right)$. A series expansion up to order $O(m^5) $ adopts the form of Eq.~\eqref{eq:m_general} with~\cite{qvoter}
\begin{subequations}\label{eq:q-voter}
\begin{align} \label{eq:q-voter_A}
 A=&2^{1-q}(q-1)-2\epsilon(1-2^{1-q}),\\
 B=&-\frac{2^{-q}}{3}(q-1)(q-2)\left(q-3\right) \nonumber\\ \label{eq:q-voter_B}
 &+2\epsilon[1-2^{-q}(2-q+q^2)].
\end{align}
\end{subequations}
Note that the nonlinear $q$-voter model with $\epsilon=0 $ is equivalent to a nonlinear voter model with $\alpha =q$, an integer value. In Fig.~\ref{fig:ab_diagram_qvoter}, we plot the paths generated in the parameter space $(B,A)$ generated from Eqs.~\eqref{eq:q-voter} by varying $\epsilon $ for several values of $q$. The case $q=1$ corresponds to the voter model, $A=B=0 $ for any value of $\epsilon$. For $q>1$, the system exhibits a GV transition at the critical value
\begin{equation} \label{eq:qvoter_epsilonc}
 \epsilon_\text{c}(q)=\frac{(q-1)(q^2-5 q+24)}{6 \left(-q^2+q+2^{q+2}-8\right)},
\end{equation}
obtained by setting $A=B/3$ in Eqs.\eqref{eq:q-voter}. The transition occurs at the VM point $A=B=0 $ for $q=2,3 $ while it crosses the GV line at negative values of $A,\,B $ for $q\ge 4$. The critical value $\epsilon_c $ tends to zero as $q $ increases. For $q\to\infty$, the imitation mechanism is not active and $A=-B=-2\epsilon $. In this situation, for $\epsilon>0$, random state changes dominate the dynamics leading to a coexistence situation where $m=0 $ is the stable solution, whereas for $\epsilon=0 $ the system is frozen in the initial condition.

\subsubsection{Voter model with aging} \label{sec:vm_aging}
Other possible models are those that allow agents to have memory effects. This is the case of the voter model with \textit{aging}~\cite{aging_juan}, where each agent $i$ holds, in addition to its state $s_i$, an internal time variable $\tau_i=0,1,2,\dots $ understood as the number of successive unsuccessful update attempts, those that did not result in a change of the state variable $s_i$. To be more precise, after setting initially all internal times to $\tau_i=0$, the updating rules of the voter model are modified in the following way: a randomly selected agent $i$ first chooses one of its neighbors at random and then copies its state with probability $p(\tau_i)$. If, as a result of this interaction, agent $i$ changes state $s_i\to -s_i$, then its internal time is reset to zero $\tau_i\to0$, otherwise it is increased in one unit $\tau_i\to\tau_i+1$. Among the functional forms of the probability $p(\tau) $ that have been previously considered, we choose the following~\cite{vm_aging}:
\begin{equation} \label{eq:p_tau_aging}
p(\tau)=\frac{p_\infty\tau+p_0\tau^*}{\tau+\tau^*},
\end{equation}
where $\tau^*>0 $ controls the rate of change of the activation probability as the age increases, i.e., the larger $\tau^*$, the slower the decay and $p_0,p_\infty\in[0,1] $ correspond to the asymptotic values for $\tau=0 $ and $\tau\to\infty$, respectively. If $p_\infty<p_0$, the probability decreases with the age $\tau$. This is the typical aging situation. On the other hand, for $p_\infty>p_0 $ the probability increases with age, i.e. the longer an agent is in the same state, the more likely it is to change its state. This is known as anti-aging~\cite{vm_aging}.

\begin{figure}[t]
 \centering
 \includegraphics[width=\columnwidth]{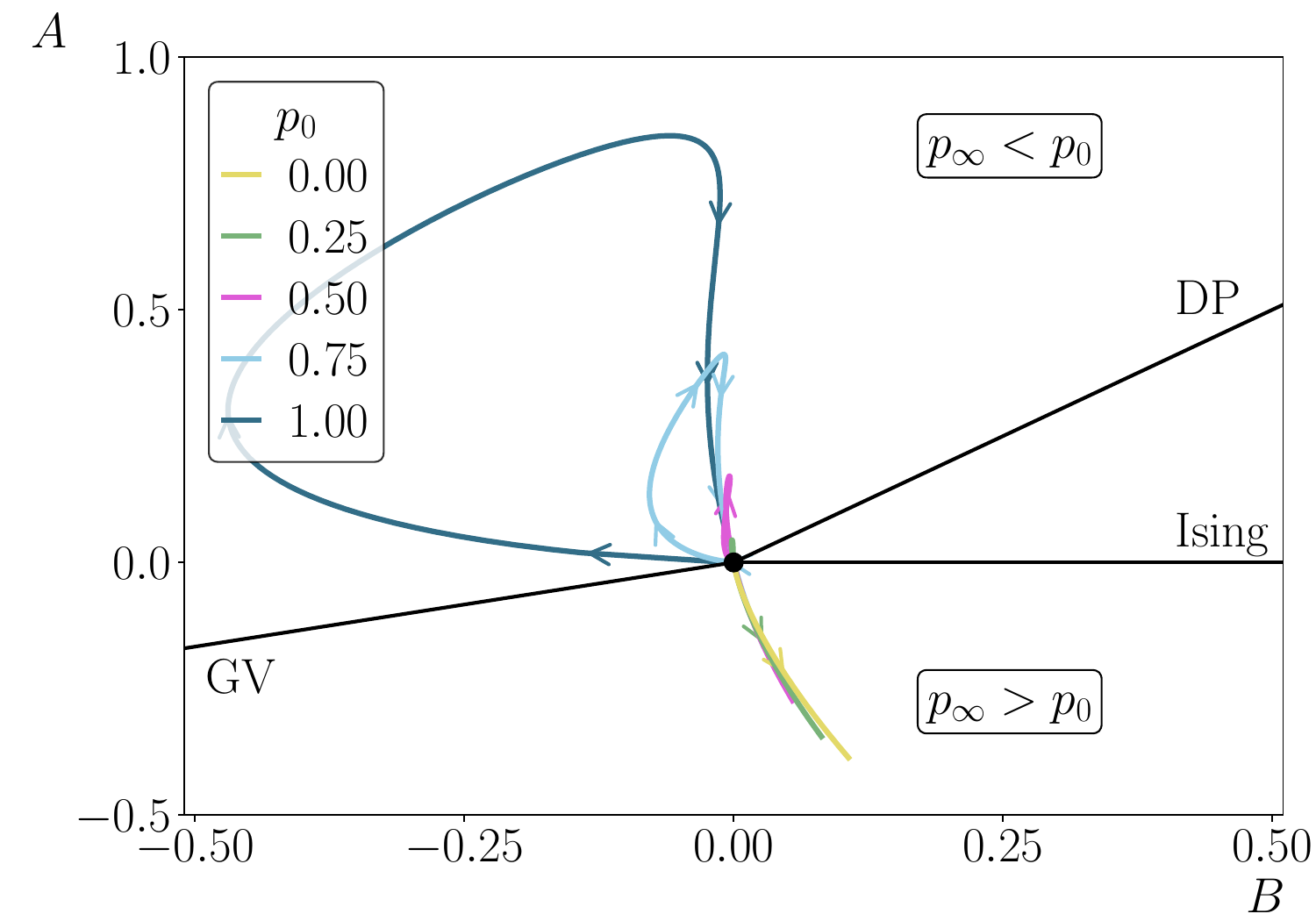}
 \caption{Paths in the parameter diagram $(B,A)$ for the voter model with aging, obtained by increasing $p_\infty $ as indicated by the arrows, for $\tau^*=1 $ and several values of $p_0$. For $p_\infty>p_0$, all curves nearly overlap.}
 \label{fig:ab_diagram_NLPVM_aging}
\end{figure}

This model is non-Markovian if one considers only the set of binary states $\{s_i\}$, since the transition rates depend on the agents' ages $\{\tau_i\}$ and hence on their history. However, the process becomes Markovian when formulated in the extended state space $\{s_i,\tau_i\}$~\cite{vm_aging}. Additionally, an adiabatic approximation technique allows us to obtain an equivalent Markovian model for the set states $\{s_i\}$ with nonlinear effective rates~\cite{nonmarkovian_to_markovian, aging_Jaume_Sara}. In this way, it is possible to write down in the mean-field limit the rate equation for the magnetization as
\begin{equation} \label{eq:m_aging}
 \frac{dm}{dt}=\frac{1-m^2}{2}\left[\Phi_0\left(\frac{1+m}{2}\right)-\Phi_0\left(\frac{1-m}{2}\right)\right],
\end{equation}
where the explicit expression of $\Phi_0(x) $ is given in Appendix~\ref{app:vm_aging}. A series expansion up to order $O(m^3)$ around $m=0$ of $\Phi_0(x)$ allow us to compute numerically the parameters $A,B$. Fig.~\ref{fig:ab_diagram_NLPVM_aging} shows paths generated in the parameter space $(B,A)$ by varying $p_\infty$, for fixed $\tau^*=1 $ and several values of $p_0$. Irrespective of the value of $\tau*$, the system exhibits a GV transition from a paramagnetic to an absorbing phase through the VM point at $p_\infty=p_0$. The paramagnetic phase is stable for $p_\infty>p_0$ (anti-aging), while the absorbing phase is stable for $0<p_\infty<p_0$ (aging)~\footnote{The case $p_\infty=0$ corresponds to a critical functional form for which the system orders, as demonstrated in Ref.~\cite{vm_aging}. However, the canonical model fails to capture this behavior, yielding $A=B=0$. It is expected that higher-order terms in Eq.\eqref{eq:m_general} are required to account for this phenomenon.}.

\subsubsection{Other models}
Other variations of the voter model have been proposed, such as that of the \textit{vacillating} voter model~\cite{Lambiotte_2007}, where agents are doubtful about their state. This model is very similar to the $q$-voter model with $q=2 $ and $\epsilon=1$, exhibiting a stable coexistence paramagnetic configuration characterized by the canonical parameters $A=-1, B=0 $, with no possible transitions.

The case of \textit{non-conservative voters}~\cite{Lambiotte_2008} is a more general model that exhibits richer scenarios. It is defined on a one-dimensional lattice and characterized by the \textit{conviction} parameter $\gamma = p_2/p_1$, the ratio between the probabilities of switching state when having two and one neighbors in opposite state. For $\gamma=1$, one recovers the \textit{vacillating} voter model. The authors also analyzed the model within a mean-field approximation, where the evolution of the global magnetization can be expressed in terms of the canonical parameters $A=2(\gamma-2)$ and $B=0$. In this limit, there exists a GV transition from the paramagnetic $m=0 $ ($\gamma<2$) to the absorbing phase $m=\pm1$ ($\gamma>2$) through the VM transition point $A=B=0$ for $\gamma=2$.

The Sznajd model~\cite{Sznajd} describes a scenario where a pair of agents sharing the same opinion can influence the opinion of a third. On the complete graph, the model corresponds to the nonlinear $q$-voter model for $\epsilon=0$ with no possible repetitions in the choice of $q=2$ neighbors. In the mean-field limit, the canonical parameters are given by $A=1/2$, $B=0$~\cite{Sznajd_MF}, indicating that the system always reaches an absorbing state. 

Another interesting example is the generalized voter model with $S$ intermediate states, which is equivalent to the two-state voter model with effective nonlinear transition rates. As shown in Ref.~\cite{DallAsta_2008}, these models are characterized by the canonical parameters $A=c/2$, $B=0$, where $c\in(0,1)$, demonstrating that the system always falls in an absorbing state.

As conjectured in Ref.~\cite{Chate2d}, the GV transitions are \enquote{order-disorder transitions driven by the interfacial noise between two absorbing states possessing equivalent `dynamical roles', this symmetry being enforced either by $\mathbb{Z}_2$ symmetry of the local rules, or by the global conservation of the magnetization}. Since the updating rules of the nonlinear voter models preserve the $\mathbb{Z}_2$ symmetry—even though they do not conserve magnetization—it is not surprising that they all consistently exhibit a GV transition.

Note, however, that there exist other models that do not fit into this general framework. One such example is the \textit{confident voter model}~\cite{confident_voters}, where agents possess a level of commitment to an opinion, being either confident or unsure, resulting in a non-scalar order parameter. Similarly, there exist \textit{layered models} such as the \textit{concealed voter model}~\cite{Gastner_2018} and the \textit{q-voter model with public and private opinions}~\cite{Kasia_layer}, where each agent is assigned two coupled opinions: an \textit{external} (publicly expressed) one and an \textit{internal} (concealed or private) conviction. As a result of the interplay between the two layers, these models are characterized by a non-scalar order parameter. Another example is the \textit{noise-reduced voter model}~\cite{noise_reduced_voter_model}, which introduces memory effects differently from the VM with aging, leading to a crossover between voter-like and zero-temperature Glauber dynamics.

\begin{figure*}[t]
 \centering 
\includegraphics[width=\columnwidth]{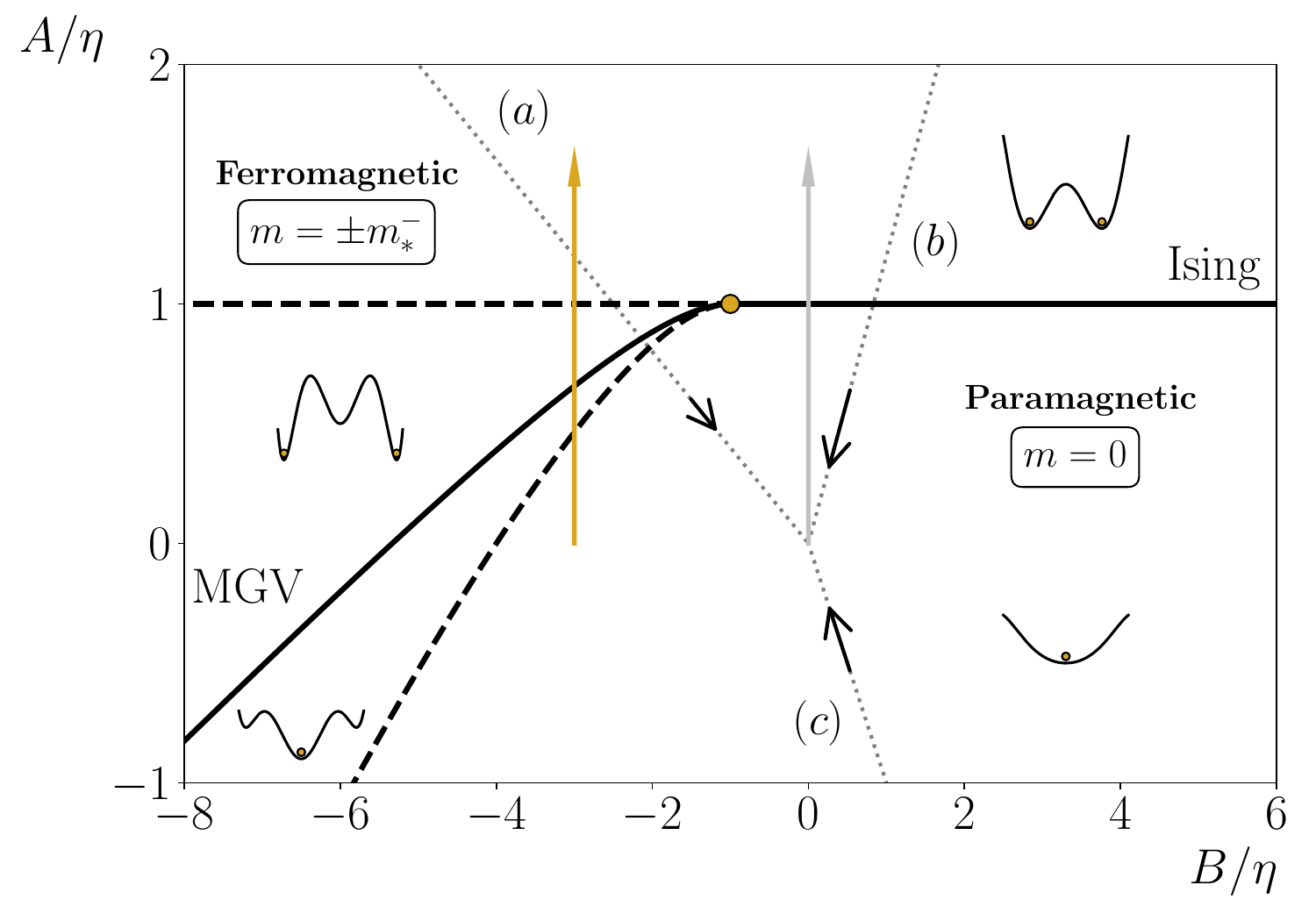}
 \includegraphics[width=\columnwidth]{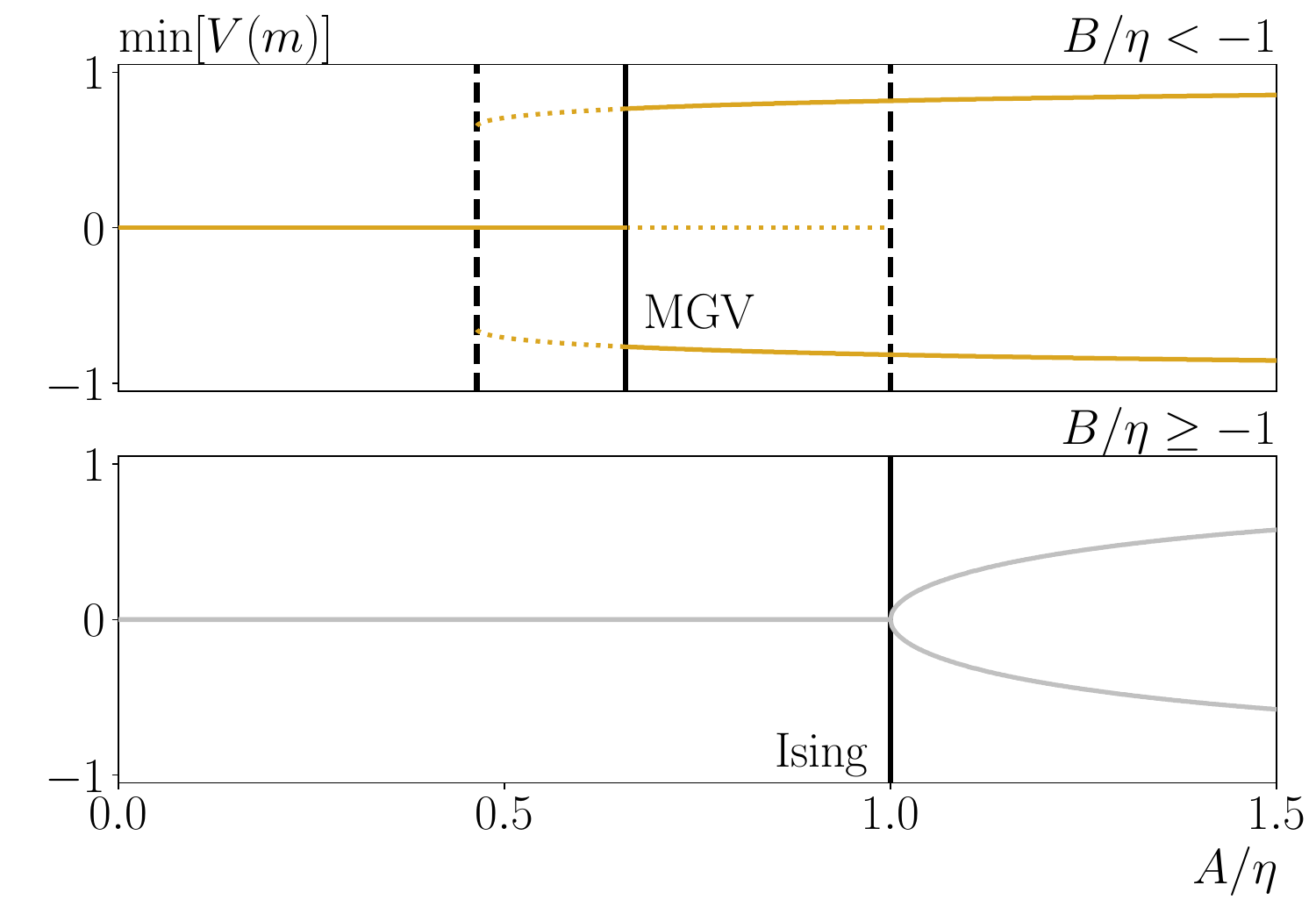}
 \caption{Left panel: Phase diagram in the parameter space $(B/\eta,A/\eta)$, valid only if $\eta>0$, for the different regimes of the potential $V(m)$ sketched at each region. Transition line: MGV: modified Generalized Voter. MGV transition line, as given by Eq.~\ref{eq:AB_eta}, corresponds to a first-order transition while the Ising transition line corresponds to a second-order transition given by $A/\eta=1$. Both lines meet at the tricritical point $(B/\eta,A/\eta)=(-1,1)$. Dashed lines correspond to $A/\eta=1$ and $A=B+2\sqrt{-B\eta}$ delimiting the region for which the potential presents three minima. The straight dotted lines are paths obtained at fixed values of $A $ and $B $ varying $\eta$ and approaching the origin as $\eta\to\infty$, as indicated by the arrows. Line (a) crosses the MGV transition line; line (b) crosses the Ising transition line; while line (c) always remains in the paramagnetic phase. Right panel: Location of the minimum of the potential $V(m)$ versus the parameter $A$ for $B/\eta=-3$ (top) and $B/\eta =0$ (bottom) corresponding to the gold and silver vertical arrows of the left panel, respectively. For $B/\eta<-1$, solid (resp. dashed) lines correspond to the absolute (resp. relative) minima of $V(m)$.}
 \label{fig:ab_eta_diagram}
\end{figure*}

\section{Phase transitions in noisy nonlinear voter models}
\label{sec:noisynonlinear_sec}
The models considered so far account only for herding behavior, i.e., agents change their state only due to interactions with their neighbors. A natural extension is to include random state changes independent of the state of the neighbors. These changes have been interpreted in various ways, including as manifestations of idiosyncrasy ~\cite{Kirman1993}. Following standard practice, and at risk of confusion with other common interpretations of the term, we will refer to these individual random changes as ``noise". One general effect of this noisy contributions is that they make the dynamics ergodic, allowing the system to reach any microscopic configuration from any other. Consequently, the consensus states are no longer absorbing. In this section, we focus on these noisy nonlinear voter models that lack absorbing states due to noise effects. Note however that there are nonlinear voter models with individual random changes that preserve the existence of absorbing states, as for example the \nl$q$VM presented in Sec.~\ref{sec:q-voter}. 

We follow the same structure as in the previous section, beginning with the introduction of a canonical model for systems that lack absorbing states due to the presence of noise and analyzing its associated phase diagram. Then, we examine specific examples by mapping the parameters of various models onto those of the canonical form. Our goal is to classify the types of phase transitions exhibited by a set of nonlinear noisy voter models.

\subsection{General mean-field formulation} \label{sec:noisynonlinear_sec_ABC}
We propose a canonical model for systems that no longer have two symmetric absorbing states due to a noise contribution. In the mean-field limit, this noise contribution manifests as a drift that drives the system toward a state of coexistence of the two opinions, $m=0$. The rate equation is 
\begin{equation}\label{eq:m_noisy_general}
 \frac{dm}{dt}=(Am-Bm^3)(1-m^2)-\eta \,m=F(m),
\end{equation}
where $\eta>0 $ modulates the intensity of the noise and $A,B$ are again two free parameters. The corresponding potential giving rise to this relaxation dynamics is
\begin{equation}
 V(m)=-\frac{A-\eta}{2} m^2 +\frac{A+B}{4} m^4 - \frac{B}{6}m^6,
\end{equation}
allowing us again to identify the equilibrium states as the absolute minima of the potential. In addition to the paramagnetic phase $m=0$, this potential can have up to four other real extrema: $\pm m_*^+$ and $\pm m_*^-$, where
\begin{equation}
 m_*^\pm = \sqrt{\frac{A + B \pm \sqrt{(A - B)^2 + 4B\eta}}{2B}}
\end{equation}
provided they exist, i.e., when they are real and the conditions $0 \le m_*^+ \le 1$ or $0 \le m_*^- \le 1$ are satisfied. Note that for $\eta > 0$, $m = \pm1$ are no longer fixed points of the dynamics. 

By analyzing the existence and stability of the solutions $\pm m_*^\pm$, it is possible to construct the phase diagram in the $(B,A)$ parameter space for a fixed value of $\eta$, which extends the one discussed in Sec.~\ref{sec:Langevin_eq}. However, in order to provide a unified description, we find it more convenient to present in Fig.~\ref{fig:ab_eta_diagram} the phase diagram in terms of the noise-rescaled parameters $(B/\eta,A/\eta)$.

The following scenarios are possible:
\begin{itemize}
\item $B>-\eta$: For $A<\eta$, the 
equilibrium state is given by $m=0$, the only minimum of the potential. Increasing the value of $A$, the system exhibits an Ising transition, changing at $A=\eta$ its equilibrium state continuously from the paramagnetic state $m=0$ to one of the two ferromagnetic solutions $\pm m_*^-$, spontaneously breaking the $\mathbb{Z}_2$ symmetry. In this case, a further increase of $A$ does not lead to a DP transition, since noise has destroyed the absorbing states.
\item $B<-\eta$: For $A< B + 2\sqrt{-B\,\eta}$, the system is in the paramagnetic phase, with $m=0$ the only minimum of the potential. When $A$ is increased beyond that value (lower dashed line in Fig.~\ref{fig:ab_eta_diagram}), the four extrema $ \pm m^{\pm}$ become real: two maxima located at $m=\pm m^+_*$ and two minima at $m=\pm m^-_*$ although $m=0$ is still the absolute minimum. A further increase of $A$ beyond the value
\begin{equation} \label{eq:AB_eta}
 A = \frac{1}{3} \left(5B+4 \sqrt{B(B -3 \eta)}\right)
\end{equation}
turns $\pm m^-_*$ into the absolute minima, inducing a first-order, discontinuous, transition from the paramagnetic $m=0$ to one of the ferromagnetic phases $m=\pm m^-_*\neq \pm 1$, breaking the $\mathbb{Z}_2$ symmetry. We refer to this transition as a \textit{Modified Generalized Voter} (MGV) transition. Note that Eq.~\eqref{eq:AB_eta} for $\eta=0$ ($B<0$) recovers the GV transition line $A=B/3$. Finally, the two maxima at $\pm m_*^{+}$ merge at $m=0$ for $A=\eta$ (upper dashed line in the figure) and a unique maximum exists at this point for $A>\eta$.
\end{itemize}

In summary, the introduction of noise eliminates the absorbing states, with deep consequences. The directed percolation (DP) transition disappears, and the resulting phase transitions are instead governed by $\mathbb{Z}_2$ symmetry breaking. These transitions can be either continuous (Ising-type) or discontinuous (MGV-type), merging at the tricritical point $A = -B = \eta$. 

Additionally, it is possible to study the possible transitions for fixed values of $A$ and $B$ varying the noise intensity $\eta$, see dotted lines in Fig.~\ref{fig:ab_eta_diagram}. An increase of $\eta$ moves the corresponding point in phase diagram asymptotically toward the origin. Consequently, the crossing of the transition lines can only occur if, in the absence of noise, the location in the phase diagram $(B,A)$ either belongs to the absorbing or the ferromagnetic phases of Fig.~\ref{fig:ab_diagram}. If, in the absence of noise, the parameters of the system satisfy $A(\eta=0)<-B(\eta=0)$, the system undergoes an MGV transition, while an Ising transition occurs otherwise. This means that an increase of noise will always induce an Ising transition if the system started in the ferromagnetic phase, while absorbing states can undergo Ising or MGV transitions depending on the exact location in phase diagram.

\begin{figure*}[t]
 \centering
 \includegraphics[width=\columnwidth]{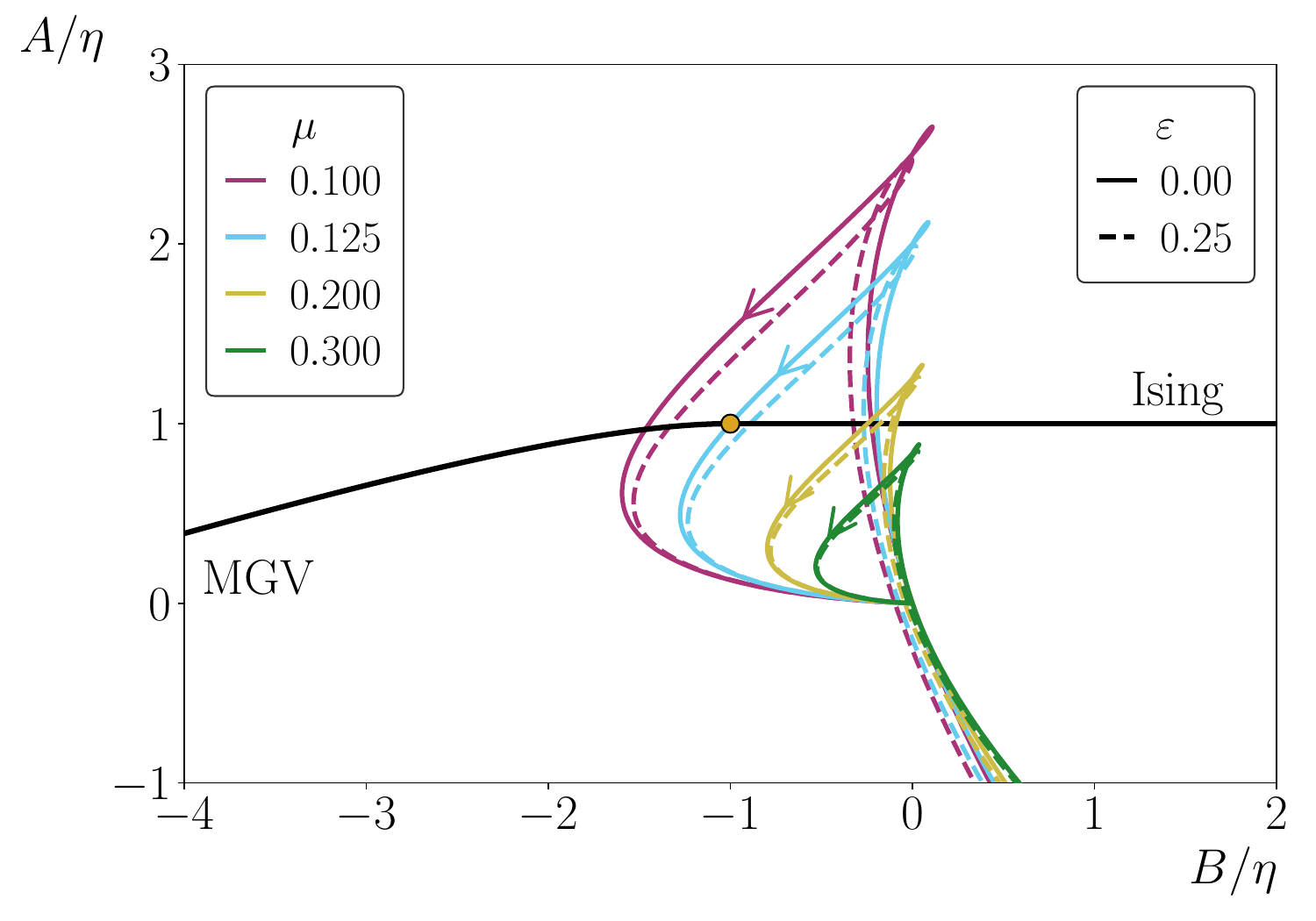}
 \includegraphics[width=\columnwidth]{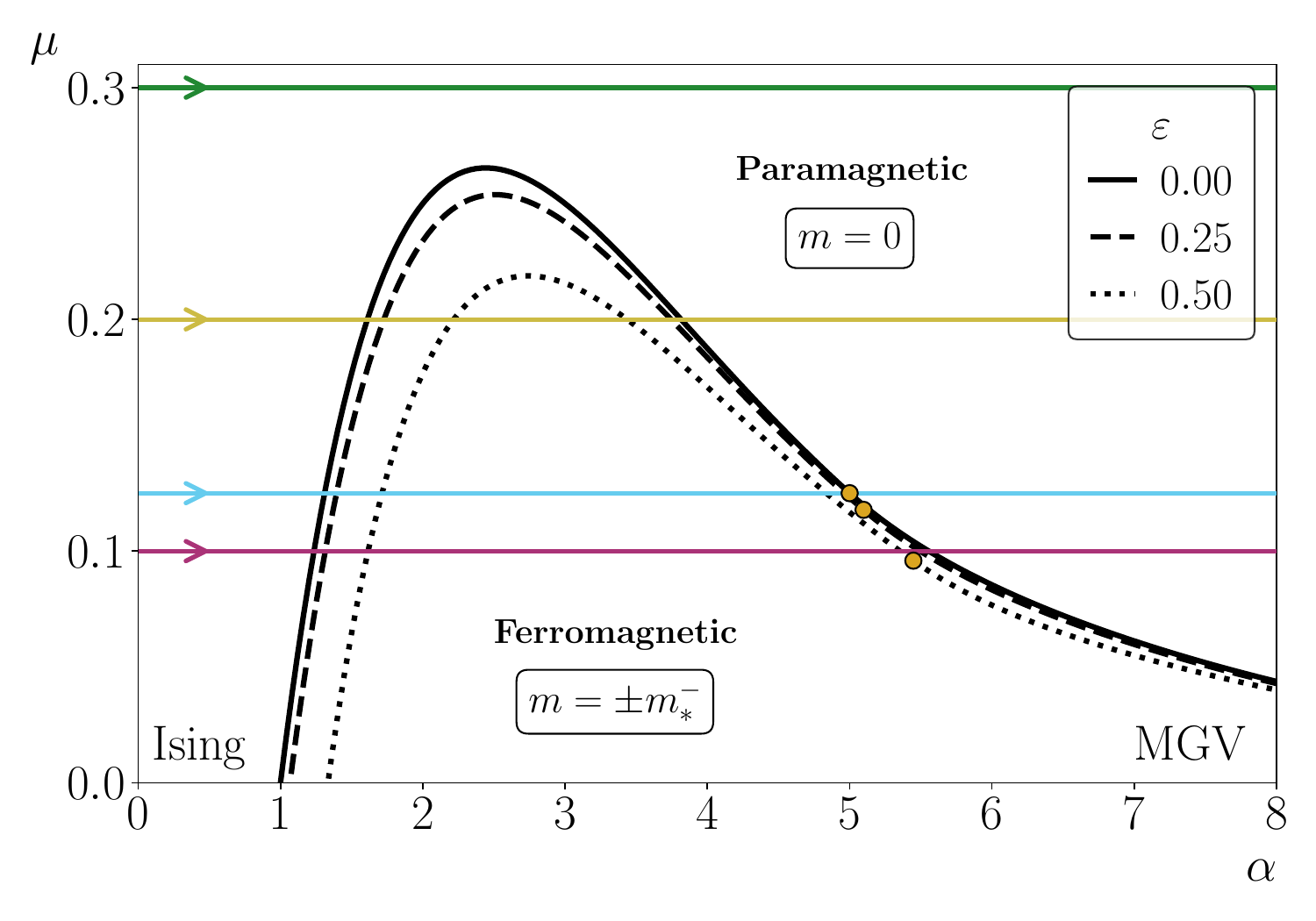}
 \caption{Left panel: Paths in the parameter space $(B/\eta, A/\eta)$ for the nonlinear noisy voter model ($\varepsilon=0$) and the nonlinear partisan noisy voter model ($\varepsilon>0$), obtained by varying the value of $\alpha$ for several fixed values of $\varepsilon, \mu$, as indicated in the legend. Lines are generated using Eqs.~\eqref{eq:NLNPVM_parameters}, which reduce to Eqs.~\eqref{eq:NLNVM_parameters} for $\varepsilon=0$ illustrating the distinct cases discussed in the main text. Arrow indicates the direction of the path as $\alpha$ increases. Transition line: MGV: Modified Generalized Voter. The dot corresponds to the tricritical point where the MGV and the Ising transition lines merge. Right panel: Phase diagram in the parameter space $(\alpha,\mu)$. Ising and MGV transition lines for $\varepsilon=0$ are given by Eqs.~(\ref{eq:NLNVM_ac_2nd_order}, \ref{eq:NLNVM_ac_1st_order}), respectively. Different line styles denote different values of $\varepsilon$, as indicated in the legend. Dots indicate the tricritical points for each value of $\varepsilon$. The horizontal lines represent the equivalent paths, originally plotted in the parameter space $(B/\eta, A/\eta)$, but now displayed in the parameter space $(\alpha, \mu)$, where the colors correspond to the same values of $\mu$ used in the left panel.}
 \label{fig:NLNPVM_AB}
\end{figure*}

\subsection{Noisy voter models with nonlinearities} \label{sec:noisy_models}
In this section, we analyze the noisy versions of some of the models studied in Sec.~\ref{sec:nonlinear_models}. In this setting, agents can change their state by either of these two mechanisms:
\begin{itemize}
 \item Random or idiosyncratic behavior: With probability $a$ an agent adopts one of the two possible states at random, independently of their neighbors. 
 \item Herding behavior: with complementary probability $1-a$, an agent is influenced by their neighbors and changes its state following the model specific rules.
\end{itemize}

Following standard practice, we quantify the effect of the random changes using the noise-herding intensity ratio parameter~\footnote{The ratio $a/(2(1-a))$ was denoted as $\varepsilon$ in Ref.~\cite{Peralta_2018}. However, in this paper this letter denotes the strength of the preference of partisan voter models and it has already appeared in the $q$-voter.}
\begin{equation} \label{eq:mu_def}
\mu\equiv \frac{a}{2(1-a)},
\end{equation}
which varies in the $[0,\infty)$ interval.

\subsubsection{Noisy nonlinear voter model}
In the presence of noise and in the mean-field limit, the rate equation for the magnetization Eq.~\eqref{eq:rateeq_NLVM} becomes
\begin{align} \label{eq:rateeqm_NLNVM}
 \frac{dm}{dt}=&- a m\\ 
 &+2^{-\alpha}(1-a)(1-m^2)\left[(1+m)^{\alpha-1}-(1-m)^{\alpha-1}\right],\nonumber
\end{align}
which, when expanded in power-series around $m=0$ to $O(m^5)$, yields the following canonical parameters
\begin{subequations}\label{eq:NLNVM_parameters} 
\begin{align}\label{eq:NLNVM_A}
A&=2^{1-\alpha}\,(1-a)(\alpha-1),\\
 B&=-\frac{2^{-\alpha}}{3}\,(1-a)(\alpha-1)(\alpha-2)(\alpha-3),\label{eq:NLNVM_B}\\
 \eta&=a.\label{eq:NLNVM_eta}
\end{align} 
\end{subequations}
These specific forms indicate that the rescaled parameters $A/\eta, B/\eta$ are functions of the normalized noise $\mu$, defined in Eq.~\eqref{eq:mu_def}. Therefore, paths in the parameter space $(B/\eta,A/\eta)$ can be generated by varying $\alpha$ or $\mu$ reproducing the known phenomenology of the noisy nonlinear voter model (N\nl VM)~\cite{Peralta_2018}. By way of example, in Fig.~\ref{fig:NLNPVM_AB} we plot some of those paths varying $\alpha$ for different values of $\mu$. 
At $\alpha = 0$, the system starts in the paramagnetic phase, and as $\alpha$ increases, four distinct scenarios emerge depending on the value of $\mu$ .
\begin{itemize}
 \item[i)] $0<\mu<\mu_\text{t}\equiv 1/8$. An Ising transition from the paramagnetic to the ferromagnetic phase occurs at $A/\eta=1$. Replacing Eqs.~(\ref{eq:NLNVM_A},\ref{eq:NLNVM_eta}), it is found that the transition occurs at the smallest $\alpha$ that satisfies the condition
 \begin{equation} \label{eq:NLNVM_ac_2nd_order}
 \mu=2^{-\alpha}(\alpha-1).
 \end{equation}
 A further increase in $\alpha$, brings the system back to the paramagnetic phase through an MGV transition. It occurs at the condition that follows from Eq.~\eqref{eq:AB_eta}, or
 \begin{equation} \label{eq:NLNVM_ac_1st_order}
 \mu=2^{-\alpha}(\alpha-1)\frac{ \left(\alpha ^2-5 \alpha +8\right) \left(\alpha ^2-5 \alpha +24\right)}{32(\alpha-2)(\alpha-3)}.
 \end{equation} 
 \item[ii)] $\mu=\mu_\text{t}$. There exists an Ising transition from the paramagnetic to the ferromagnetic phase given by Eq.~\eqref{eq:NLNVM_ac_2nd_order}. Aditionally, a re-entrant tricritical transition occurs through the tricritical point $(B/\eta,A/\eta)=(-1,1)$ which, in terms of the parameters of our model, occurs at $\mu_t=1/8$ and $\alpha=5$. 
 \item[iii)] $\mu_\text{t}<\mu<\mu_L\equiv \frac{1}{2 e \ln2}\approx 0.265$. There exist two Ising transitions, the first from the paramagnetic to the ferromagnetic phase and the second returns the system to the paramagnetic phase. Both transitions are given by the two solutions of Eq.~\eqref{eq:NLNVM_ac_2nd_order}. 
 \item[iv)] $\mu>\mu_L$. No transition is possible and the system remains always in the paramagnetic phase.
\end{itemize}
A representative example of each case is plotted in the left panel of Fig.~\ref{fig:NLNPVM_AB}. From the pairs of values $\alpha,\mu$ that yield either the Ising or MGV transitions, the equivalent phase diagram $(\alpha,\mu)$ can be constructed, see right panel of Fig.~\ref{fig:NLNPVM_AB}.

\subsubsection{Noisy nonlinear partisan voter model} \label{sec:NLNPVM}
As a second example, we introduce the \textit{noisy nonlinear partisan voter model} (N\nl PVM) as the noisy version of the \nl PVM. It is important to note that in this model noisy changes do not depend on the preference of the agents. 

In order to recover the N\nl VM in the limit of no preference, $\varepsilon=0$, we need to modify the noise mechanism. We consider that the new random value is adopted with a probability $a/2$. In this way, the time rescaling affects both noise and herding and can be absorbed in a overall change $t\to t/2$, as in the noiseless version.

Just as in the \nl PVM, the global description of this model involves two macroscopic variables. Following the procedure in Sec.~\ref{sec:nlPVM}, we extend the result demonstrated in Ref.~\cite{llabres} for $\alpha=1$ by applying an adiabatic elimination technique for $\alpha\neq1$. As detailed in Appendix~\ref{app:sec:NLNPVM}, the reduced model is governed by the following rate equation for the magnetization
\begin{align} \label{eq:rateeqm_NLNPVM}
 \frac{dm}{dt}&=- a m +2^{-\alpha} (1-a) (1-m^2)\times\nonumber\\
 &\Biggl[ (1+m)^{\alpha-1}-(1-m)^{\alpha-1} \nonumber\\
 &-(1-a) \varepsilon ^2 \left[(1+m)^{\alpha }-(1-m)^{\alpha }\right] \times\nonumber \\
 &\frac{
 (1+m)^{\alpha-1 }+ (1-m)^{\alpha-1 } }{2^{1+\alpha} a+(1-a) \left[(1+m)^{\alpha
 }+(1-m)^{\alpha }\right]}\, \Biggr],
\end{align}
where we have performed the aforementioned time rescaling $t\to t/2$. Expanding around $m=0$ to $O(m^5)$ the parameters of the general model are identified as
\begin{subequations}\label{eq:NLNPVM_parameters} 
\begin{align} \label{eq:NLNPVM_A} 
A&=2^{1-\alpha}\,(1-a) \left((\alpha-1)- \varepsilon^2 \frac{ (1-a) \alpha}{2^{\alpha}a+(1-a)}\right),\\
 B&=-\frac{2^{-\alpha}}{3}\,(1-a) (\alpha-1) \Biggl[ (\alpha-2)(\alpha-3)  \label{eq:NLNPVM_B}\\
 &-\varepsilon ^2\frac{ (1-a) \alpha \left[2^{2+\alpha} a (\alpha -2)+(1-a)(\alpha -8) \right]}{ \left[2^{\alpha } a+(1-a)\right]^2} \Biggr],\nonumber\\
 \eta&=a\label{eq:NLNPVM_eta}.
\end{align}
\end{subequations}
For fixed $\varepsilon>0$ and $\mu$, paths in the parameter space $(B/\eta,A/\eta)$ are generated by varying $\alpha$, see Fig.~\ref{fig:NLNPVM_AB}. The same phenomenology of the N\nl VM is observed, indicating that the preference has little impact on determining the possible types of transitions. The four scenarios exhibited by the N\nl VM are also present, although the boundary values of the intervals for each scenario $\mu_\text{t}(\varepsilon)$, $\mu_\text{L}(\varepsilon)$ now depend on $\varepsilon$. Unfortunately, these values, as well as the MGV transition line given by the condition of Eq.~\eqref{eq:AB_eta}, must be determined numerically. In contrast, the Ising transition occurs at
\begin{equation}\label{eq:NLNPVM_ac_2nd_order}
 \mu_\mathrm{c}(\varepsilon)=
     2^{-1-\alpha } \left(\alpha -2+\sqrt{\alpha(\alpha -4 \varepsilon ^2)}\right).
\end{equation}
but only if, in the noiseless case, the system begins in the absorbing region of phase diagram $(B,A)$, i.e., $\alpha>\dfrac{1}{1-\varepsilon^2}$, see Fig.~\ref{fig:NLPVM_ae_diagram}. We also note that the ferromagnetic phase in the parameter space $(\alpha, \mu)$ is reduced as the preference $\varepsilon$ increases, see right panel of Fig.~\ref{fig:NLNPVM_AB}.

\begin{figure*}[t]
 \centering
 \includegraphics[width=\columnwidth]{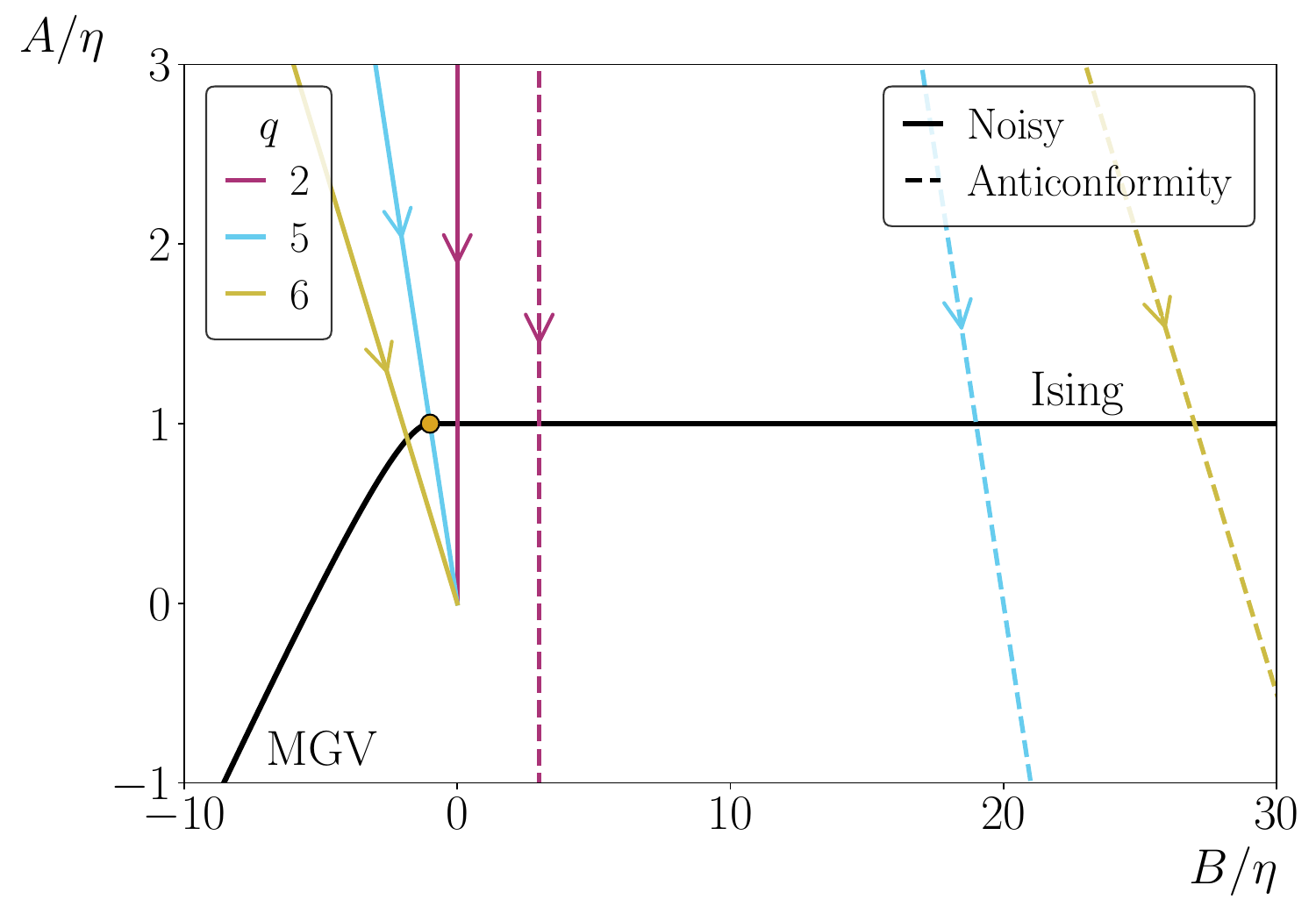}
 \includegraphics[width=\columnwidth]{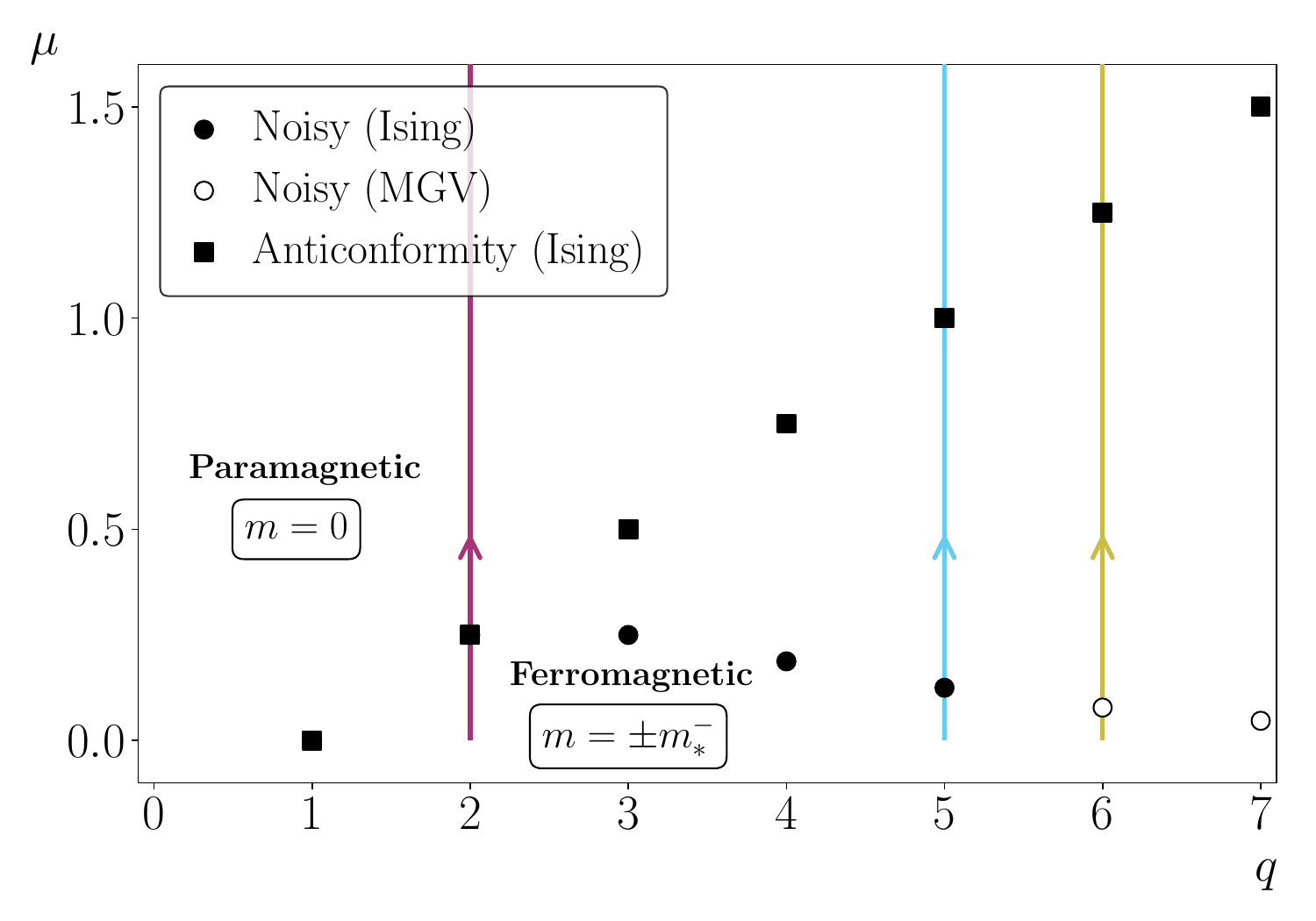}
 \caption{Left panel: Paths in the parameter space $(B/\eta, A/\eta)$ for the noisy $q$-voter model and the $q$-voter model with anticonformity, obtained by varying the value of $\mu$ for several fixed values of $q$, as indicated in the legend. Lines are generated using Eqs.~\eqref{eq:NLNVM_parameters} and Eqs.~\eqref{eq:noisyq-voter_anti_params}, respectively. Arrows indicate the direction of the path as $\mu$ increases. Transition line: MGV: modified Generalized Voter. The dot corresponds to the tricritical point where the MGV and the Ising transition line merge. Right panel: Phase diagram in the parameter space $(q,\mu)$. For $q=1,2$, both models ovelap. Vertical lines represent the equivalent paths, originally plotted in the parameter space $(B/\eta, A/\eta)$, but now displayed in the parameter space $(q, \mu)$, where the colors correspond to the same values of $q$ used in left panel. These lines overlap for both models. }
 \label{fig:nq-voter}
\end{figure*}

\subsubsection{Nonlinear $q$-voter model with noise}\label{sec:noisyq-voter}
In this section, we review the two extensions of the \nl$q$VM considered in Ref.~\cite{q-voter_noise}. We note that in Ref.~\cite{q-voter_noise} repetitions in the choice of the $q$ neighbors are not allowed, contrary to the \nl$q$VM studied in Sec.~\ref{sec:q-voter}. However, both versions of the model, with or without repetitions, are equivalent in the mean-field limit.

The first variation is the \textit{noisy $q$-voter model} (N$q$VM), a natural extension in the same spirit as the previously introduced noisy nonlinear voter model (N\nl VM) and noisy nonlinear partisan voter model (N\nl PVM). For simplicity, we restrict our analysis to the case $\epsilon=0$. In the mean-field limit, the rate equation for the magnetization when considering noise effects reads
\begin{align}\label{eq:nqvoter_m_noisy}
\frac{dm}{dt}=&- a m\\ 
&+2^{-q}(1-a)(1-m^2)\left[(1+m)^{q-1}-(1-m)^{q-1}\right],\nonumber
\end{align}
which coincides with rate equation of the N\nl VM Eq.~\eqref{eq:rateeqm_NLNVM} for integer values of $\alpha$ and evidences that both models are equivalent for $\alpha=q$. Consequently, the parameters of the canonical model $A,B,\eta$ are also given by Eqs.~(\ref{eq:NLNVM_A},\ref{eq:NLNVM_B},\ref{eq:NLNVM_eta}) with $\alpha=q$. The key difference in this model is that the nonlinear parameter $q$ takes integer values, meaning that continuous transitions can only be observed by varying $\mu$. For this reason, paths in the parameter space $(B/\eta,A/\eta)$ are generated by varying the normalized noise intensity $\mu$ fixing a value of $q$, see left panel of Fig.~\ref{fig:nq-voter}. The system exhibits a unique phase transition whose nature depends on $q$. For $q\leq5$, there exists an Ising transition, given by Eq.~\eqref{eq:NLNVM_ac_2nd_order}, while for $q>5$, the system undergoes an MGV transition given by Eq.\eqref{eq:NLNVM_ac_1st_order}. For the particular case $q=5$, the transition presents tricritical properties. The transition points in the parameter space $(q,\mu)$ are displayed in the right panel of Fig.~\ref{fig:nq-voter}.

The second extension is the \textit{nonlinear $q$-voter model with anticonformity} (A$q$VM), which does not explicitly include the type of noise considered in our study so far but is still worth mentioning. Instead, the model incorporates contrarian behavior, which eliminates absorbing states and effectively acts as a form of noise. The dynamics of the model is governed by the following rules: An agent $i$ and $q$ of their neighbors are selected at random (repetitions are not allowed). If all $q$ neighbors share the same state and they do not agree with agent $i$, agent $i$ adopts the state of the group with probability $p_1$, behaving as a conformist. On the other hand, if the agent's state coincides with that of the group, the agent switches state with probability $p_2$, exhibiting anticonformist behavior. Note that for $p_2=0$ and $p_1$=1, the updating rules of the $q$-voter model for $\epsilon=0$ (without repetition) are recovered. Since, the results only depend on the ratio of $p_1/p_2$, we set $p_1=1$, $p_2=a$ without loss of generality from this point forward. These possible interactions give rise to the following rate equation for the magnetization:
\begin{align}\label{eq:AqVM_rateeqm}
\frac{dm}{dt}&=2^{-q} \Bigl( -a\left[(1+m)^{q+1}-(1-m)^{q+1}\right] \nonumber \\
&+(1-m^2)\left[(1+m)^{q-1}-(1-m)^{q-1}\right]\Bigr),
\end{align}
which after the series expansion around $m=0$ up to $O(m^5)$ together with the condition $\eta(a=0)=0$ yields the canonical parameters
\begin{subequations}\label{eq:noisyq-voter_anti_params}
\begin{align}
 A&=2^{1-q} \bigl[(1-a) q-1\bigr], \label{eq:noisyq-voter_anti_A} \\
 B&= \frac{2^{-q}}{3}\Bigl[q \left((a-1) q^2+6 q+5 a-11\right)+6\Bigr]\label{eq:noisyq-voter_anti_B} \\
 \eta&=2^{1-q}a. \label{eq:noisyq-voter_anti_eta}
\end{align}
\end{subequations}
Paths in the parameter space $(B/\eta,A/\eta)$ are generated again by varying $\mu$, as defined in Eq.~\eqref{eq:mu_def}, fixing a value of $q$, see Fig.~\ref{fig:nq-voter}. Differently from the N$q$VM, only Ising transitions are possible at the critical value
\begin{equation}
\mu_\text{c}=\frac{q-1}{4},
\end{equation}
which coincides with that of the N$q$VM for $q=1,2$. The set of critical points is shown in right panel of Fig.~\ref{fig:nq-voter}. This example demonstrates that the social rules governing agent interactions can fundamentally influence the nature of the phase transition.

Our analysis encompasses the particular case $ q = 2 $, corresponding to the Sznajd model with anticonformity~\cite{Sznajd_anti}. The canonical parameters are identified as $ A = 1/2 - a $, $ B = 3a/2 $, and $ \eta = a/2 $. The model exhibits an Ising transition from the ferromagnetic to the paramagnetic phase at $ a = 1/3 $, or equivalently $\mu = 1/4$.

\begin{figure*}[t]
 \centering
 \includegraphics[width=\columnwidth]{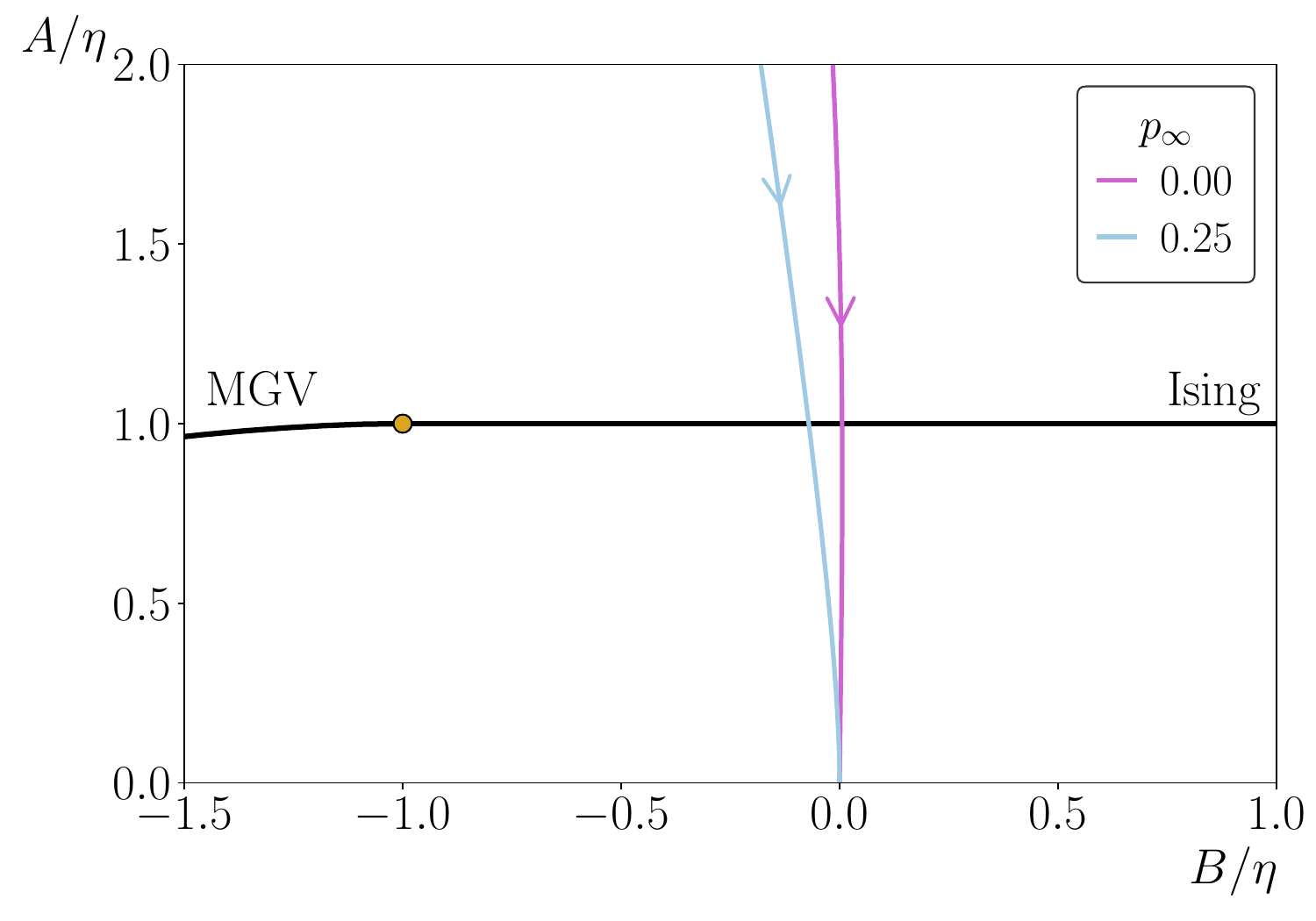}
 \includegraphics[width=\columnwidth]{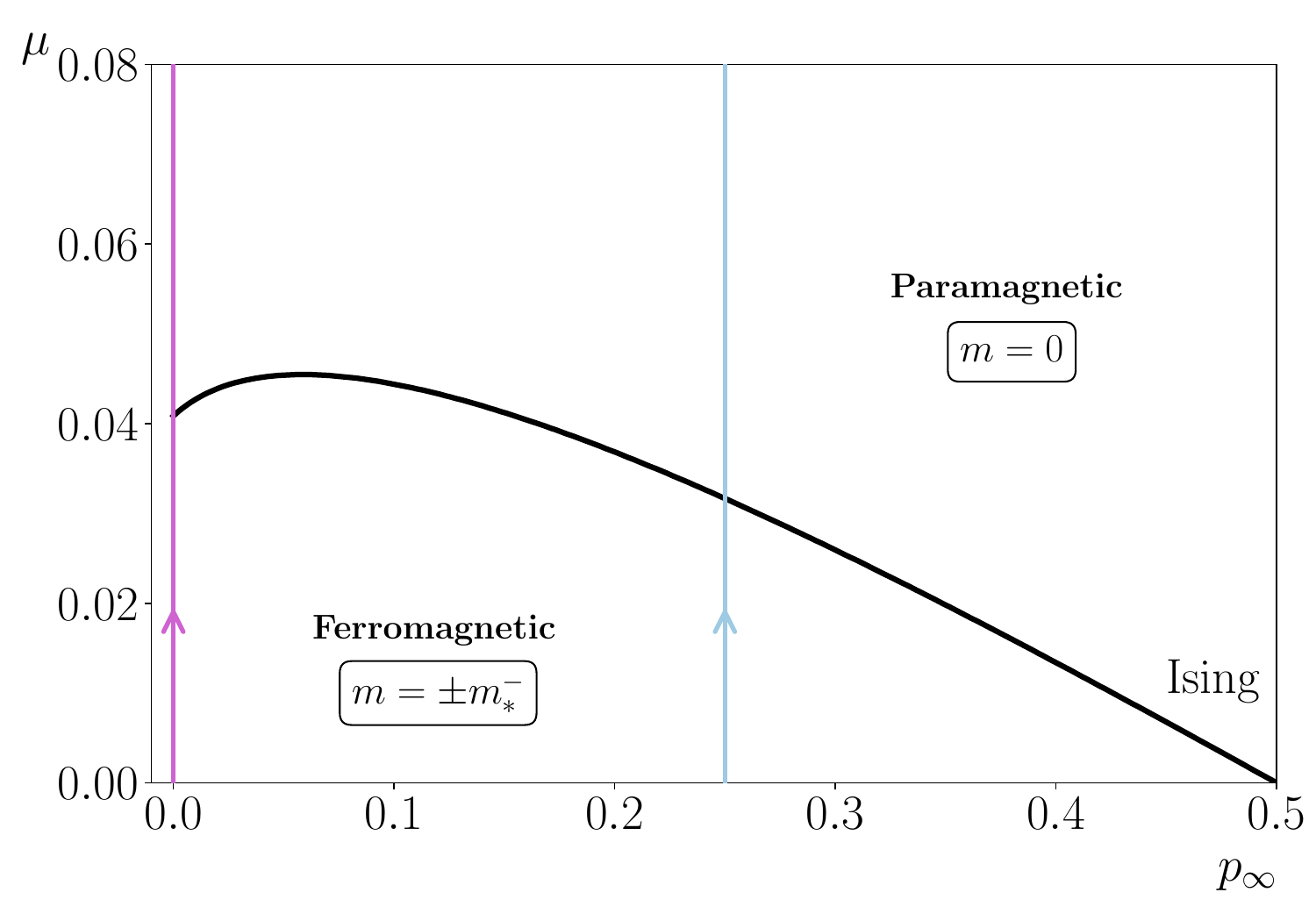}
 \caption{Left panel: Paths in the parameter space $(B/\eta, A/\eta)$ for the noisy voter model with aging, obtained by varying the value of $\mu$ while fixing $p_0=0.5, \tau^*=2$ for $p_\infty=0, 0.25$, as indicated in the legend. Arrows indicates the direction of the path as $\mu$ increases. Right panel: Phase diagram in the parameter space $(p_\infty,\mu)$ for $p_0=0.5, \tau^*=2$. Vertical lines represent the equivalent paths, originally plotted in the parameter space $(B/\eta, A/\eta)$, but now displayed in the parameter space $(p_\infty,\mu)$,where the colors correspond to the same values of \pinf.}
 \label{fig:ab_diagram_NVM_aging}
\end{figure*}

\subsubsection{Noisy voter model with aging}\label{sec:nvm_aging}
Next, we review the \textit{noisy voter model with aging}, which incorporates age-independent idiosyncratic changes to the \textit{voter model with aging} studied in Sec.~\ref{sec:vm_aging}. Following an analogous adiabatic approximation technique, we obtain the noisy version of the mean-field limit rate equation for the magnetization Eq.~\eqref{eq:m_aging}, which reads~\cite{aging_Jaume_Sara}
\begin{align} \label{eq:nvm_aging}
 \frac{dm}{dt}&=-am \nonumber\\
 &+(1-a)\,\frac{1-m^2}{2}\left[\Phi_a\left(\frac{1+m}{2}\right)-\Phi_a\left(\frac{1-m}{2}\right)\right],
\end{align}
where the same aging kernel $p(\tau)$ has been considered, given by Eq.~\eqref{eq:p_tau_aging}, and the explicit expression for $\Phi_a(x)$ is given in Appendix~\ref{app:vm_aging}. In Sec.~\ref{sec:vm_aging}, we showed that the system dynamically approaches a consensus state in the typical aging scenario, $p_\infty<p_0$, while coexistence is reached for the anti-aging case, $p_\infty>p_0$. With the addition of noise, the anti-aging case does not exhibit any notable phenomena since random choices introduce a drift that drives the system towards the coexistence solution $m=0$. Therefore, in this section we focus exclusively on the aging case, $p_\infty<p_0$, where a competition between ordering and idiosyncratic behavior emerges.

From Eq.~\eqref{eq:nvm_aging}, we readily identify $\eta=a$. However, a series expansion of the second term in powers around $m=0$ up to $O(m^5)$ leads to overly complex expressions, and deriving explicit forms of $A$ and $B$ for arbitrary values of the parameters $a, p_0,\,p_\infty,\,\tau^*$ is not straightforward. For further progress, $A$ and $B$ are calculated numerically by fixing $p_0$ and $\tau^*$, while treating $0<p_\infty<p_0$ and $\mu$, as defined in Eq.~\eqref{eq:mu_def}, as free parameters. In the left panel of Fig.~\ref{fig:ab_diagram_NVM_aging}, paths in the parameter space $(B/\eta,A/\eta)$ are generated by varying $\mu$ for a fixed value of \pinf, and an Ising transition from the ferromagnetic to the paramagnetic phase is observed. For $p_0=0.5$, $\tau^*=2$, $p_\infty=0$, the critical value reported in Ref.~\cite{Artime_2018} is recovered. Neither tricritical nor MGV transitions are possible for any combination of the parameters $p_0$, \pinf, $\tau^*$, $\mu$. This model only presents Ising phase transitions. The phase diagram $(p_\infty,\mu)$ of the right panel of Fig.~\ref{fig:ab_diagram_NVM_aging} is constructed numerically by determining the pairs of values $p_\infty,\mu$ at which the Ising transition line is crossed, thereby reproducing the phenomenology reported in Ref.~\cite{complete_aging}. 

Alternatively, the possible transitions of the system can be analyzed by fixing $\mu$ and varying $p_\infty$ leading to the following scenarios depending on the values of $\mu, p_0$ and $\tau^*$:
\begin{itemize}
 \item Unique Ising transition. If the system resides in the ferromagnetic phase for $p_\infty=0$, an Ising transition to the paramagnetic region occurs by increasing \pinf.
 \item Double Ising transition. Starting in the paramagnetic phase, the system undergo two successive Ising transitions: first, a transition from the paramagnetic to the ferromagnetic phase, and then another transition that brings the system back to the paramagnetic phase.
 \item No possible transition. The system always remains in the paramagnetic phase.
\end{itemize}

\section{Finite size effects in noisy nonlinear voter models}\label{sec:finite}

Strictly speaking, thermodynamic phases and phase transitions only occur in the thermodynamic limit, where $N \to \infty $. However, one can still observe transitions for finite $ N$, manifested as shifts in the maxima of the stationary probability distribution $P_{\mathrm{st}}(m)$ of the order parameter $m$. In this section, we develop a stochastic framework to analyze how fluctuations arising from finite-size effects modify the behavior seen in the infinite-size limit. As in previous sections, we limit the analysis to the complete graph.

Let us denote by $P(m;t)$ the probability distribution of the order parameter $m$ at time $t$. Using a standard approach~\cite{vanKampen:2007, Toral2014StochasticNM}, its dynamical evolution is governed by the Fokker-Planck equation \begin{equation} \label{eq:FP_1D}
\frac{\partial P(m;t)}{\partial t}=\mathbf{H} P(m;t),
\end{equation}
where the operator is $\mathbf{H}$ defined as
\begin{equation} \label{eq:NVM_H}
\mathbf{H}=-\frac{\partial}{\partial m}F(m)+\frac 1N\frac{\partial^2}{\partial m^2} D(m).
\end{equation}
Here $F(m)$ is the drift, which coincides with the function appearing in the rate equation of the process, Eq.~\eqref{eq:m_noisy_general}, 
and $D(m)\ge 0$ is the diffusion function. Explicit expressions for $F(m)$ and $D(m)$ are provided in Appendices \ref{app:sec:NLNPVM} and \ref{app:sec:qVMC} for the N\nl VM, the N\nl PVM and the A$q$VM, respectively.

The stationary probability distribution can be written in terms of an exponential function
\begin{equation} \label{eq:Pstform}
 P_{\mathrm{st}}(m)=\mathcal{Z}^{-1}\cdot\exp{\left[-N\,V(m,N)\right]},
\end{equation}
where $\mathcal{Z}$ is the normalization constant and $V(m,N)$ is a ``potential function" given by
\begin{equation} \label{eq:potential}
 V(m,N)=\int\frac{-F(m)+D'(m)/N}{D(m)}dm.
\end{equation}
In the thermodynamic limit, $P_{\mathrm{st}}(m)$ consists of a sum of Dirac-delta functions centered around the absolute minima of $V(m,\infty)$. These, in turn, coincide with the zeros of the function $F(m)$. However, for finite $N$, the absolute minima of $V(m,N)$ correspond only to the most probable states. In both cases, transitions are identified as changes of the location of the most probable states when varying some system's parameters. These changes can occur continuously or discontinuously, identifying the transition as Ising-type or MGV-type, respectively. By an abuse of language, the transition points in the finite-size case will still be referred to as ``critical'' points. 

In systems with $\mathbb{Z}_2$ symmetry, the drift and diffusion functions satisfy the relations $F(-m) = -F(m)$ and $D(-m) = D(m)$. These symmetries guarantee that $m = 0$ is a trivial extremum of the potential $V(m, N)$. In the case of a continuous, Ising-type transition, the character of $m = 0$ changes at the critical point: it switches from a minimum in the paramagnetic phase to a maximum in the ferromagnetic phase. The critical point is thus determined by the condition $\left.V''(m, N)\right|_{m = 0} = 0.$

Applying this criterion to the N\nl PVM, and using the expressions for the drift and diffusion given in Eqs.~(\ref{eq:rateeqm_NLNPVM}, \ref{app:NPVM_diff}), one finds—through a straightforward but lengthy calculation—that the system undergoes an Ising-type transition at the critical value
\begin{align}\label{eq:ac_finite_size}
 \mu_\text{c}(\varepsilon, N)&=2^{-1-\alpha}\left(\alpha -2+\sqrt{\alpha(\alpha -4 \varepsilon ^2)}\right)\nonumber \\
 &+\frac{2^{-\alpha}}{N} \left[\alpha(3-\alpha)+\frac{(\alpha -1)^2}{\alpha}\,\varepsilon^2 \right. \nonumber \\
 &\left. +\frac{3+\alpha(\alpha -2)}{\alpha^2}\,\varepsilon^4 +O(\varepsilon^6)\right]+O(N^{-2}).
\end{align}
This transition lines separates the ferromagnetic phase ($\mu < \mu_\text{c}(\varepsilon, N)$) from the paramagnetic phase ($\mu > \mu_\text{c}(\varepsilon, N)$).

We begin by analyzing the case $\varepsilon = 0$. In this limit, the critical point of the N\nl VM reported in Ref.~\cite{Peralta_2018}, is recovered
\begin{equation}
\mu_\text{c}(0, N) = 2^{-\alpha}\left(\alpha - 1 + \frac{\alpha(3 - \alpha)}{N}\right).
\end{equation}
When $\alpha = 1$, the model corresponds to the standard NVM, for which the critical value simplifies to $\mu_\text{c} = 1/N$~\cite{Kirman1993,Carro2016,Peralta_sto_2018}. Although in this case the system undergoes an MGV-type transition, the condition $\left.V''(m, N)\right|_{m = 0} = 0$ still correctly identifies the critical point, as $m = 0$ changes stability at $\mu_\text{c}$. In the thermodynamic limit, the critical value vanishes, $\mu_\text{c} \to 0$, and the system remains in the paramagnetic phase for any $\mu > 0$.

\begin{figure}[t]
 \centering
 \includegraphics[width = \columnwidth]{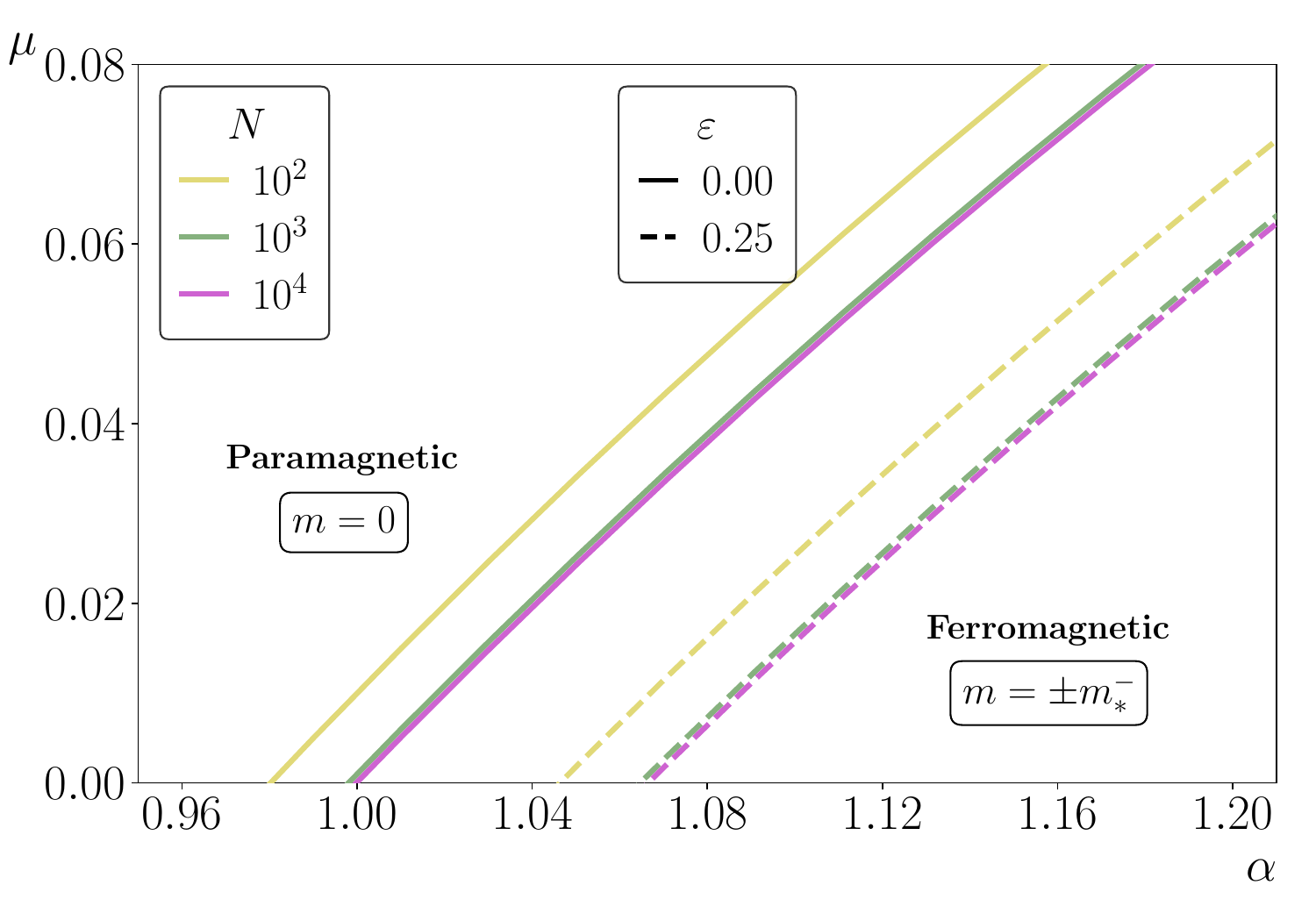}
 \caption{Transition line $\mu_\mathrm{c}(\varepsilon,N)$ generated with Eq.~\eqref{eq:ac_finite_size} by varying $\alpha$ for several values of $N,\varepsilon$, as indicated in the legend.}
 \label{fig:critical_finite_size}
\end{figure}

We now consider the case $\varepsilon > 0$. In this regime, the critical point $\mu_\text{c}(\varepsilon, N)$ decreases with both increasing $\varepsilon$ and $N$, indicating that the ferromagnetic phase is progressively narrowing, see Fig.~\ref{fig:critical_finite_size}. In the linear case $\alpha = 1$, corresponding to the standard NPVM, the system undergoes a finite-size MGV-type transition. However, this transition is not captured by the condition $\left.V''(m, N)\right|_{m = 0} = 0$, since the point $m = 0$ remains a minimum of the potential even at the transition. Unlike the NVM, the preference in the NPVM induces a stable disordered state. In the ferromagnetic phase, the potential $V(m, N)$ presents three minima: one at $m = 0$ and two symmetric ones at $m \pm 1$. 

Regarding the A$q$VM, a simpler calculation obtained from Eqs.~(\ref{eq:AqVM_rateeqm},\ref{app:eq:AqVM_FP_diffusion}) gives the critical point
\begin{equation}
\mu_\mathrm{c}(N)=\frac{q-1}{4}-\frac{q(q+1)(q-2)}{2N}.
\end{equation}
In particular, for $q = 1$, the system behaves analogously to the NVM: it exhibits only a finite-size, noise-induced transition for $\mu_\mathrm{c} = 1/N$, and no phase transitions occur in the thermodynamic limit. 

Both cases illustrate how finite-size transitions can be transformed into bona-fide phase transitions through the inclusion of nonlinear rates

\section{Universality classes} \label{sec:Universality_classes}
In Sec.~\ref{sec:noisy_models}, we mapped the parameters of several noisy voter models onto the canonical mean-field model introduced in Sec.~\ref{sec:noisynonlinear_sec_ABC}, enabling a unified classification of their phase diagrams and transition types. In this section, we investigate the universality classes of these transitions, extending the analysis beyond the mean-field case to complete graph, regular lattices and complex networks. We focus exclusively on the continuous transitions that lie along the Ising line of the phase diagram (left panel of Fig.~\ref{fig:ab_eta_diagram}), including those passing through the tricritical point.

The universality classes are mainly determined by the set of critical exponents characterizing the behavior of different quantities, such as the magnetization $\tm\equiv\langle | m | \rangle_\mathrm{st}$, near the transition point. It is also common to use the susceptibility $\chi$, defined as the variation of the magnetization with respect to a parameter breaking the $\mathbb{Z}_2$ symmetry. Note that the models lack Hamiltonian dynamics, therefore it is not straightforward to define other quantities, usual in statistical mechanics, such as the internal energy or the specific heat. If $T$ is a measure of the distance in parameter space to the transition point, the relations $\tm\sim T^\beta$ and $\chi\sim|T|^{-\gamma}$, define the critical exponents $\beta$ and $\gamma$.

In the mean-field limit, corresponding to all-to-all coupling and the thermodynamic limit, the critical behavior follows from the Landau theory of phase transitions~\cite{Huang87}, applied to the potential in Eq.~\eqref{eq:m_noisy_general}. This leads to the so-called mean-field or classical values of the critical exponents $\beta=1/2$ and $\gamma=1$ for those transitions crossing the critical line, and $\beta=1/4$, $\gamma=1$ for the transitions crossing through the tricritical point. As all noisy models discussed in previous sections map into this potential, they belong to this universality class.

These mean-field exponents are appropriate to describe many models, irrespective of the microscopic details or symmetries, when their spatial dimension $d$ is greater than an \textit{upper critical dimension} $\dc$. According to the Ginzburg criterion~\cite{Huang87}, for $\mathbb{Z}_2$ symmetric models, it is $\dc=4$ for the standard continuous transitions, and $\dc=3$ for tricritical transitions.

To characterize the universality class, a useful tool is that of the finite-size scaling theory~\cite{Cardy:1988}. This theory predicts that the behavior of the magnetization and the susceptibility for finite systems follows the general relations~\cite{deutsch:92}
\begin{subequations}\label{eq:scalingmchi}
\begin{align}
\tm &= N^{-\beta/\bar{\nu}}f_\tm(T N^{1/\bar{\nu}}),\label{eq:scalingmchi-a}\\
 \chi &= N^{\gamma/ \bar{\nu}} f_\chi(T N^{1/\bar{\nu}}),
\end{align}
\end{subequations}
where 
\begin{align}
\bar{\nu} =
\begin{cases}
d\nu & \text{if } d<d_\mathrm{c}, \\
d_\mathrm{c}\nu & \text{if } d>d_\mathrm{c}.
\end{cases}
\end{align}
Here $f_\tm(x)$, and $f_\chi(x)$ are scaling functions, $d$ is the spatial dimension, $\dc$ the upper-critical dimension and $\nu$ the critical exponent of the correlation length. For the Landau theory mentioned before it is $\nu=1/2$ both for critical and tricritical points. By using appropriate limits of the scaling functions, it is possible to determine that in the thermodynamic limit one recovers the critical behaviour $\tm\sim T^\beta$ and $\chi\sim |T|^{-\gamma}$. Our intention is to determine the set of exponents $\beta,\gamma,\nu$ and the upper-critical dimension $\dc$ that best fit those general relations to data coming from numerical simulations of the noisy models.
A special family of universality classes is that of the Ising model, with a upper-critical dimension $\dc=4$. Above the upper-critical dimension $d>\dc$, the exponents of the Ising universality class $\beta,\gamma,\nu$ are equal to those predicted by the Landau theory and they do not depend on $d$ anymore. The Ising model is expected to be the relevant one for systems with $\mathbb{Z}_2$ symmetry and our results will confirm this expectation.

In the numerical simulations of the noisy agent-based models, 
we use the microscopic rules to simulate the time evolution of the system variables. After a transient, we compute in the stationary regime moments of the order parameter $m$ as a function of the noise intensity $\mu$ and the system size $N$, $\tm_n(\mu,N) \equiv\langle|m^n|\rangle_\mathrm{st}$, keeping other parameters constant. From there, we compute the magnetization $\tm(\mu,N)$ and use the fluctuation-dissipation relation $\chi(\mu,N)=N\left(\tm_2-\tm^2\right)$ to compute the susceptibility. In the case of an all-to-all interaction where a single variable, $m$, suffices to represent the whole system, we have used a very precise numerical scheme, detailed in Appendix~\ref{app:pst} to determine the stationary probability distribution $P_\text{st}(m)$ and from there compute the necessary averages. The distance to the transition point is quantified as $T\equiv 1-\mu/\mu_\text{c}$ for the critical points, and $T\equiv 1-\mu/\mu_\text{t}$ for the tricritical point.

For the numerical determination of the transition points, usually unknown, it is useful to use the so-called Binder cumulant~\cite{Binder:cumulant}
\begin{equation}
 U_4(\mu,N)=1-\frac{\tm_4}{3(\tm_2)^2}.
\end{equation}
whose predicted scaling near the transition is
\begin{equation}\label{eq:scalingum}
U_4 = f_U(T N^{1/\bar{\nu}}),
\end{equation}
with $f_U(x)$ a scaling function. The usefulness of the Binder cumulant is that at the transition point $T=0$ it takes the same value independently of the system size, making it very convenient for the determination of the value of $\mu$ at the transition point.

We now consider separately the results for the critical and tricritical points.

\begin{figure*}[t]
 \centering
 \includegraphics[width = \textwidth]{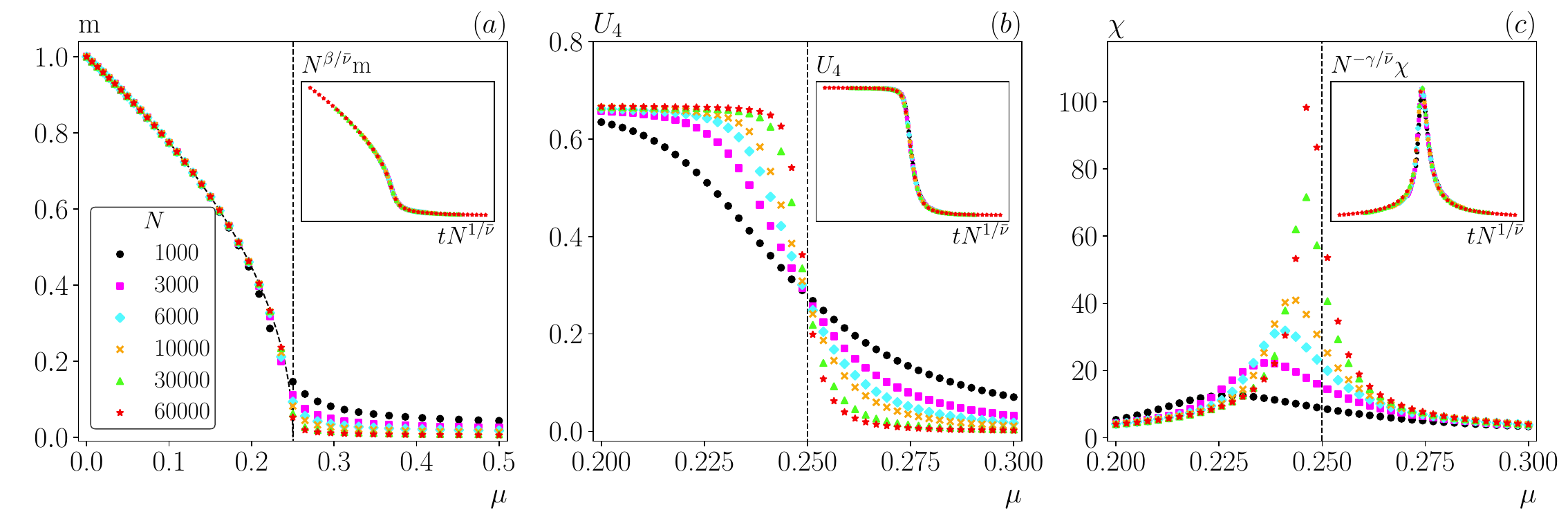}
 \includegraphics[width = \textwidth]{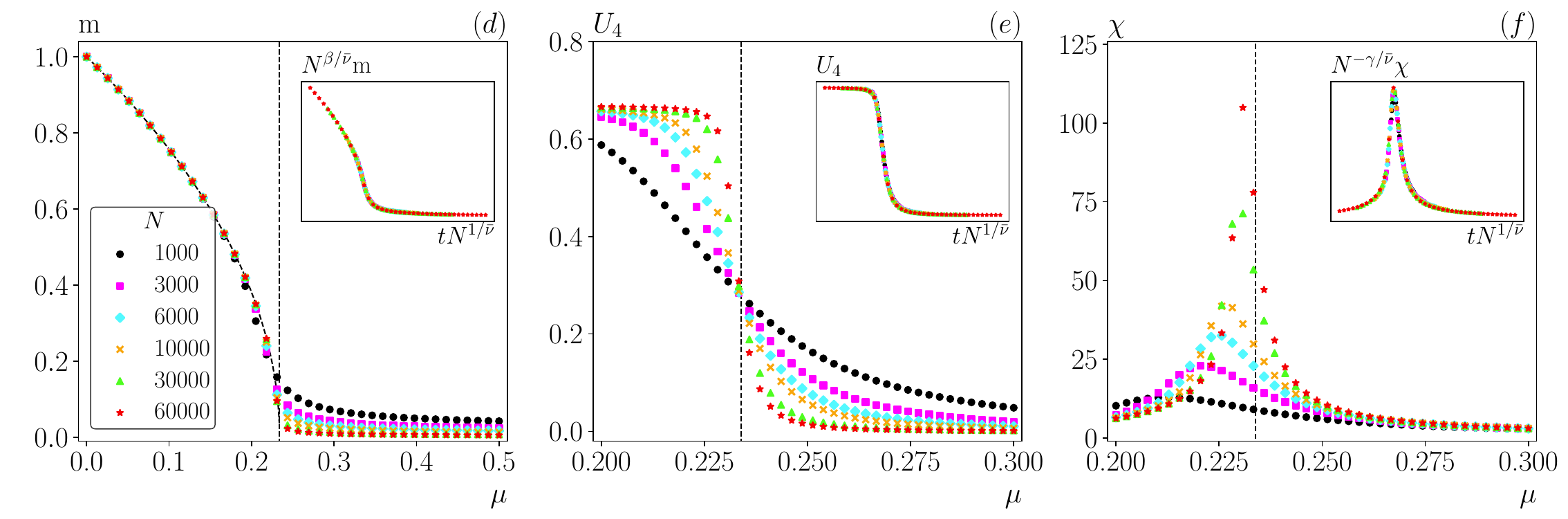}
 \includegraphics[width = \textwidth]{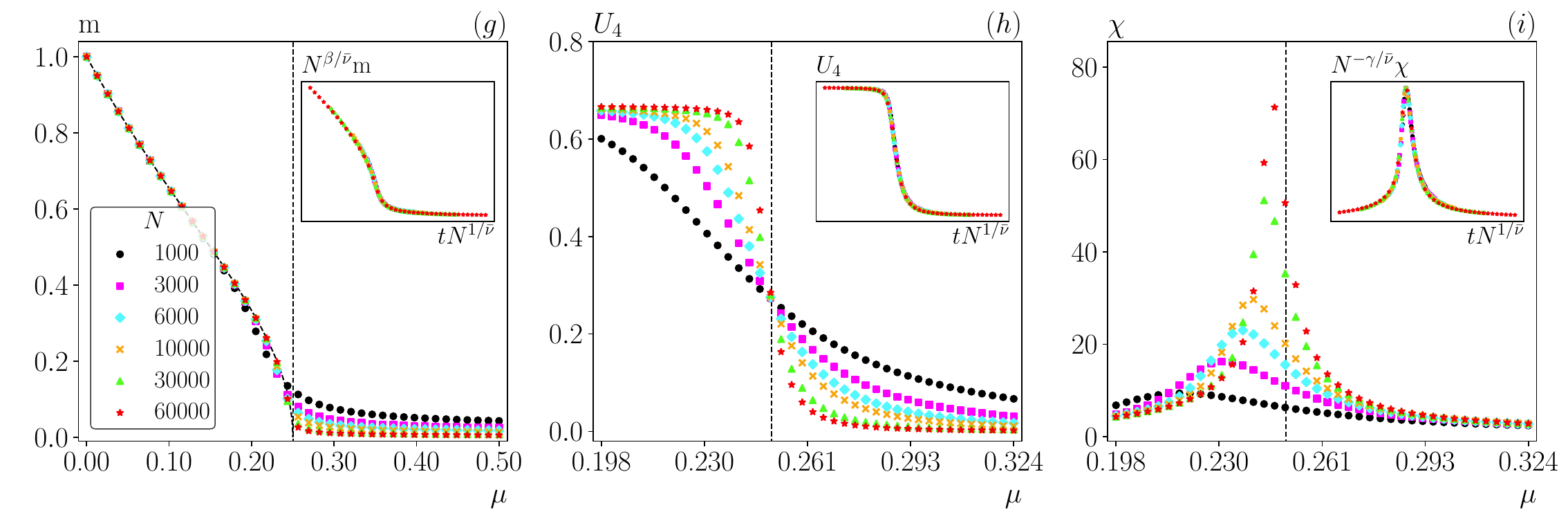}
 \caption{($a,d,g$) Magnetization, ($b,e,h$) Binder cumulant, and, ($c,f,i$) susceptibility versus the noise-herding intensity parameter $ \mu$ for $\alpha=q=2$ on the complete graph for different system sizes. Top row corresponds to the N\nl VM, middle row to the N\nl PVM with $\varepsilon=0.25$ and bottom row to the A$q$VM. The insets show the collapses with the corresponding mean-field critical exponents~\cite{yeomans1992statistical}. The vertical lines indicate the critical point.}
 \label{fig:scaling_alpha=2_CG}
\end{figure*}

\begin{figure*}[t]
 \centering
 \includegraphics[width =\textwidth]{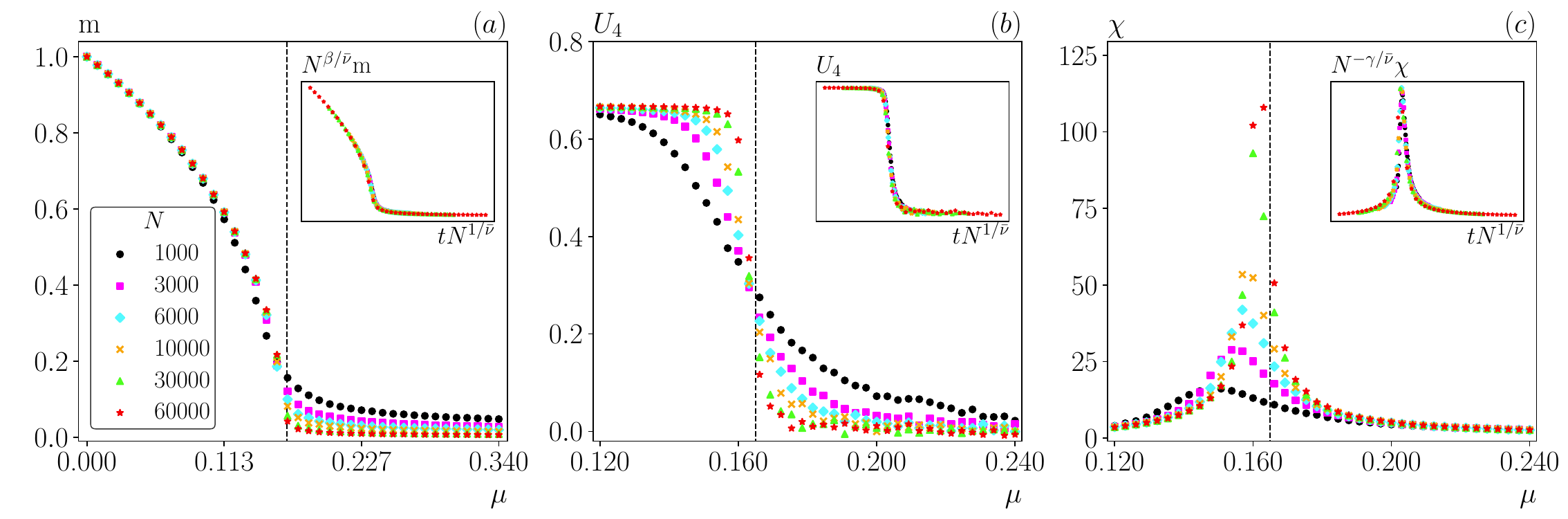}
 \includegraphics[width = \textwidth]{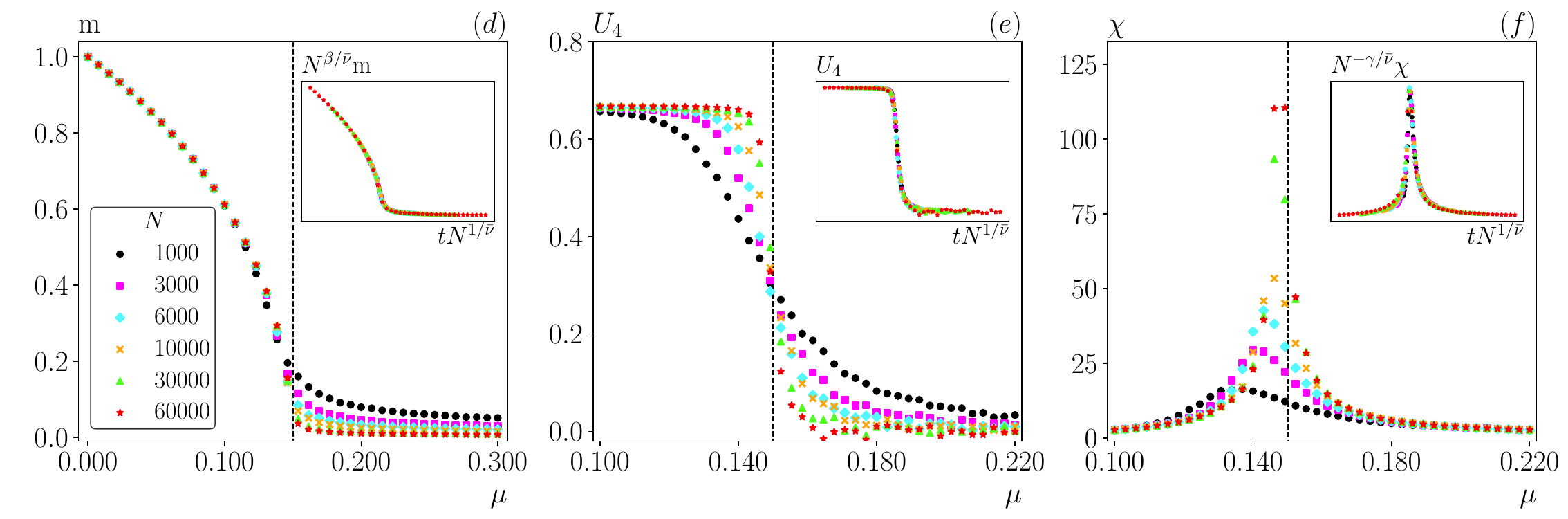}
 \includegraphics[width = \textwidth]{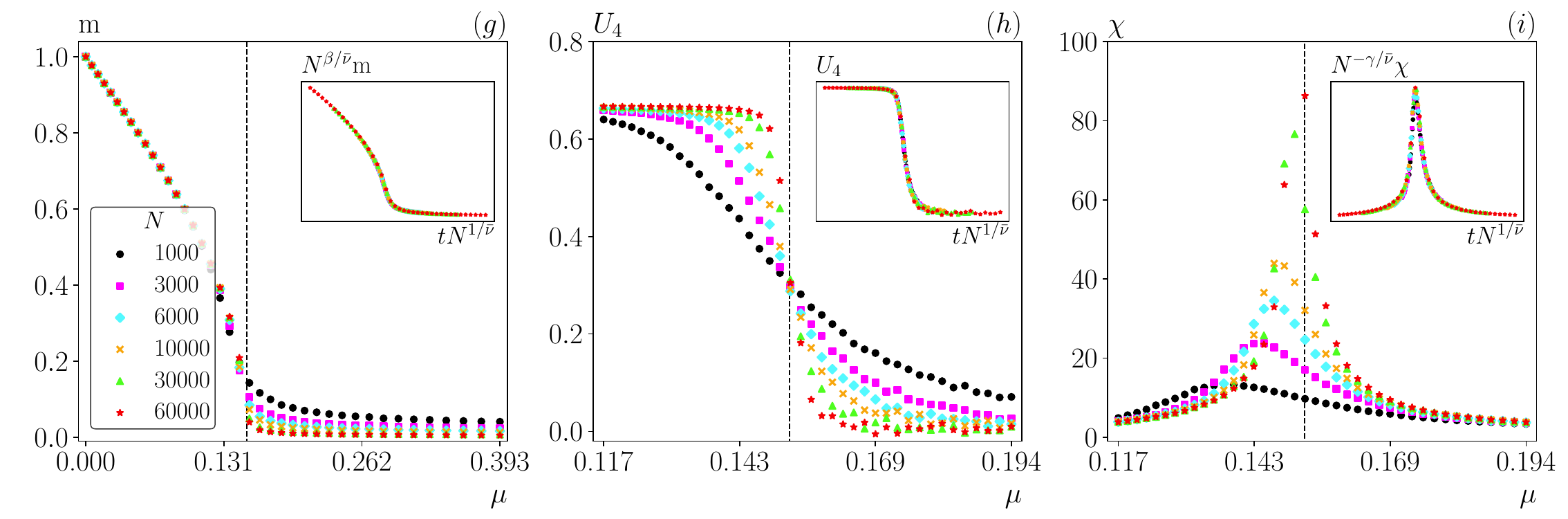}
 \caption{($a,d,g$) Magnetization, ($b,e,h$) Binder cumulant, and, ($c,f,i$) susceptibility versus the noise-herding intensity parameter $ \mu$ for $\alpha=2$ on Erd\H{o}s-R\'enyi networks with $\langle k \rangle=8$ for different system sizes. Top row corresponds to the N\nl VM, middle row to the N\nl PVM with $\varepsilon=0.25$ and bottom row to the A$q$VM. The insets show the collapses with the corresponding Ising critical exponents~\cite{yeomans1992statistical}. The vertical lines indicate the critical point $\mu_\mathrm{c}$.}
 \label{fig:scaling_alpha=2_ER}
\end{figure*}

\begin{figure*}[t]
 \centering
 \includegraphics[width = \textwidth]{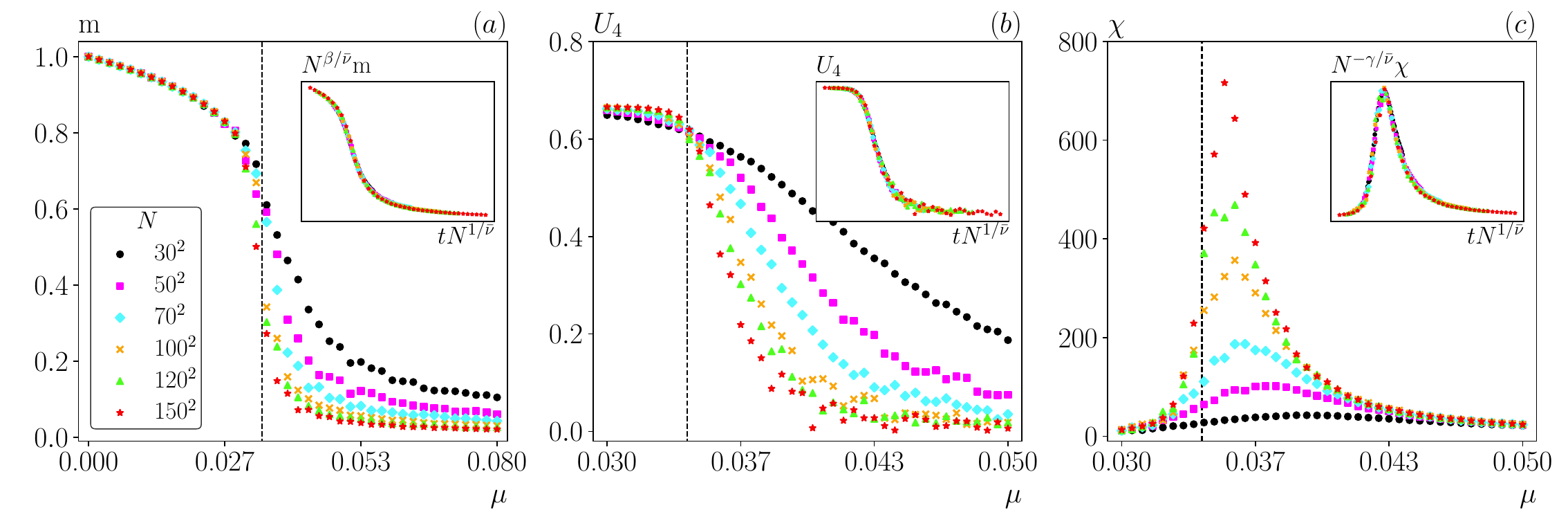}
 \includegraphics[width = \textwidth]{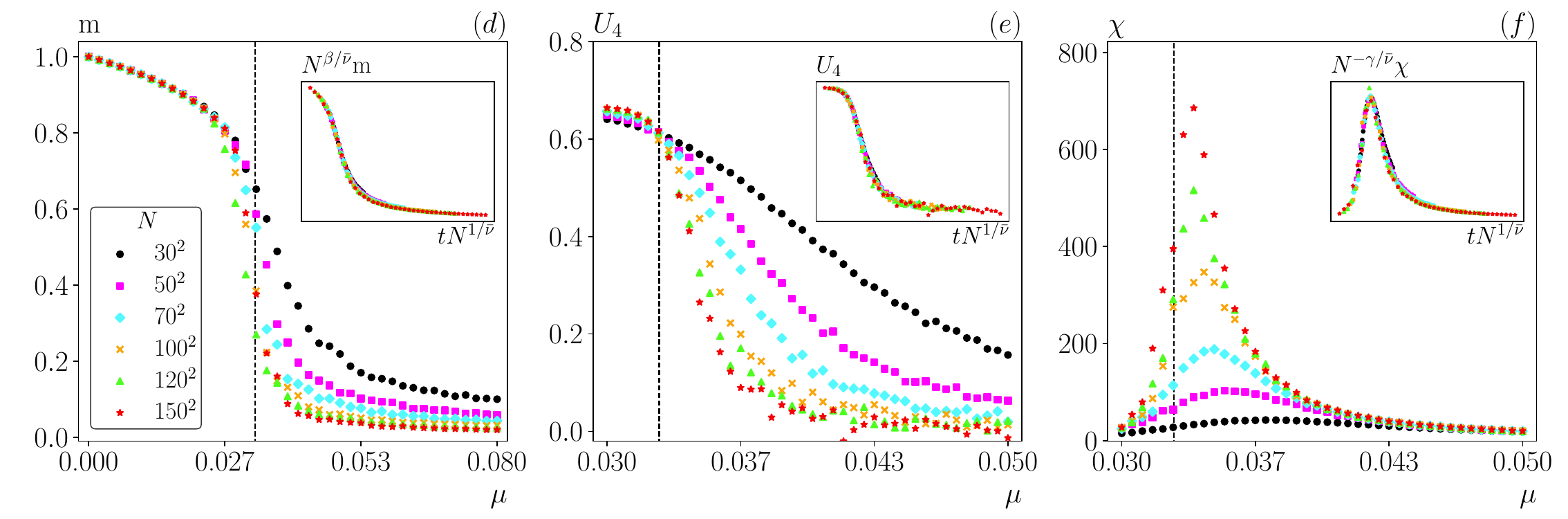}
 \includegraphics[width = \textwidth]{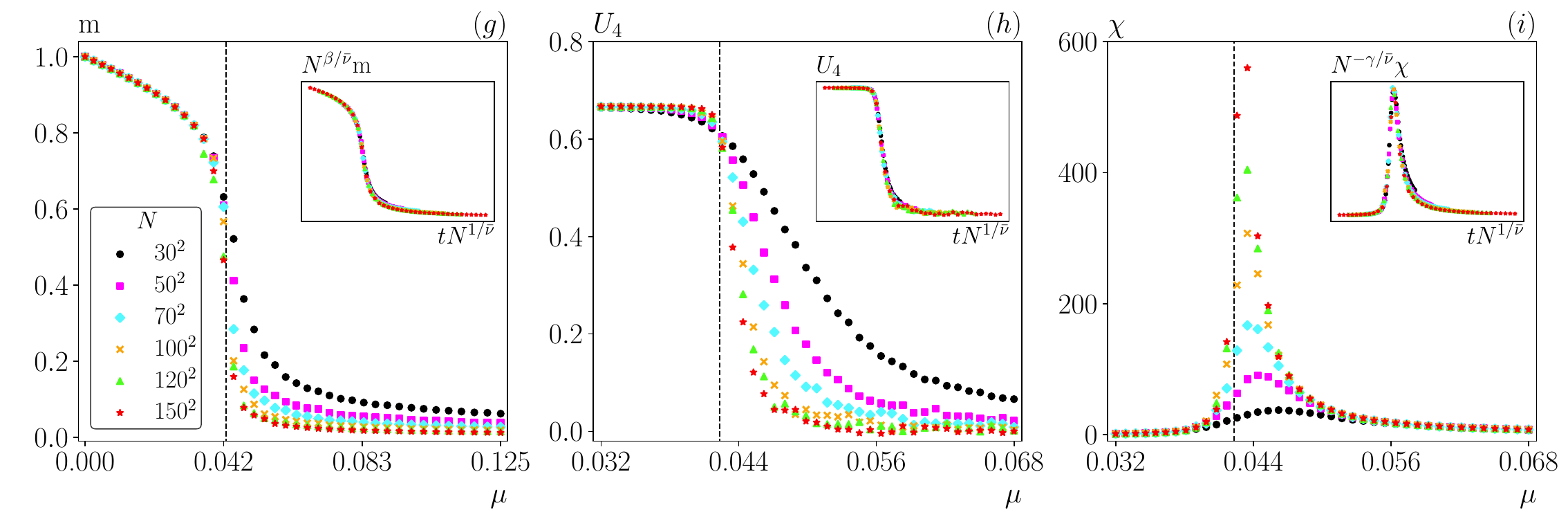}
 \caption{($a,d,g$) Magnetization, ($b,e,h$) Binder cumulant, and, ($c,f,i$) susceptibility versus the noise-herding intensity parameter $ \mu$ for $\alpha=2$ on a two-dimensional regular lattice for different system sizes. Top row corresponds to the N\nl VM, middle row to the N\nl PVM with ferromagnetic distribution of preferences and $\varepsilon=0.25$ and bottom row to the A$q$VM. The insets show the collapses with the corresponding Ising critical exponents~\cite{yeomans1992statistical}. The vertical lines indicate the critical point $\mu_\mathrm{c}$.}
 \label{fig:scaling_alpha=2_2d}
\end{figure*}

\subsection{Universality class at the critical point} \label{sec:Universality_critical}
We focus on the N\nl VM, N\nl PVM and A$q$VM, excluding from the analysis the noisy voter model with aging since it has already been shown to belong to the Ising universality class~\cite{Artime_2019}. Although our results are rather general, and for the sake of concreteness, we focus here on the specific case $\alpha=q=2$. 

In the fully connected topology, the critical point is exactly determined from the mean-field analysis presented in Sec.\ref{sec:noisynonlinear_sec_ABC}. We validate this prediction using the Binder cumulant and confirm the expected scaling behavior given in Eqs.~(\ref{eq:scalingmchi},~\ref{eq:scalingum}), as shown in Fig.\ref{fig:scaling_alpha=2_CG}. We obtain excellent overlaps for the three models when using the mean-field critical exponents $\beta = 1/2$, $\gamma = 1$, and $\nu = 1/2$, and the upper critical dimension $\dc = 4$.

For Erd\H{o}s-R\'enyi networks, we determine the critical point numerically via the Binder cumulant. Given that these networks have an effective infinite dimensionality~\cite{Eguiluz2003}, we again use the mean-field critical exponents. The resulting finite-size scaling collapse, presented in Fig.~\ref{fig:scaling_alpha=2_ER}, is accurate for all models, indicating that the Ising model mean-field universality class is a solid candidate for the models in those networks.

For further analysis, we have performed the finite-size scaling analysis for a regular lattice of dimension $d=2$ using the corresponding critical exponents of the Ising model, which are $\beta=1/8$, $ \gamma=7/4$ and $\nu=1$. The critical points are again determined via the Binder cumulant. For the N\nl VM and the A$q$VM, the data collapse is consistent with the expected scaling, see Fig.~\ref{fig:scaling_alpha=2_2d}, further supporting their inclusion in the 2D Ising universality class.

The N\nl PVM on a two-dimensional lattice exhibits more subtle behavior due to the presence of quenched disorder in agent preferences. Initially, preferences must be assigned, and a natural choice is to sample them randomly and average over many realizations. However, this approach leads to a breakdown of self-averaging: near the critical point, the magnetization exhibits large sample-to-sample fluctuations that do not vanish with increasing system size. Further details about the self-averaging issue are provided in Appendix~\ref{sec:app:self}.

To resolve this, we consider a checkerboard configuration where each agent is surrounded by agents with opposite preferences. Under this setup, the scaling laws using the 2D Ising critical exponents provide an excellent fit to the data (see Fig.~\ref{fig:scaling_alpha=2_2d}), further supporting the claim that all the models belong to the Ising universality class.

\begin{figure*}[t]
 \centering
 \includegraphics[width = \textwidth]{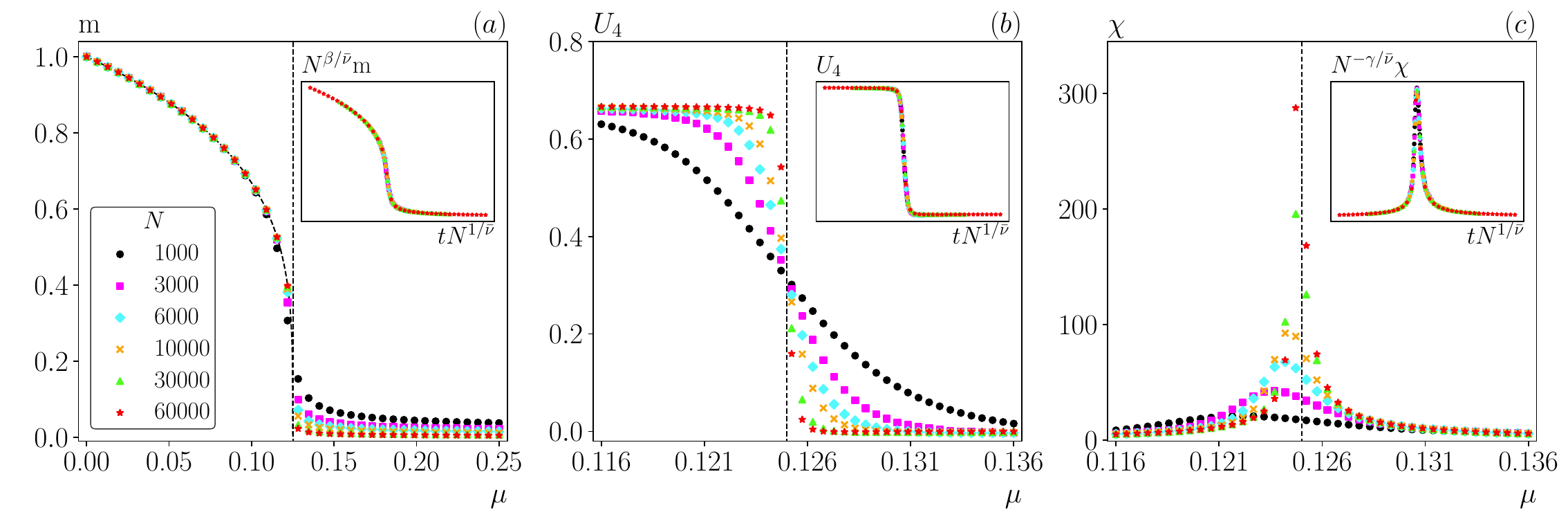}
 \includegraphics[width = \textwidth]{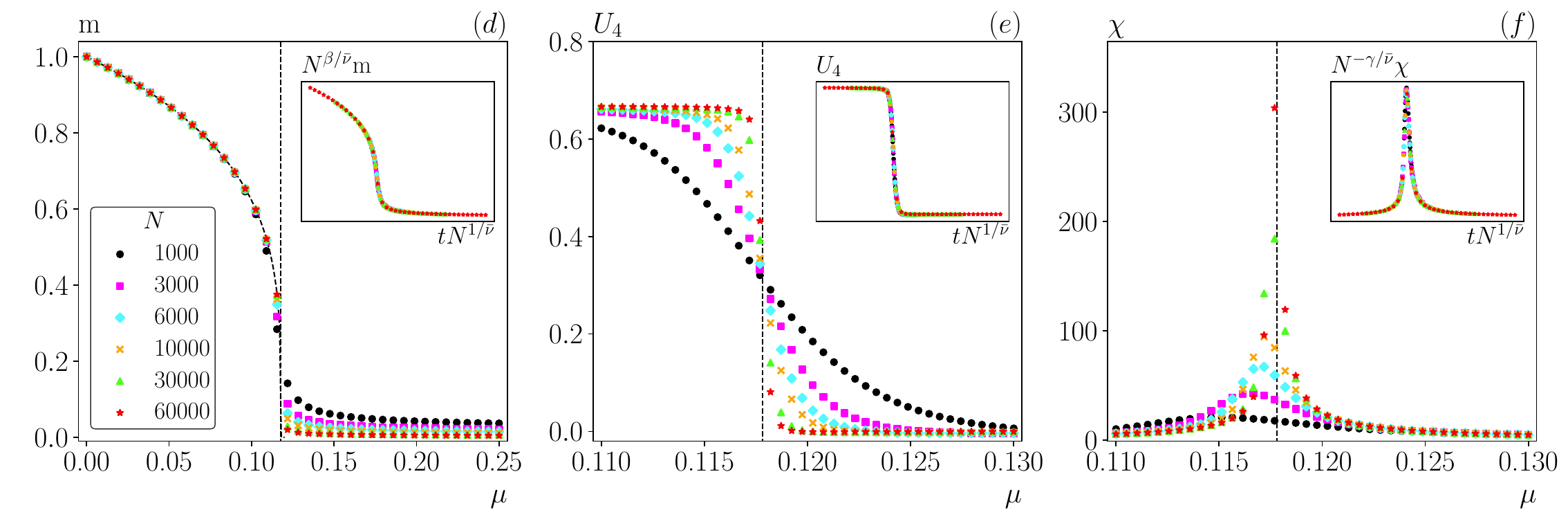}
 \caption{($a,d$) Magnetization, ($b,e$) Binder cumulant, and, ($c,f$) susceptibility versus the noise-herding intensity parameter $ \mu$ at the tricritical point on the complete graph for different system sizes. Top (resp. bottom) row corresponds to the N\nl VM (resp. N\nl PVM), with $\alpha_\mathrm{t}=5$ (resp. $\alpha_\mathrm{t}(\varepsilon=0.25)=5.0979$). In ($a,d$), the solid black line indicates the solution of Eq.~\eqref{eq:rateeqm_NLNPVM}. The insets show the collapses with the corresponding mean-field tricritical exponents~\cite{yeomans1992statistical}. The vertical lines indicate the tricritical point $\mu_\mathrm{t}$.}
 \label{fig:scaling_tri_CG}
\end{figure*}

\begin{figure*}[t]
 \centering
 \includegraphics[width = \textwidth]{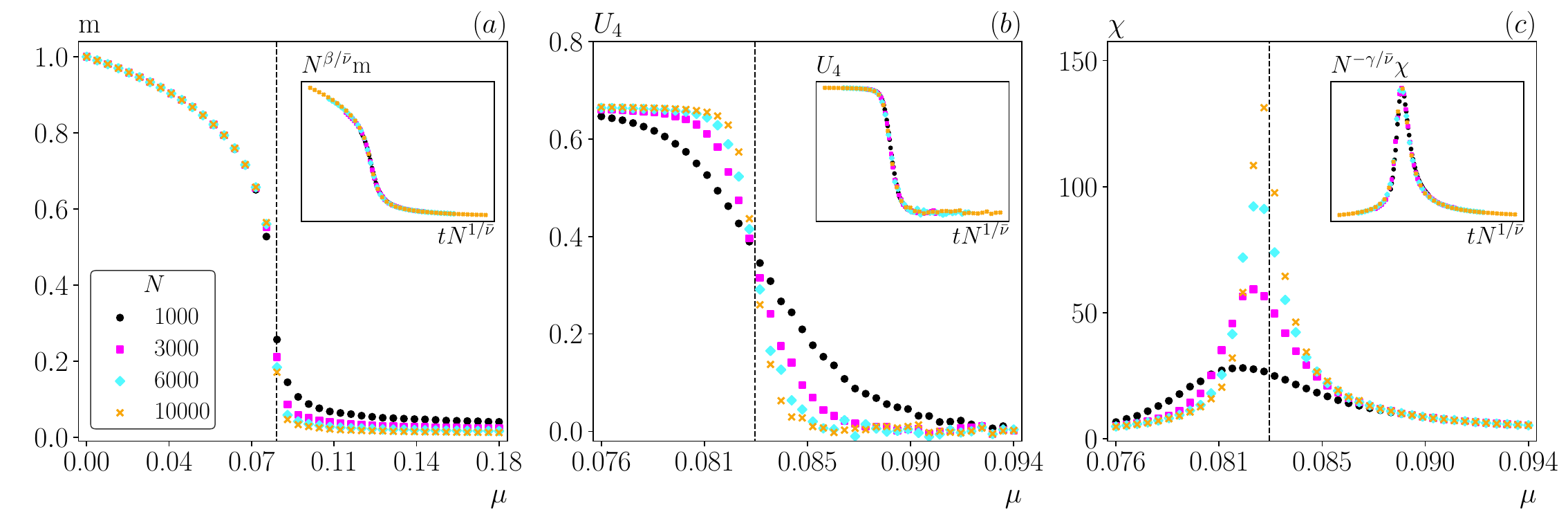}
 \caption{($a$) Magnetization, ($b$) Binder cumulant, and, ($c$) susceptibility versus the noise-herding intensity parameter $\mu$ at the tricritical point, $\alpha_\mathrm{t} = 6.07$ and $\mu_\mathrm{t} = 0.083$, for the N\nl VM on Erd\H{o}s-R\'enyi networks with $\langle k \rangle=15$ for different system sizes. The insets show the collapses with the corresponding mean-field tricritical exponents~\cite{yeomans1992statistical}. The vertical lines indicate the tricritical point $\mu_\mathrm{t}$.}
 \label{fig:scaling_tri_ER}
\end{figure*}

\subsection{Universality class at the tricritical point} \label{sec:Universality_tricritical}
We study here the universality class of the tricritical transitions exhibited by both the N\nl VM and the N\nl PVM. We begin with the N\nl VM on the complete graph. The mean-field analysis in Sec.~\ref{sec:noisy_models} yields he tricritical point $\alpha_\mathrm{t} = 5$, $\mu_\mathrm{t} = 1/8$. We study the scaling behavior fixing $\alpha=\alpha_\text{t}$ while varying the noise intensity $\mu$ crossing the value $\mu_\text{t}=1/8$. As shown in Fig.~\ref{fig:scaling_tri_CG}, we observe an excellent agreement with the scaling laws Eqs.~(\ref{eq:scalingmchi},\ref{eq:scalingum}) using the mean-field tricritical exponents $\beta=1/4$, $\gamma=1$, $\nu=1/2$ and the upper-critical dimension $\dc=3$.

We next consider the N\nl PVM on the complete graph. For a fixed value of $\varepsilon$, the tricritical point $\alpha_\text{t},\mu_\text{t}$ is obtained by solving $A-\eta=A+B=0$ using Eqs.~\eqref{eq:NLNPVM_parameters}. Although it is not possible to obtain an analytical expression, the numerical analysis yields, for instance, $\alpha_\text{t}=5.0979$ and $\mu_\text{t}=0.1178$ for $\varepsilon=0.25$. Again, we find excellent agreement with the scaling relations using the mean-field tricritical exponents, see Fig.~\ref{fig:scaling_tri_CG}.

We further expect that these mean-field tricritical exponents remain valid for both models on networks with high dimensionality, such as Erd\H{o}s-R\'enyi networks. However, verifying this prediction requires determining the tricritical point $(\alpha_\mathrm{t}, \mu_\mathrm{t})$ with sufficient precision.

As a first attempt, one could employ the ``pair approximation" (PA) method~\cite{VazquezPA, lucia_NLVM, Peralta_2018}, which has been already applied to the N\nl VM, to describe accurately the stationary value of the order parameter $\tm$. Using this approach, the approximate tricritical point for an Erd\H{o}s-R\'enyi network with average degree $\langle k \rangle = 15$ can be estimated as $\alpha_\mathrm{t} = 6.1547$ and $\mu_\mathrm{t} = 0.08168$ (see Appendix~\ref{app:PA_NLNVM} for details). However, this method yields insufficient precision in determining the tricritical point and so it fails to satisfy finite-size-scaling relations with the required accuracy. 
 
To overcome this limitation, we introduce an alternative method for accurately locating the tricritical point, which we detail in Appendix~\ref{sec:app:tri}. Owing to its general applicability to models exhibiting tricritical behavior, this method provides a reliable tool for studying scaling properties. Applying it to the N\nl VM on an ER network with $\langle k \rangle = 15$, we obtain refined estimates $\alpha_\mathrm{t} = 6.07$ and $\mu_\mathrm{t} = 0.083$, significantly improving upon PA predictions. With this accurate value, the scaling analysis shown in Fig.~\ref{fig:scaling_tri_ER} supports the conclusion that this system belongs to the mean-field tricritical universality class.

\begin{table*}[t]
\renewcommand{\arraystretch}{1.5}
\centering
\begin{tabular}{ccccc}
\toprule
\textbf{Model} & \textbf{Parameters}  & \textbf{Absorbing states} & \textbf{Canonical $(A,B)$} & \textbf{Universality Class} \\ 
\midrule
Nonlinear Voter Model (\nl VM) & $\alpha$ & Yes & Eqs.~\eqref{eq:NLVM} & GV  \\
Nonlinear Partisan Voter Model (\nl PVM)& $\alpha$, $\varepsilon$ & Yes & Eqs.~\eqref{eq:NLPVM} & GV \\
Nonlinear $q$-Voter Model (\nl$q$VM) & $q,\epsilon$ & Yes & Eqs.~\eqref{eq:q-voter} &  GV \\
Voter Model with Aging & $p_0$, $p_\infty$ & Yes & Numerical  & GV \\
\addlinespace
Noisy Nonlinear Voter Model (N\nl VM) & $\alpha$, $\mu$ & No & Eqs.~\eqref{eq:NLNVM_parameters} & Ising / MGV  \\
Noisy Nonlinear Partisan Voter Model (N\nl PVM) & $\alpha$, $\varepsilon$, $\mu$ & No & Eqs.~\eqref{eq:NLNPVM_parameters} & Ising / MGV  \\
Noisy $q$-Voter Model (N$q$VM) & $q$, $\epsilon=0$, $\mu$ & No & Eqs.~\eqref{eq:NLNPVM_parameters} with $\alpha=q$ & Ising / MGV  \\
$q$-Voter Model with Anticonformity (A$q$VM) & $q$, $\epsilon=0$, $\mu$ & No & Eqs.~\eqref{eq:noisyq-voter_anti_params} & Ising  \\
Noisy Voter Model with Aging  & $p_0$, $p_\infty$, $\mu$ & No & Numerical & Ising \\
\bottomrule
\end{tabular}
\caption{Summary of the nonlinear voter models analyzed in this work. For each model, we provide its full name, acronym, relevant parameters, presence of absorbing states, mapping parameters $(A,B)$ of the canonical models Eq.~\eqref{eq:m_general} or Eq.~\eqref{eq:m_noisy_general}, and universality class of the phase transitions.}
\label{tab:models_summary}
\end{table*}

\section{Summary and conclusions}\label{sec:conclusions}
In this paper, we addressed the question of the universality classes of phase transitions in nonlinear voter models. As a reference framework, we first studied how a variety of different models with two symmetric absorbing states can be described by the canonical model introduced in Ref.~\cite{Hammal}. This canonical model accounts for three different phase transitions: (1) An Ising transition, associated with $\mathbb{Z}_2$ symmetry breaking, where the stable state of the system shifts continuously from $m=0$ to one of the two stable solutions $\pm m^*$. (2) A directed percolation (DP) transition, in which the stable state becomes one of the two absorbing states $m=\pm1$. (3) A generalized voter transition (GV), understood as a superposition of the former two, characterized by a discontinuous transition from $m=0$ to $m=\pm1$. We demonstrated that several models, such as the nonlinear voter model (\nl VM), the nonlinear partisan voter model (\nl PVM), the nonlinear $q$-voter model (qVM), the Sznajd model and the voter model with aging can be mapped, at the mean field level, onto the canonical model of Ref.~\cite{Hammal}. We also found that the $\mathbb{Z}_2$ symmetry inherent in their local rules consistently leads to a GV transition.

We then extended this framework to include noise, modeling imperfect imitation where agents can change state independently of their neighbors. This noise destroys the absorbing states and introduces a drift toward the disordered state $m = 0$. In this proposed canonical mean-field model, DP transitions are no longer present, and the GV transition becomes a modified generalized voter (MGV) transition: a discontinuous transition from $m = 0$ to a partially ordered state $\pm m^* \ne \pm 1$. The phase diagram features a line of MGV transitions and a line of continuous Ising transitions merging at a tricritical point.

We showed that many noisy nonlinear models from the literature, such as the noisy nonlinear voter model (N\nl VM), the nonlinear voter model with anticonformity (A$q$VM), the noisy voter model (NVM) with aging, can be mapped onto this canonical model. We also introduced the noisy nonlinear partisan voter model (N\nl PVM) as a novel model combining nonlinearity and preference under noise, and showed that it also maps consistently onto the same model. Considering a stochastic framework to account for finite-size effects we show that these models exhibit well defined transitions in the thermodynamic limit, while their linear counterparts display finite-size transitions that disappear in the thermodynamic limit. 

Beyond the mean-field analysis, we applied finite-size scaling techniques to investigate the universality classes of these transitions. We established that the continuous Ising-like transitions exhibited by the N\nl VM, N\nl PVM, and A$q$VM indeed belong to the Ising family of universality classes. The analysis of these models on a complete graph, a two-dimensional regular lattice and Erd\H{o}s–R\'enyi networks confirm the critical exponents of the Ising universality class below and above the upper-critical dimension. On the other hand, for both the N\nl VM and N\nl PVM, we also characterized the phase transition at the tricritical point on the complete graph and confirmed that the associated critical exponents match those of the mean-field tricritical universality class with $d_c = 3$.

The validation of scaling laws needs an accurate determination of the transition point. However, determining the exact location of the tricritical point in both complex and regular networks is more challenging due to the lack of theoretical predictions. To address this, we developed a high-precision numerical method for localizing the tricritical point. Applying it to the N\nl VM on Erd\H{o}s–R\'enyi networks, we again confirmed the expected mean-field tricritical exponents.

This work is a step forward toward a unified classification of phase transitions in nonequilibrium models of opinion dynamics.

\acknowledgments{
We thank Tobias Galla and Christopher Kitching for useful comments on the Partisan Voter Model. We are also grateful to Antonio F. Peralta for insightful discussions on the determination of the tricritical point. Partial financial support has been received from Grants PID2021-122256NB-C21/C22 and PID2024-157493NB-C21/C22 funded by MICIU/AEI/10.13039/501100011033 and by “ERDF/EU”, and the María de Maeztu Program for units of Excellence in R\&D, grant CEX2021-001164-M.

\begin{widetext}
\appendix
\section{The nonlinear noisy partisan voter model} \label{app:sec:NLNPVM}
In addition to the binary state variable $s_i\in\{-1,1\}$, each agent has a preference for one of the two states, $p_i\in\{-1,1\}$. This gives rise to four distinct types of agents, labeled $i^+_+$, $ i^+_-$, $ i^-_+ $ and $ i^-_-$, where the superscript indicates the preference and the subscript indicates the state of the agent. The variables $ n^p_s,\,x^p_s$; $ s,p=\pm1$, denote the number and the density of agents of each type, respectively. For simplicity, we consider that the number of supporters of each state is the same, half of the population, such that the relations $x^+_++x^+_-=1/2$, $ x^-_++x^-_-=1/2$ hold throughout the dynamics. The model evolves according to the following rules: Initially, the states $\{s_i\}$ are randomly assigned. At the same time, half of the agents are randomly designated with a preference $ p_i = +1 $, while the other half are assigned $p_i=-1$. Then,
\begin{enumerate}
\item Select an agent, say, $i$, with state $ s_i $.
\item With probability $ \displaystyle\frac a2$, agent $i$ randomly selects one of the two possible states.
\item Otherwise, with probability $(1-a)$, agent $i$ is persuaded to change its state with a probability equal to a power $ \alpha $ of the fraction of neighbors holding the opposite state. Then, the change is accepted with probability $\displaystyle\frac{1}{2}(1-s_i p_i\varepsilon)$.
\end{enumerate}
These updating rules, lead to the following individual transition probabilites,
\begin{subequations} \label{app:eq:i_rates}
 \begin{align}
 \omega^{++}_i&\equiv\omega(i^+_-\to i^+_+)=\frac a4+(1-a)\left(\frac{1}{k_i}\sum_{j\in \nu(i)}\frac{1+s_j}{2}\right)^{\alpha}\left(\frac{1+\varepsilon}{2}\right), \\
 \omega^{+-}_i&\equiv\omega(i^+_+\to i^+_-)=\frac a4+(1-a)\left(\frac{1}{k_i}\sum_{j\in \nu(i)}\frac{1-s_j}{2}\right)^{\alpha}\left(\frac{1-\varepsilon}{2}\right),\\
 \omega^{-+}_i&\equiv\omega(i^-_-\to i^-_+)=\frac a4+(1-a)\left(\frac{1}{k_i}\sum_{j\in \nu(i)}\frac{1+s_j}{2}\right)^{\alpha}\left(\frac{1-\varepsilon}{2}\right),\\
 \omega^{--}_i&\equiv\omega(i^-_+\to i^-_-)=\frac a4 +(1-a)\left(\frac{1}{k_i}\sum_{j\in \nu(i)}\frac{1-s_j}{2}\right)^{\alpha}\left(\frac{1+\varepsilon}{2}\right),
 \end{align} 
\end{subequations}
where $k_i$ is the degree of agent $i$ and $\nu(i)$ is its set of neighbors.

In the complete graph, the global variables are sufficient to describe the state system. Since the preference is fixed, i.e., $x^+_++x^+_-=1/2$, $ x^-_++x^-_-=1/2$ and the number of agents is conserved $x^+_++x^+_-+x^-_++x^-_-=1$, two independent variables are needed to describe the system. For convenience they have been chosen to be the difference $\Delta\equiv x^+_+-x^-_- $ and the sum $\Sigma\equiv x^+_++x^-_-$ of the density of agents that are in their preferred state. $\Delta$ is related with the magnetization as $m=2\Delta$, while $\Sigma$ can be interpreted as the density of ``satisfied" agents. The global transition rates in terms of these variables are given by
\begin{subequations} \label{app:eq:NLNPVM_global_rates}
\begin{align}
 \Omega^{++}&\equiv \Omega\left(\Delta \to \Delta+ 1/N;\Sigma \to \Sigma + 1/N \right)=\frac{N}{2}\left[\frac a4+(1-a)\left(\frac{1}{2}+\Delta\right)^{\alpha}\left(\frac{1+\varepsilon}{2}\right)\right]\left(1-\Delta-\Sigma\right),\\
 \Omega^{+-}&\equiv \Omega\left(\Delta \to \Delta + 1/N;\Sigma \to \Sigma - 1/N \right)=\frac{N}{2}\left[\frac a4+(1-a)\left(\frac{1}{2}+\Delta\right)^{\alpha}\left(\frac{1-\varepsilon}{2}\right)\right]\left(\Sigma-\Delta\right),\\
 \Omega^{-+}&\equiv \Omega\left(\Delta \to \Delta - 1/N;\Sigma \to \Sigma + 1/N \right)=\frac{N}{2}\left[\frac a4+(1-a)\left(\frac{1}{2}-\Delta\right)^{\alpha}\left(\frac{1+\varepsilon}{2}\right)\right]\left(1+\Delta-\Sigma\right),\\ 
 \Omega^{--}&\equiv \Omega\left(\Delta \to \Delta - 1/N;\Sigma \to \Sigma - 1/N \right)=\frac{N}{2}\left[\frac a4+(1-a)\left(\frac{1}{2}-\Delta\right)^{\alpha}\left(\frac{1-\varepsilon}{2}\right)\right]\left(\Sigma+\Delta\right).
\end{align}
\end{subequations}

From these transition rates, one can write down the rate equations for $ \Delta,\Sigma $ in the mean-field limit, meaning by that complete graph and thermodynamic limit, as
\begin{align} \label{app:eq:nonlinearNPVM_rateeq_D} \frac{d\Delta}{dt}&=-\frac a2 \Delta + 2^{-2-\alpha} (1-a)\left[(1+2 \Delta )^{\alpha } (1+ \varepsilon -2 \Delta -2 \Sigma \varepsilon) -(1-2 \Delta )^{\alpha } (1 + \varepsilon+2 \Delta -2 \Sigma \varepsilon)\right], \\ 
\label{app:eq:nonlinearNPVM_rateeq_S} \frac{d\Sigma}{dt}&= \frac a4 (1-2 \Sigma )+ 2^{-2 -\alpha}(1-a) \left[(1+2 \Delta )^{\alpha } (1+\varepsilon -2 \Delta \varepsilon -2 \Sigma ) +(1-2 \Delta )^{\alpha } (1+\varepsilon +2 \Delta \varepsilon -2 \Sigma)\right].
\end{align}
With the aim of simplifying the description of the system, we now obtain an equivalent one-variable model with effective rates for the $\Delta$ variable. This is a further approximation inspired by the result obtained in Ref.~\cite{llabres} for $\alpha=1$, where an adiabatic elimination technique was applied to obtain an equivalent reduced model. Importantly, this reduction does not alter the location of the fixed points. Even if the adiabatic approximation does not hold globally for all values of $\alpha$, the fixed points remain unchanged. By setting the time derivative of $\Sigma$, Eq.~\eqref{app:eq:nonlinearNPVM_rateeq_S} to zero and substituting $ \Sigma(\Delta) $ into Eqs.~\eqref{app:eq:i_rates} one obtains the global rates of the reduced model
\begin{subequations} \label{app:eq:NLNPVM_eff_rates}
\begin{align} 
 \Omega^+(\Delta)&\equiv \Omega^{++}+\Omega^{+-} = \frac{N}{4}\Biggl[ (1-2 \Delta )
 \left(\frac a2 +(1-a) \left(\frac{1}{2}+\Delta\right)^{\alpha}\right) \nonumber \\
 &-(1-a)^2 \varepsilon ^2\left(\frac{1}{2}+\Delta\right)^{\alpha } \left(1-4\Delta^2\right)\frac{(1-2 \Delta)^{\alpha -1}+ (1+2\Delta )^{\alpha-1 }}{2^{\alpha} a + (1-a)\left[ (1+2 \Delta )^{\alpha}+(1-2 \Delta)^{\alpha}\right]} \Biggr] \label{app:NPVM_eff_wp},\\
 \Omega^-(\Delta)&\equiv \Omega^{-+}+\Omega^{--} = \frac{N}{4}\Biggl[ (1+2 \Delta )
 \left(\frac a2 +(1-a) \left(\frac{1}{2}-\Delta\right)^{\alpha}\right) \nonumber \\
 &-(1-a)^2 \varepsilon ^2\left(\frac{1}{2}-\Delta\right)^{\alpha } \left(1-4\Delta^2\right)\frac{(1-2 \Delta)^{\alpha -1}+ (1+2\Delta )^{\alpha-1 }}{2^{\alpha} a + (1-a)\left[ (1+2 \Delta )^{\alpha}+(1-2 \Delta)^{\alpha}\right]} \Biggr] \label{app:NPVM_eff_wm}.
\end{align}
\end{subequations}
After using a standard approach~\cite{vanKampen:2007,Toral2014StochasticNM}, one can write down a Fokker-Planck equation, Eq.~\eqref{eq:FP_1D}, for the probability density function $ P(m,t)$ where the drift $ F(m)=\left(\Omega^+-\Omega^-\right)/N $ and the diffusion $ D(m)=~\left(\Omega^++\Omega^-\right)/2N $ are given by
\begin{align}
 F(m)=&-\frac a2 m\nonumber \\
 + 2^{-1-\alpha}& (1-a) \left(1-m^2\right) \Biggl[ (1+m)^{\alpha -1}-(1-m)^{\alpha -1}-\frac{(1-a) \left[(1+m)^{\alpha -1}+(1-m)^{\alpha -1}\right]
 \left[(1+m)^{\alpha }-(1-m)^{\alpha }\right]}{2^{\alpha} a+(1-a)
 \left[(1+m)^{\alpha }+(1-m)^{\alpha }\right]} \varepsilon ^2\Biggr]
 \label{app:NPVM_drift} \\
 D(m)&=\frac a2+2^{-1-\alpha } (1-a) \left(1-m^2\right) \left[(1+m)^{\alpha -1}+(1-m)^{\alpha -1}\right]
 \left[1-\frac{(1-a) \left[(1+m)^{\alpha }+(1-m)^{\alpha }\right]}{2^{\alpha}
 a+(1-a) \left[(1+m)^{\alpha }+(1-m)^{\alpha }\right]} \varepsilon ^2\right]\label{app:NPVM_diff}
\end{align}
where we have used the definition of the magnetization $ m=2\Delta $ and the relation $ |dm/d\Delta|=2 $ to express the rates and hence the functions $ F(m) $ and $ D(m) $ in terms of $ m$.

The rate equation of the reduced model for the magnetization is given by the drift, $\displaystyle \frac{dm}{dt}=F(m)$. In the limiting case of vanishing preference $ \varepsilon=0$, the rate equation for the N\nl VM, Eq.~\eqref{eq:rateeq_NLVM}, is recovered with an overall time-rescaling factor $1/2$. Additionally, by taking $a=0$, one obtains all the corresponding equations for the \nl PVM.

\section{Noisy voter model with aging. Series expansion of the rate equation}\label{app:vm_aging}
As detailed in Ref.~\cite{aging_Jaume_Sara}, in the mean-field limit the rate equation for the density $x$ of agents in state $+1$, is given by 
\begin{equation}
\frac{dx}{dt}=\frac{a}{2}(1-2x) + (1-a) x (1-x) [\Phi_a(x) - \Phi_a(1-x)],
\end{equation}
where the function $\Phi_a(x)$ for the particular functional form of Eq.~\eqref{eq:p_tau_aging} is given by
\begin{equation}
\Phi_a(x)=\frac{p_0 \,_2F_1\left(1,\tau^*\xi(x,a);1+\tau^*;\gamma(p_\infty x, a) \right)+\displaystyle \frac{p_\infty }{1+\tau^*}\gamma(p_0 x,a)\, _2F_1\left(2,1+\tau^*\xi(x,a);2+\tau^*;\gamma(p_\infty x, a)\right)}{_2F_1\left(1,\tau^*\xi(x,a);\tau^*;\gamma(p_\infty x, a) \right)},
\end{equation}
in terms of the hypergeometric function $_2F_1$, and where we have introduced the functions 
\begin{equation} 
\label{eq:Xi_def}
\xi(x,a)\equiv\frac{\gamma(p_0 x,a)}{\gamma(p_\infty x,a)}, \quad \gamma(z,a)= \frac{a}{2}+(1-a)(1-z).
\end{equation}
By using the relation $m=2x-1$, it is possible to write down the rate equation for the magnetization as
\begin{equation} \label{eq:app:nvm_aging}
 \frac{dm}{dt}=-am+(1-a)\,\frac{1-m^2}{2}\left[\Phi_a\left(\frac{1+m}{2}\right)-\Phi_a\left(\frac{1-m}{2}\right)\right].
\end{equation}
In order to identify the parameters $ A,B $, a power-series expansion of the second term of the equation around $ m=0 $ up to $ O(m^3) $ is needed. Unfortunately, it is not possible to derive manageable expressions for $A,B$ for arbitrary values of $p_0, p_\infty,\tau^*, a$, and we need to compute them numerically. In the limit $a=0$, Eq.~\eqref{eq:app:nvm_aging} becomes the rate equation for the noiseless voter model with aging, Eq.~\eqref{eq:m_aging}.

\section{Numerical determination of the stationary distribution for the magnetization}\label{app:pst}

Let $n$ indicate the number of agents in state $+1$. In the all-to-all coupling scheme the rates at which $n$ increases or decreases by $1$ can be determined from the microscopic rates of the process. For example, for the noisy nonlinear voter model, we have
\begin{subequations}
\begin{align}
\Omega(n\to n+1)&=(N-n)\left[\frac a2 + (1-a) \left(\frac nN\right)^\alpha\right],\\
\Omega(n\to n-1)&=n\left[\frac a2 + (1-a) \left(\frac{N-n}{N}\right)^\alpha\right].
\end{align}
\end{subequations}
The master equation of the process providing the evolution of the probability distribution $P(n,t)$ is~\cite{vanKampen:2007, Toral2014StochasticNM}
\begin{equation}\label{app:eq:masterNLNVM}
\frac{\partial P(n;t)}{\partial t}= \left( E^{-1} - 1\right)\left[ \Omega(n\to n+1)P(n;t)\right] +\left( E - 1\right) \left[ \Omega(n\to n-1)P(n;t)\right].
\end{equation}
Here $E$ is the step operator acting on any function $f(j)$ as $E^l[f(j)]=f(j+l)$. By setting the time derivative to zero of Eq.~\eqref{app:eq:masterNLNVM}, one arrives at the recurrence relation
\begin{equation}
P_\mathrm{st}(n)=P_\mathrm{st}(0)\prod_{k=0}^{n-1}\frac{\Omega(k\to k+1)}{\Omega(k+1\to k)}.
\end{equation}
Starting from an initial arbitrary value $P_\mathrm{st}(0)$ we can compute numerically using the previous relation all the stationary probabilities $P_\mathrm{st}(n)$. Finally, the value of $P_\mathrm{st}(0)$ is obtained from the normalization condition $\sum_{n=0}^NP_\mathrm{st}(n)=1$. The stationary distribution for the magnetization $m$ follows readily from the relation $m=\dfrac{2n}{N}-1$.

\section{$q$-voter model with anticonformity}\label{app:sec:qVMC}
From the updating rules of the A$q$VM described in Sec.~\ref{sec:noisyq-voter}, it is possible to write down the global rates of the process as
\begin{subequations}
\begin{align}
\Omega(n \rightarrow n+1)&=(N-n)\left[\left(\frac{n}{N}\right)^q+a\left(\frac{N-n}{N}\right)^q\right], \\
\Omega(n+n-1)&=n\left[\left(\frac{N-n}{N}\right)^q+ a\left(\frac{n}{N}\right)^q\right],
\end{align}
\end{subequations}
which leads to the following drift and diffusion functions of a Fokker-Planck equation, Eq.~\eqref{eq:FP_1D},
\begin{align} \label{app:eq:AqVM_FP_drift}
F(m)&=2^{-q} \Bigl(-a \left[(1+m)^{q+1}-(1-m)^{q+1}\right]+(1-m^2)\left[(1+m)^{q-1}-(1-m)^{q-1}\right]\Bigr)\\ \label{app:eq:AqVM_FP_diffusion}
 D(m)=&2^{1-q}\Bigl(a\left[(1+m)^{q+1}+(1-m)^{q+1}\right]+(1-m^2)\left[(1+m)^{q-1}+(1-m)^{q-1}\right]\Bigr),
\end{align}
where we have used the relations $ m=2n/N-1 $ and $ |dm/dn|=2/N $ to express the rates and hence the functions $ F(m) $ and $ D(m) $ in terms of $ m$.
\section{The self-averaging problem in the nonlinear noisy partisan voter model} \label{sec:app:self}
When simulating the N\nl PVM on networks other than the complete graph, preferences are initially distributed randomly across the nodes of the network and remain fixed throughout the dynamics. Then, initial values of the state variables are randomly assigned and the evolution of the system begins. The distribution of preferences is hence a quenched disorder, whose relevance may vary depending on the network topology. Macroscopic variables, such as the magnetization $\tm$, follow a double average: over the stationary state for a given distribution of preferences, and over the distribution of preferences. The concept of self-averaging emerges when calculating those averages and their fluctuations~\cite{lack_SA,lack_SA_lett}. A system is self-averaging if macroscopic observables converge to a unique value when averaging over different realizations of the disorder. In such systems, fluctuations decrease as the system size increases, vanishing in the thermodynamic limit. On the other hand, non-self-averaging systems are characterized by macroscopic variables that fail to converge to a single value. In these cases, significant fluctuations persist regardless of the growth in system size. Additionally, quenched disorder significantly amplifies fluctuations near the critical point. The degree of self-averaging can be quantified by the relative variance $R[m]$ of the order parameter $m$, defined as
\begin{equation} \label{app:eq:relative_var}
R[m]=\frac{\sigma^2[m]}{\langle \langle |m|\rangle \rangle_\text{st}^2}=\frac{\langle \langle m^2 \rangle \rangle_\text{st}}{\langle \langle |m|\rangle \rangle_\text{st}^2}-1,
\end{equation}
where $\langle\langle \cdot \rangle\rangle_\text{st}$ denotes the double average taken in the stationary regime. In a self-averaging system, $ R[m] $ vanishes in the thermodynamic limit, i.e., $R[m] \to 0 $ as $ N \to \infty $, indicating that fluctuations between different disorder realizations become negligible for large system sizes.

For the N\nl PVM, the degree of self-averaging is highly sensitive to the underlying network topology, as shown in Fig.~\ref{fig:app:self-av}. In Erd\H{o}s-R\'enyi networks, which effectively have an infinite dimensionality, the system exhibits clear self-averaging behavior. The relative variance decays with system size as a power law, $R[m] \sim N^{-z}$, with an exponent $ z = 1.01(1)$, obtained via linear regression in log-log scale, see inset of panel $(a)$.

In contrast, on a two-dimensional square lattice, self-averaging behavior depends on the spatial distribution of agent preferences. If we set a checkerboard configuration, in which each agent is surrounded by neighbors with opposite preferences, the system is self-averaging. In this case, the relative variance again follows a power-law decay, $ R[m] \sim N^{-z} $, with a fitted exponent $ z = 1.06(7) $, as illustrated in panel $(b)$. However, when preferences are randomly distributed, the system exhibits non-self-averaging behavior, evidenced by an increase in $ R[m] $ with increasing system size, see inset of panel $(c)$. 

\begin{figure}[h]
\centering
\includegraphics[width = \textwidth]{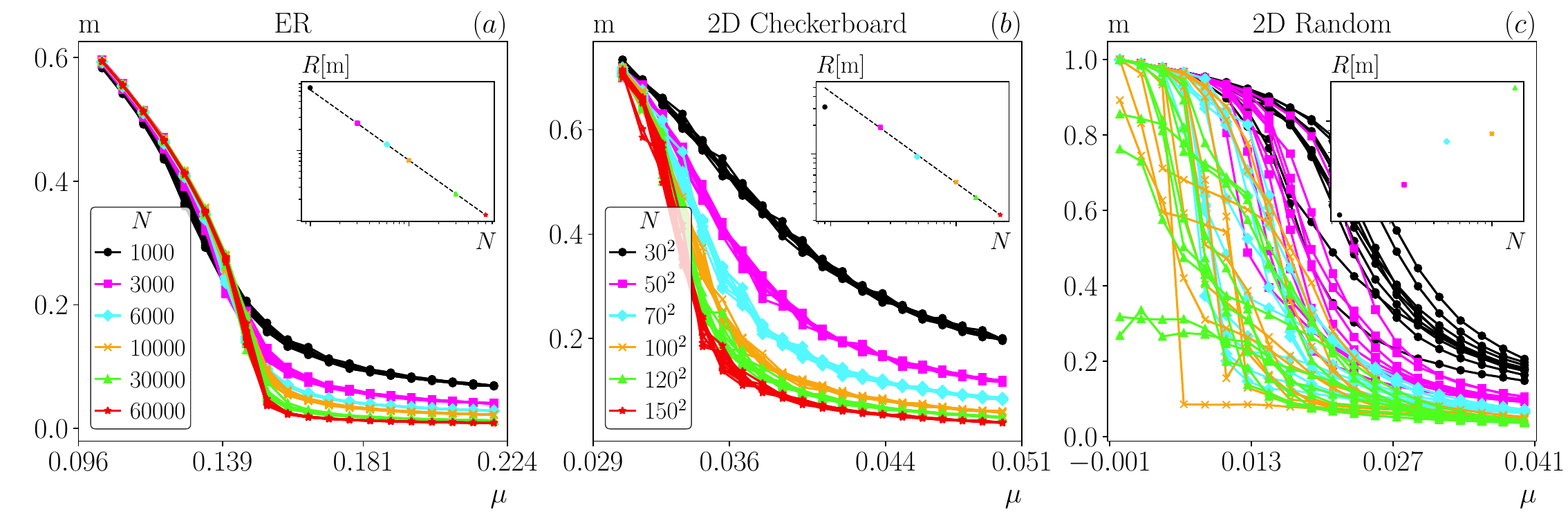}
\caption{Realizations of the magnetization, \tm, versus the noise-herding intensity ratio parameter $\mu$ for $\alpha=2$ and different preferences distribution and topologies: $(a)$ Erd\H{o}s-R\'enyi networks with $\langle k \rangle=8$ (left), two-dimensional square lattice with checkerboard $(b)$ or with random $(c)$ preferences distribution. Insets: Relative variance $R[m]$, as defined in Eq.~\ref{app:eq:relative_var} at the critical point versus the system size $N$. Insets are in log-log scale making evident a power-law behavior $R\sim N^{-z}$ and lines represents the linear fit with slope $z=1.01(1)$ in $(a)$ and $z=1.06(7)$ in $(b)$. For random preference distribution $(c)$ the system becomes non-self averaging.}
\label{fig:app:self-av}
\end{figure}

\section{Pair-approximation in the noisy nonlinear voter model} \label{app:PA_NLNVM}

We apply the \textit{pair approximation} developed in Ref.~\cite{Peralta_2018} to study the dynamics of the N\nl VM. In this approach, the density of active links $\rho$, understood as the density of links connecting agents in opposite state, is used to describe the state of the system. Additionally, it is convenient to define the number of agents with degree $k$, $N_k$, of which $\textbf{n}_k$ are in state $+1$, the number of links coming out of nodes in state $+1$, $n_L=\frac 12\sum_{i=1}^{N}k_i (1+s_i)$ and the intensive variables, $m_k=2\textbf{n}_k/N_k-1$ and $m_L=2 n_L/N\langle k \rangle -1$, with $\langle k \rangle$ the average degree of the network.

Since $m_k=m_L=0$ is always a fixed point of the dynamics, the approximate mean-field evolution equation for $\rho$ reads
\begin{equation} \label{app:eq:drhodt}
\frac{d \rho}{d t}=2\mu (1-2 \rho)+\frac{2}{\langle k \rangle } \sum_k P_k\left\langle(k-2 q)\left(\frac{q}{k}\right)^\alpha\right\rangle_{-1}, 
\end{equation}
where $q$ is the number of neighbors with state $+1$, $P_k=N_k/N$ is the degree distribution and $\langle \dots \rangle_{-1}$ denotes the average over the probability $P_{-1}(k,q)$ of selecting an agent in state $-1$ with degree $k$ and $q$ neighbors in state $+1$. Within the pair approximation, this probability is assumed to be binomial $P_{-1}(k,q)\approx \binom{k}{q}\rho ^q (1-\rho)^{k-q}$.

Next, we write the rates of the process in the $m_L$ variable as
\begin{align}
\pi^+_k&= (N_k-n_k)\left[\frac a2 +(1-a) \left\langle \left(\frac{q}{k}\right)^\alpha \right\rangle_{-1} \right],\\
\pi^-_k&= n_k\left[\frac a2 +(1-a) \left\langle \left(\frac{k-q}{k}\right)^\alpha \right\rangle_1 \right],
\end{align}
where $\langle \dots \rangle_{-1/1}$ are the averages over the binomial distributions with single event probabilities $p_{-1}=\frac{\rho}{1-m_L}$ and $1-p_1=\frac{\rho}{1-m_L}$, where $m_L\neq0$. These rates allow us to obtain the drift $\displaystyle F_L=\sum_k \frac{2k}{\langle k \rangle N}\left[\pi^+_k-\pi^-_k\right]$ of a Fokker-Planck equation. After an adiabatic elimination $m_k\approx m_L$, $\rho \approx \xi (1-m_L^2)$ and a series expansion in powers of $m_L$, we obtain
\begin{equation}
F_L=-a m_L+\frac{1-a}{\langle k \rangle}\left(c_1 m_L+c_3 m_L^3+O(m_L^5)\right),
\end{equation}
where the coefficients $c_1$ and $c_3$ are given by
\begin{align}
c_1&=\frac 12 \sum_{k}P_k k^{1-\alpha}\left[ f_k'(0)-f_k(0)\right],\\
c_3&=\sum_{k}P_k k^{1-\alpha} \left[\frac 13 f_k'''(0)- \frac 12 f_k''(0)\right] ,\\
f_k(m_L)&=\sum_{q=0}^k\binom{k}{q}q^\alpha\left[\xi(1+m_L)\right]^q \left[1-\xi(1+m_L)\right]^{k-q}.
\end{align}
The tricritical point is calculated numerically, by setting Eq.~\eqref{app:eq:drhodt} to zero, together with the conditions $-a+c_1 (1-a)  / \langle k \rangle = 0$ and $c_3=0$ and finding $(\xi, \mu, \alpha)$. 

For Erd\H{o}s-R\'enyi networks, the degree distribution is given by 
\begin{equation}
P_k=\frac{\langle k \rangle^k e^{- \langle k \rangle} }{k!},
\end{equation}
and the obtained tricritical point is $(\alpha_\mathrm{t},\mu_\mathrm{t})=(6.1547, 0.08168)$.

\section{Accurate determination of the tricritical point} \label{sec:app:tri}

The tricritical point in the mean-field limit can be determined analytically using the canonical model introduced in Sec.~\ref{sec:noisynonlinear_sec_ABC}. Here, we present a numerical procedure for arbitrary network topologies to identify the tricritical point via finite-size scaling analysis. The core idea is to traverse the Ising transition line in the parameter space until tricritical behavior is observed.

As discussed in Sec.~\ref{sec:noisy_models}, both the N\nl VM and the N\nl PVM exhibit a tricritical point. We consider the two-dimensional parameter space $(\alpha, \mu)$ and move along the Ising transition line, keeping all other model parameters fixed (e.g., $\varepsilon$ in the case of the N\nl PVM).

For each value of $\alpha$, the location of the continuous (Ising or tricritical) transition is identified by the crossing point of the Binder cumulant $U(\alpha, \mu, N)$ computed for different system sizes $N$. This crossing occurs independently of the specific values of the critical exponents. Once the transition line is determined, the tricritical point $(\alpha_\mathrm{t}, \mu_\mathrm{t})$ is identified by analyzing the scaling behavior of the order parameter. According to Eq.~\eqref{eq:scalingmchi-a}, the rescaled order parameter $\tm(\alpha, \mu, N)\, N^{\beta_\mathrm{t}/\bar{\nu}_\mathrm{t}}$ becomes independent of $N$ when evaluated at the tricritical point, provided the correct tricritical exponents $\beta_\mathrm{t}$ and $\bar{\nu}_\mathrm{t}$ are used. Therefore, plotting this quantity as a function of $\alpha$ for various $N$ yields a crossing point precisely at the tricritical point $(\alpha_\mathrm{t}, \mu_\mathrm{t})$.

To test this method, we consider the N\nl VM on the complete graph. Using the approach described in Appendix~\ref{app:pst}, it is possible to compute the magnetization $\tm$ as a function of the parameters $\alpha$, $\mu$, and the system size $N$ with high numerical accuracy. In this case, the transition point is given by Eq.\eqref{eq:NLNVM_ac_2nd_order}, and thus the Binder cumulant crossing is not required. As shown in Fig.\ref{fig:app:tri_method}, the plots of $\tm(\alpha, \mu) N^{\beta_\mathrm{t}/\bar{\nu}_\mathrm{t}}$ for several values of $N$ intersect at a single point, which coincides with the known tricritical value $\alpha_\mathrm{t} = 5$, thereby validating our method.

In principle, an alternative strategy for locating the tricritical point involves analyzing the scaling behavior of the order parameter along the second-order transition line using the critical exponents. One may plot $\tm(\alpha, \mu) N^{\beta_\mathrm{c}/\bar{\nu}_\mathrm{c}}$ versus $\alpha$, where $\beta_\mathrm{c}$ and $\bar{\nu}_\mathrm{c}$ are the critical exponents, and $\mu$ denotes the Binder cumulant crossing point for each $\alpha$. Along the second-order transition line, this quantity should remain approximately constant for different system sizes $N$, as critical scaling holds. However, as the tricritical point is approached, this scaling breaks down and deviations from the constant behavior emerge. Thus, the onset of these deviations could signal the location of the tricritical point. However, as illustrated in the inset of Fig.~\ref{fig:app:tri_method}, the crossover between the critical and tricritical regimes is smooth rather than abrupt, making it difficult to determine the tricritical point accurately with this method.

Finally, we apply our method to the N\nl VM on Erd\H{o}s–R\'enyi networks. For average degree $\langle k \rangle = 15$, we obtain refined estimates $\alpha_\mathrm{t} = 6.07$ and $\mu_\mathrm{t} = 0.082956$, significantly improving upon predictions from pair approximation, as shown in Fig.~\ref{fig:app:tri_method}.

\begin{figure}[h]
\centering
\includegraphics[width=0.49\textwidth]{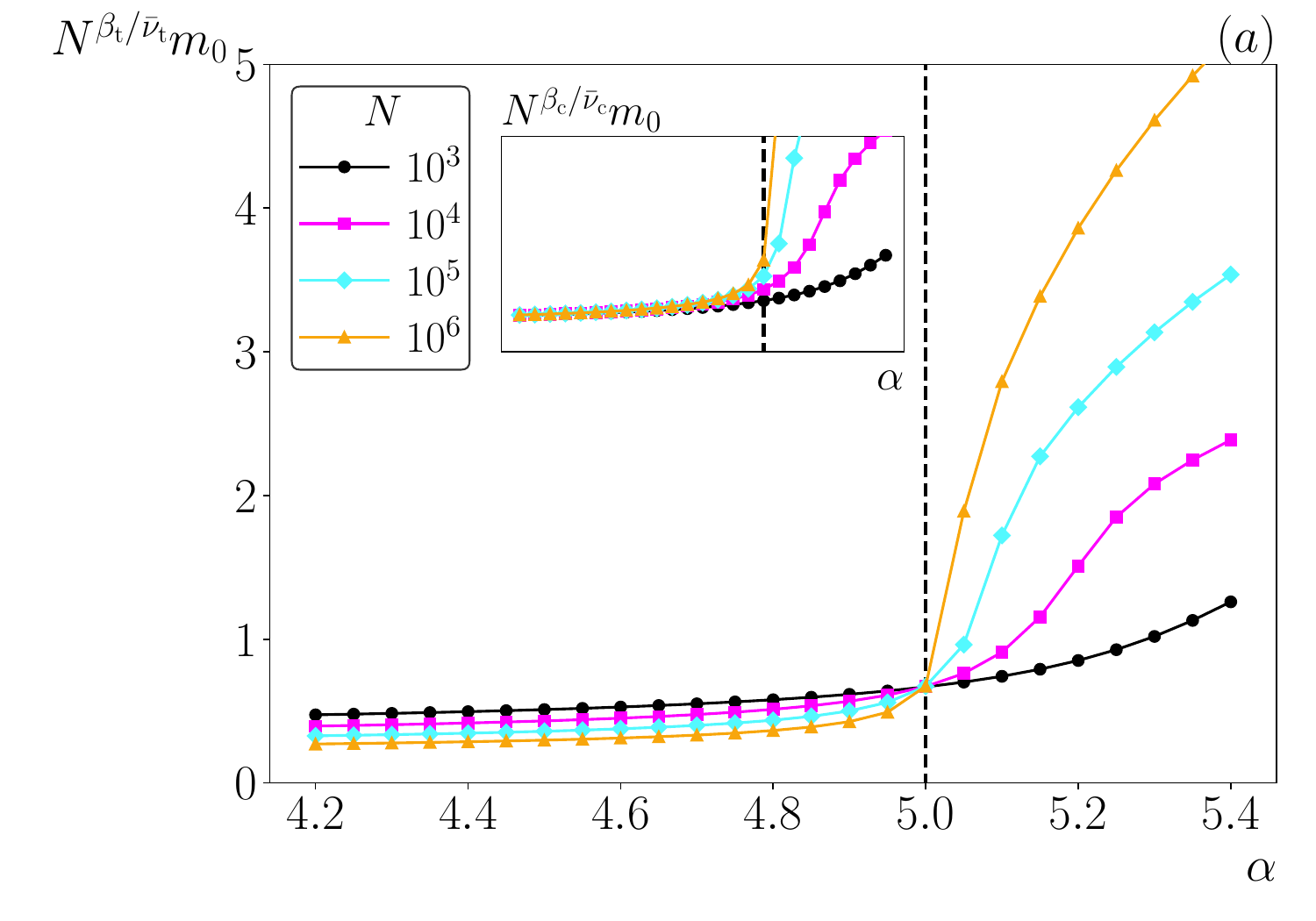}
\includegraphics[width=0.49\textwidth]{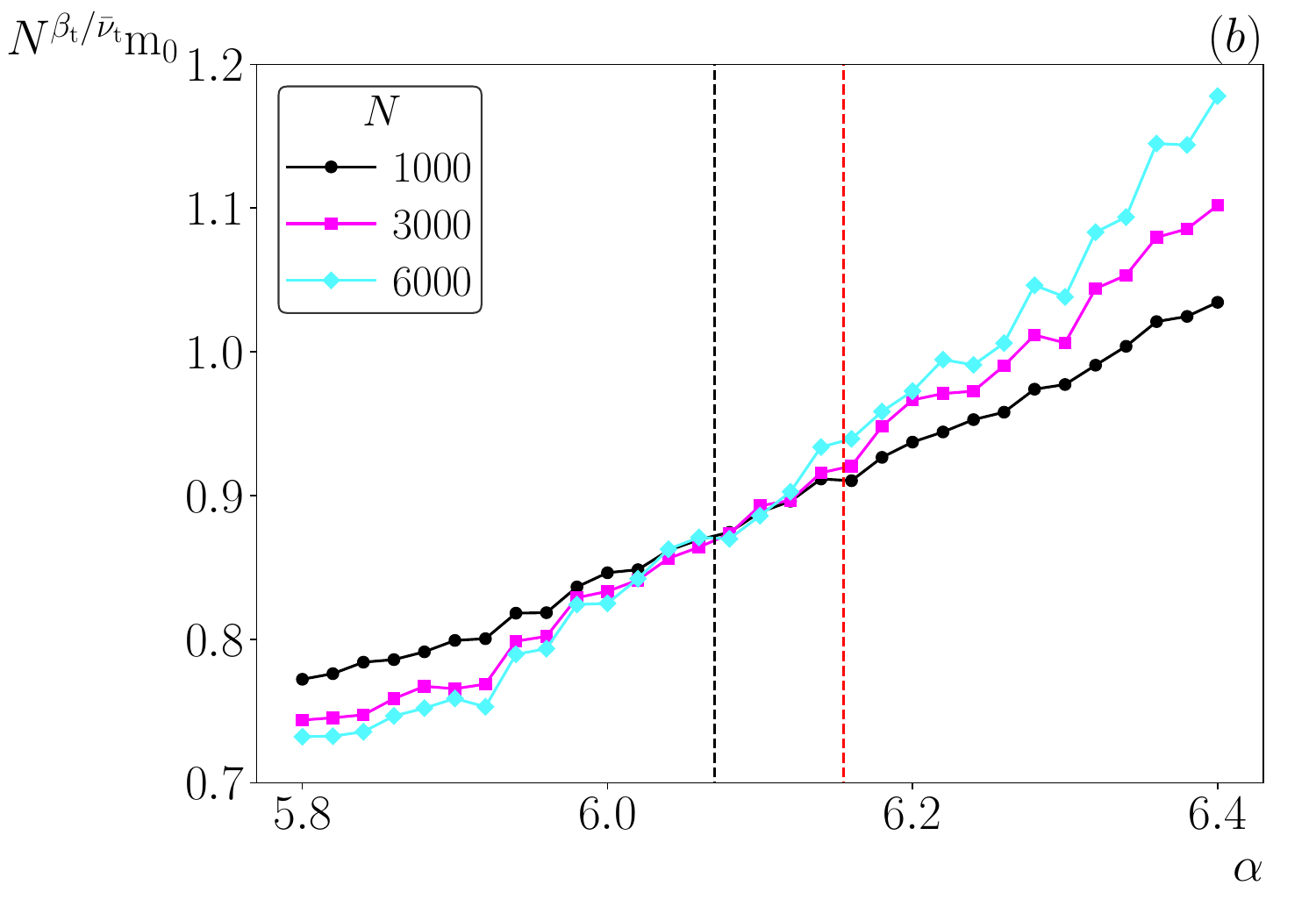}
\caption{$N^{\beta_\mathrm{t}/\bar{\nu}_\mathrm{t}} m_0$ versus $\alpha$ using the tricritical exponents for the N\nl VM on $(a)$ complete graph and $(b)$ Erd\H{o}s-R\'enyi networks for several system sizes. Black vertical dashed lines indicate the tricritical point obtained with $(a)$ mean-field theoretical prediction, and $(b)$ the scaling method explanined in Appendix~\ref{sec:app:tri}. The red vertical dashed line in $(b)$ denotes the tricritical point obtained from the pair approximation, as explained in Appendix~\ref{app:PA_NLNVM}. In $(a)$, the inset shows the same scaling using the critical exponents.}
\label{fig:app:tri_method}
\end{figure}
\end{widetext}
\bibliography{references}

\begin{thebibliography}{68}%
\makeatletter
\providecommand \@ifxundefined [1]{%
 \@ifx{#1\undefined}
}%
\providecommand \@ifnum [1]{%
 \ifnum #1\expandafter \@firstoftwo
 \else \expandafter \@secondoftwo
 \fi
}%
\providecommand \@ifx [1]{%
 \ifx #1\expandafter \@firstoftwo
 \else \expandafter \@secondoftwo
 \fi
}%
\providecommand \natexlab [1]{#1}%
\providecommand \enquote  [1]{``#1''}%
\providecommand \bibnamefont  [1]{#1}%
\providecommand \bibfnamefont [1]{#1}%
\providecommand \citenamefont [1]{#1}%
\providecommand \href@noop [0]{\@secondoftwo}%
\providecommand \href [0]{\begingroup \@sanitize@url \@href}%
\providecommand \@href[1]{\@@startlink{#1}\@@href}%
\providecommand \@@href[1]{\endgroup#1\@@endlink}%
\providecommand \@sanitize@url [0]{\catcode `\\12\catcode `\$12\catcode `\&12\catcode `\#12\catcode `\^12\catcode `\_12\catcode `\%12\relax}%
\providecommand \@@startlink[1]{}%
\providecommand \@@endlink[0]{}%
\providecommand \url  [0]{\begingroup\@sanitize@url \@url }%
\providecommand \@url [1]{\endgroup\@href {#1}{\urlprefix }}%
\providecommand \urlprefix  [0]{URL }%
\providecommand \Eprint [0]{\href }%
\providecommand \doibase [0]{https://doi.org/}%
\providecommand \selectlanguage [0]{\@gobble}%
\providecommand \bibinfo  [0]{\@secondoftwo}%
\providecommand \bibfield  [0]{\@secondoftwo}%
\providecommand \translation [1]{[#1]}%
\providecommand \BibitemOpen [0]{}%
\providecommand \bibitemStop [0]{}%
\providecommand \bibitemNoStop [0]{.\EOS\space}%
\providecommand \EOS [0]{\spacefactor3000\relax}%
\providecommand \BibitemShut  [1]{\csname bibitem#1\endcsname}%
\let\auto@bib@innerbib\@empty
\bibitem [{\citenamefont {Clifford}\ and\ \citenamefont {Sudbury}(1973)}]{vm_clifford}%
  \BibitemOpen
  \bibfield  {author} {\bibinfo {author} {\bibfnamefont {P.}~\bibnamefont {Clifford}}\ and\ \bibinfo {author} {\bibfnamefont {A.}~\bibnamefont {Sudbury}},\ }\bibfield  {title} {\bibinfo {title} {A model for spatial conflict},\ }\href {https://doi.org/10.1093/biomet/60.3.581} {\bibfield  {journal} {\bibinfo  {journal} {Biometrika}\ }\textbf {\bibinfo {volume} {60}},\ \bibinfo {pages} {581} (\bibinfo {year} {1973})}\BibitemShut {NoStop}%
\bibitem [{\citenamefont {Holley}\ and\ \citenamefont {Liggett}(1975)}]{liggett}%
  \BibitemOpen
  \bibfield  {author} {\bibinfo {author} {\bibfnamefont {R.~A.}\ \bibnamefont {Holley}}\ and\ \bibinfo {author} {\bibfnamefont {T.~M.}\ \bibnamefont {Liggett}},\ }\bibfield  {title} {\bibinfo {title} {Ergodic theorems for weakly interacting infinite systems and the voter model},\ }\href@noop {} {\bibfield  {journal} {\bibinfo  {journal} {The Annals of Probability}\ }\textbf {\bibinfo {volume} {3}},\ \bibinfo {pages} {643 } (\bibinfo {year} {1975})}\BibitemShut {NoStop}%
\bibitem [{\citenamefont {Marro}\ and\ \citenamefont {Dickman}(1999)}]{marroBOOK}%
  \BibitemOpen
  \bibfield  {author} {\bibinfo {author} {\bibfnamefont {J.}~\bibnamefont {Marro}}\ and\ \bibinfo {author} {\bibfnamefont {R.}~\bibnamefont {Dickman}},\ }\href@noop {} {\emph {\bibinfo {title} {Nonequilibrium Phase Transitions in Lattice Models}}},\ Collection Alea-Saclay: Monographs and Texts in Statistical Physics\ (\bibinfo  {publisher} {Cambridge University Press},\ \bibinfo {year} {1999})\BibitemShut {NoStop}%
\bibitem [{\citenamefont {Krapivsky}\ \emph {et~al.}(2010)\citenamefont {Krapivsky}, \citenamefont {Redner},\ and\ \citenamefont {Ben-Naim}}]{rednerBOOK}%
  \BibitemOpen
  \bibfield  {author} {\bibinfo {author} {\bibfnamefont {P.~L.}\ \bibnamefont {Krapivsky}}, \bibinfo {author} {\bibfnamefont {S.}~\bibnamefont {Redner}},\ and\ \bibinfo {author} {\bibfnamefont {E.}~\bibnamefont {Ben-Naim}},\ }\href@noop {} {\emph {\bibinfo {title} {A Kinetic View of Statistical Physics}}}\ (\bibinfo  {publisher} {Cambridge University Press},\ \bibinfo {year} {2010})\BibitemShut {NoStop}%
\bibitem [{\citenamefont {Starnini}\ \emph {et~al.}(2025)\citenamefont {Starnini}, \citenamefont {Baumann}, \citenamefont {Galla}, \citenamefont {Garcia}, \citenamefont {Iñiguez}, \citenamefont {Karsai}, \citenamefont {Lorenz},\ and\ \citenamefont {Sznajd-Weron}}]{tobias_opinion}%
  \BibitemOpen
  \bibfield  {author} {\bibinfo {author} {\bibfnamefont {M.}~\bibnamefont {Starnini}}, \bibinfo {author} {\bibfnamefont {F.}~\bibnamefont {Baumann}}, \bibinfo {author} {\bibfnamefont {T.}~\bibnamefont {Galla}}, \bibinfo {author} {\bibfnamefont {D.}~\bibnamefont {Garcia}}, \bibinfo {author} {\bibfnamefont {G.}~\bibnamefont {Iñiguez}}, \bibinfo {author} {\bibfnamefont {M.}~\bibnamefont {Karsai}}, \bibinfo {author} {\bibfnamefont {J.}~\bibnamefont {Lorenz}},\ and\ \bibinfo {author} {\bibfnamefont {K.}~\bibnamefont {Sznajd-Weron}},\ }\href {https://arxiv.org/abs/2507.11521} {\bibinfo {title} {Opinion dynamics: Statistical physics and beyond}} (\bibinfo {year} {2025}),\ \Eprint {https://arxiv.org/abs/2507.11521} {arXiv:2507.11521 [physics.soc-ph]} \BibitemShut {NoStop}%
\bibitem [{\citenamefont {Fern\'andez-Gracia}\ \emph {et~al.}(2014)\citenamefont {Fern\'andez-Gracia}, \citenamefont {Suchecki}, \citenamefont {Ramasco}, \citenamefont {San~Miguel},\ and\ \citenamefont {Egu\'{\i}luz}}]{IstheVM}%
  \BibitemOpen
  \bibfield  {author} {\bibinfo {author} {\bibfnamefont {J.}~\bibnamefont {Fern\'andez-Gracia}}, \bibinfo {author} {\bibfnamefont {K.}~\bibnamefont {Suchecki}}, \bibinfo {author} {\bibfnamefont {J.~J.}\ \bibnamefont {Ramasco}}, \bibinfo {author} {\bibfnamefont {M.}~\bibnamefont {San~Miguel}},\ and\ \bibinfo {author} {\bibfnamefont {V.~M.}\ \bibnamefont {Egu\'{\i}luz}},\ }\bibfield  {title} {\bibinfo {title} {Is the voter model a model for voters?},\ }\href {https://doi.org/10.1103/PhysRevLett.112.158701} {\bibfield  {journal} {\bibinfo  {journal} {Phys. Rev. Lett.}\ }\textbf {\bibinfo {volume} {112}},\ \bibinfo {pages} {158701} (\bibinfo {year} {2014})}\BibitemShut {NoStop}%
\bibitem [{\citenamefont {Dornic}\ \emph {et~al.}(2001)\citenamefont {Dornic}, \citenamefont {Chat\'e}, \citenamefont {Chave},\ and\ \citenamefont {Hinrichsen}}]{Chate2d}%
  \BibitemOpen
  \bibfield  {author} {\bibinfo {author} {\bibfnamefont {I.}~\bibnamefont {Dornic}}, \bibinfo {author} {\bibfnamefont {H.}~\bibnamefont {Chat\'e}}, \bibinfo {author} {\bibfnamefont {J.}~\bibnamefont {Chave}},\ and\ \bibinfo {author} {\bibfnamefont {H.}~\bibnamefont {Hinrichsen}},\ }\bibfield  {title} {\bibinfo {title} {Critical coarsening without surface tension: The universality class of the voter model},\ }\href {https://doi.org/10.1103/PhysRevLett.87.045701} {\bibfield  {journal} {\bibinfo  {journal} {Phys. Rev. Lett.}\ }\textbf {\bibinfo {volume} {87}},\ \bibinfo {pages} {045701} (\bibinfo {year} {2001})}\BibitemShut {NoStop}%
\bibitem [{\citenamefont {Suchecki}\ \emph {et~al.}(2005{\natexlab{a}})\citenamefont {Suchecki}, \citenamefont {Egu\'{\i}luz},\ and\ \citenamefont {San~Miguel}}]{Suchecki}%
  \BibitemOpen
  \bibfield  {author} {\bibinfo {author} {\bibfnamefont {K.}~\bibnamefont {Suchecki}}, \bibinfo {author} {\bibfnamefont {V.~M.}\ \bibnamefont {Egu\'{\i}luz}},\ and\ \bibinfo {author} {\bibfnamefont {M.}~\bibnamefont {San~Miguel}},\ }\bibfield  {title} {\bibinfo {title} {Voter model dynamics in complex networks: Role of dimensionality, disorder, and degree distribution},\ }\href {https://doi.org/10.1103/PhysRevE.72.036132} {\bibfield  {journal} {\bibinfo  {journal} {Phys. Rev. E}\ }\textbf {\bibinfo {volume} {72}},\ \bibinfo {pages} {036132} (\bibinfo {year} {2005}{\natexlab{a}})}\BibitemShut {NoStop}%
\bibitem [{\citenamefont {Vazquez}\ and\ \citenamefont {Egu{\'\i}luz}(2008{\natexlab{a}})}]{EguiluzNJP}%
  \BibitemOpen
  \bibfield  {author} {\bibinfo {author} {\bibfnamefont {F.}~\bibnamefont {Vazquez}}\ and\ \bibinfo {author} {\bibfnamefont {V.~M.}\ \bibnamefont {Egu{\'\i}luz}},\ }\bibfield  {title} {\bibinfo {title} {Analytical solution of the voter model on uncorrelated networks},\ }\href {https://doi.org/10.1088/1367-2630/10/6/063011} {\bibfield  {journal} {\bibinfo  {journal} {New Journal of Physics}\ }\textbf {\bibinfo {volume} {10}},\ \bibinfo {pages} {063011} (\bibinfo {year} {2008}{\natexlab{a}})}\BibitemShut {NoStop}%
\bibitem [{\citenamefont {Sood}\ and\ \citenamefont {Redner}(2005)}]{redner1}%
  \BibitemOpen
  \bibfield  {author} {\bibinfo {author} {\bibfnamefont {V.}~\bibnamefont {Sood}}\ and\ \bibinfo {author} {\bibfnamefont {S.}~\bibnamefont {Redner}},\ }\bibfield  {title} {\bibinfo {title} {Voter model on heterogeneous graphs},\ }\href {https://doi.org/10.1103/PhysRevLett.94.178701} {\bibfield  {journal} {\bibinfo  {journal} {Phys. Rev. Lett.}\ }\textbf {\bibinfo {volume} {94}},\ \bibinfo {pages} {178701} (\bibinfo {year} {2005})}\BibitemShut {NoStop}%
\bibitem [{\citenamefont {Sood}\ \emph {et~al.}(2008)\citenamefont {Sood}, \citenamefont {Antal},\ and\ \citenamefont {Redner}}]{redner2}%
  \BibitemOpen
  \bibfield  {author} {\bibinfo {author} {\bibfnamefont {V.}~\bibnamefont {Sood}}, \bibinfo {author} {\bibfnamefont {T.}~\bibnamefont {Antal}},\ and\ \bibinfo {author} {\bibfnamefont {S.}~\bibnamefont {Redner}},\ }\bibfield  {title} {\bibinfo {title} {Voter models on heterogeneous networks},\ }\href {https://doi.org/10.1103/PhysRevE.77.041121} {\bibfield  {journal} {\bibinfo  {journal} {Phys. Rev. E}\ }\textbf {\bibinfo {volume} {77}},\ \bibinfo {pages} {041121} (\bibinfo {year} {2008})}\BibitemShut {NoStop}%
\bibitem [{\citenamefont {Suchecki}\ \emph {et~al.}(2005{\natexlab{b}})\citenamefont {Suchecki}, \citenamefont {Egu{\'\i}luz},\ and\ \citenamefont {San~Miguel}}]{SucheckiEPL}%
  \BibitemOpen
  \bibfield  {author} {\bibinfo {author} {\bibfnamefont {K.}~\bibnamefont {Suchecki}}, \bibinfo {author} {\bibfnamefont {V.~M.}\ \bibnamefont {Egu{\'\i}luz}},\ and\ \bibinfo {author} {\bibfnamefont {M.}~\bibnamefont {San~Miguel}},\ }\bibfield  {title} {\bibinfo {title} {Conservation laws for the voter model in complex networks},\ }\href {https://doi.org/10.1209/epl/i2004-10329-8} {\bibfield  {journal} {\bibinfo  {journal} {Europhysics Letters}\ }\textbf {\bibinfo {volume} {69}},\ \bibinfo {pages} {228} (\bibinfo {year} {2005}{\natexlab{b}})}\BibitemShut {NoStop}%
\bibitem [{\citenamefont {Malte~Henkel}(2008)}]{Henkel}%
  \BibitemOpen
  \bibfield  {author} {\bibinfo {author} {\bibfnamefont {S.~L.}\ \bibnamefont {Malte~Henkel}, \bibfnamefont {Haye~Hinrichsen}},\ }\href@noop {} {\emph {\bibinfo {title} {Non-Equilibrium Phase Transitions, Volume I: \\ Absorbing Phase Transitions}}}\ (\bibinfo  {publisher} {Springer},\ \bibinfo {address} {Bristol},\ \bibinfo {year} {2008})\BibitemShut {NoStop}%
\bibitem [{\citenamefont {\'Odor}(2004)}]{Odor2004}%
  \BibitemOpen
  \bibfield  {author} {\bibinfo {author} {\bibfnamefont {G.}~\bibnamefont {\'Odor}},\ }\bibfield  {title} {\bibinfo {title} {Universality classes in nonequilibrium lattice systems},\ }\href {https://doi.org/10.1103/RevModPhys.76.663} {\bibfield  {journal} {\bibinfo  {journal} {Rev. Mod. Phys.}\ }\textbf {\bibinfo {volume} {76}},\ \bibinfo {pages} {663} (\bibinfo {year} {2004})}\BibitemShut {NoStop}%
\bibitem [{\citenamefont {Al~Hammal}\ \emph {et~al.}(2005)\citenamefont {Al~Hammal}, \citenamefont {Chat\'e}, \citenamefont {Dornic},\ and\ \citenamefont {Mu\~noz}}]{Hammal}%
  \BibitemOpen
  \bibfield  {author} {\bibinfo {author} {\bibfnamefont {O.}~\bibnamefont {Al~Hammal}}, \bibinfo {author} {\bibfnamefont {H.}~\bibnamefont {Chat\'e}}, \bibinfo {author} {\bibfnamefont {I.}~\bibnamefont {Dornic}},\ and\ \bibinfo {author} {\bibfnamefont {M.~A.}\ \bibnamefont {Mu\~noz}},\ }\bibfield  {title} {\bibinfo {title} {Langevin description of critical phenomena with two symmetric absorbing states},\ }\href {https://doi.org/10.1103/PhysRevLett.94.230601} {\bibfield  {journal} {\bibinfo  {journal} {Phys. Rev. Lett.}\ }\textbf {\bibinfo {volume} {94}},\ \bibinfo {pages} {230601} (\bibinfo {year} {2005})}\BibitemShut {NoStop}%
\bibitem [{\citenamefont {Droz}\ \emph {et~al.}(2003)\citenamefont {Droz}, \citenamefont {Ferreira},\ and\ \citenamefont {Lipowski}}]{Droz}%
  \BibitemOpen
  \bibfield  {author} {\bibinfo {author} {\bibfnamefont {M.}~\bibnamefont {Droz}}, \bibinfo {author} {\bibfnamefont {A.~L.}\ \bibnamefont {Ferreira}},\ and\ \bibinfo {author} {\bibfnamefont {A.}~\bibnamefont {Lipowski}},\ }\bibfield  {title} {\bibinfo {title} {Splitting the voter \uppercase{P}otts model critical point},\ }\href {https://doi.org/10.1103/PhysRevE.67.056108} {\bibfield  {journal} {\bibinfo  {journal} {Phys. Rev. E}\ }\textbf {\bibinfo {volume} {67}},\ \bibinfo {pages} {056108} (\bibinfo {year} {2003})}\BibitemShut {NoStop}%
\bibitem [{\citenamefont {Vazquez}\ and\ \citenamefont {L\'opez}(2008)}]{Fede_Clopez_AB}%
  \BibitemOpen
  \bibfield  {author} {\bibinfo {author} {\bibfnamefont {F.}~\bibnamefont {Vazquez}}\ and\ \bibinfo {author} {\bibfnamefont {C.}~\bibnamefont {L\'opez}},\ }\bibfield  {title} {\bibinfo {title} {Systems with two symmetric absorbing states: Relating the microscopic dynamics with the macroscopic behavior},\ }\href {https://doi.org/10.1103/PhysRevE.78.061127} {\bibfield  {journal} {\bibinfo  {journal} {Phys. Rev. E}\ }\textbf {\bibinfo {volume} {78}},\ \bibinfo {pages} {061127} (\bibinfo {year} {2008})}\BibitemShut {NoStop}%
\bibitem [{\citenamefont {Redner}(2019)}]{rednerCompteRendues}%
  \BibitemOpen
  \bibfield  {author} {\bibinfo {author} {\bibfnamefont {S.}~\bibnamefont {Redner}},\ }\bibfield  {title} {\bibinfo {title} {Reality-inspired voter models: A mini-review},\ }\href {https://doi.org/https://doi.org/10.1016/j.crhy.2019.05.004} {\bibfield  {journal} {\bibinfo  {journal} {Comptes Rendus Physique}\ }\textbf {\bibinfo {volume} {20}},\ \bibinfo {pages} {275} (\bibinfo {year} {2019})}\BibitemShut {NoStop}%
\bibitem [{\citenamefont {Schweitzer}\ and\ \citenamefont {Behera}(2009)}]{Schweitzer}%
  \BibitemOpen
  \bibfield  {author} {\bibinfo {author} {\bibfnamefont {F.}~\bibnamefont {Schweitzer}}\ and\ \bibinfo {author} {\bibfnamefont {L.}~\bibnamefont {Behera}},\ }\bibfield  {title} {\bibinfo {title} {Nonlinear voter models: the transition from invasion to coexistence},\ }\href {https://doi.org/10.1140/epjb/e2009-00001-3} {\bibfield  {journal} {\bibinfo  {journal} {The European Physical Journal B}\ }\textbf {\bibinfo {volume} {67}},\ \bibinfo {pages} {301} (\bibinfo {year} {2009})}\BibitemShut {NoStop}%
\bibitem [{\citenamefont {Min}\ and\ \citenamefont {San~Miguel}(2017)}]{Min}%
  \BibitemOpen
  \bibfield  {author} {\bibinfo {author} {\bibfnamefont {B.}~\bibnamefont {Min}}\ and\ \bibinfo {author} {\bibfnamefont {M.}~\bibnamefont {San~Miguel}},\ }\bibfield  {title} {\bibinfo {title} {Fragmentation transitions in a coevolving nonlinear voter model},\ }\href {https://doi.org/10.1038/s41598-017-13047-2} {\bibfield  {journal} {\bibinfo  {journal} {Scientific Reports}\ }\textbf {\bibinfo {volume} {7}},\ \bibinfo {pages} {12864} (\bibinfo {year} {2017})}\BibitemShut {NoStop}%
\bibitem [{\citenamefont {Castellano}\ \emph {et~al.}(2009)\citenamefont {Castellano}, \citenamefont {Mu\~noz},\ and\ \citenamefont {Pastor-Satorras}}]{qvoter}%
  \BibitemOpen
  \bibfield  {author} {\bibinfo {author} {\bibfnamefont {C.}~\bibnamefont {Castellano}}, \bibinfo {author} {\bibfnamefont {M.~A.}\ \bibnamefont {Mu\~noz}},\ and\ \bibinfo {author} {\bibfnamefont {R.}~\bibnamefont {Pastor-Satorras}},\ }\bibfield  {title} {\bibinfo {title} {Nonlinear $q$-voter model},\ }\href {https://doi.org/10.1103/PhysRevE.80.041129} {\bibfield  {journal} {\bibinfo  {journal} {Phys. Rev. E}\ }\textbf {\bibinfo {volume} {80}},\ \bibinfo {pages} {041129} (\bibinfo {year} {2009})}\BibitemShut {NoStop}%
\bibitem [{\citenamefont {Lambiotte}\ and\ \citenamefont {Redner}(2007)}]{Lambiotte_2007}%
  \BibitemOpen
  \bibfield  {author} {\bibinfo {author} {\bibfnamefont {R.}~\bibnamefont {Lambiotte}}\ and\ \bibinfo {author} {\bibfnamefont {S.}~\bibnamefont {Redner}},\ }\bibfield  {title} {\bibinfo {title} {Dynamics of vacillating voters},\ }\href {https://doi.org/10.1088/1742-5468/2007/10/L10001} {\bibfield  {journal} {\bibinfo  {journal} {Journal of Statistical Mechanics: Theory and Experiment}\ }\textbf {\bibinfo {volume} {2007}},\ \bibinfo {pages} {L10001} (\bibinfo {year} {2007})}\BibitemShut {NoStop}%
\bibitem [{\citenamefont {Lambiotte}\ and\ \citenamefont {Redner}(2008)}]{Lambiotte_2008}%
  \BibitemOpen
  \bibfield  {author} {\bibinfo {author} {\bibfnamefont {R.}~\bibnamefont {Lambiotte}}\ and\ \bibinfo {author} {\bibfnamefont {S.}~\bibnamefont {Redner}},\ }\bibfield  {title} {\bibinfo {title} {Dynamics of non-conservative voters},\ }\href {https://doi.org/10.1209/0295-5075/82/18007} {\bibfield  {journal} {\bibinfo  {journal} {Europhysics Letters}\ }\textbf {\bibinfo {volume} {82}},\ \bibinfo {pages} {18007} (\bibinfo {year} {2008})}\BibitemShut {NoStop}%
\bibitem [{\citenamefont {Volovik}\ and\ \citenamefont {Redner}(2012)}]{confident_voters}%
  \BibitemOpen
  \bibfield  {author} {\bibinfo {author} {\bibfnamefont {D.}~\bibnamefont {Volovik}}\ and\ \bibinfo {author} {\bibfnamefont {S.}~\bibnamefont {Redner}},\ }\bibfield  {title} {\bibinfo {title} {Dynamics of confident voting},\ }\href {https://doi.org/10.1088/1742-5468/2012/04/P04003} {\bibfield  {journal} {\bibinfo  {journal} {Journal of Statistical Mechanics: Theory and Experiment}\ }\textbf {\bibinfo {volume} {2012}},\ \bibinfo {pages} {P04003} (\bibinfo {year} {2012})}\BibitemShut {NoStop}%
\bibitem [{\citenamefont {Dall'Asta}\ and\ \citenamefont {Castellano}(2007)}]{noise_reduced_voter_model}%
  \BibitemOpen
  \bibfield  {author} {\bibinfo {author} {\bibfnamefont {L.}~\bibnamefont {Dall'Asta}}\ and\ \bibinfo {author} {\bibfnamefont {C.}~\bibnamefont {Castellano}},\ }\bibfield  {title} {\bibinfo {title} {Effective surface-tension in the noise-reduced voter model},\ }\href {https://doi.org/10.1209/0295-5075/77/60005} {\bibfield  {journal} {\bibinfo  {journal} {Europhysics Letters}\ }\textbf {\bibinfo {volume} {77}},\ \bibinfo {pages} {60005} (\bibinfo {year} {2007})}\BibitemShut {NoStop}%
\bibitem [{\citenamefont {Ramirez}\ \emph {et~al.}(2024)\citenamefont {Ramirez}, \citenamefont {Vazquez}, \citenamefont {San~Miguel},\ and\ \citenamefont {Galla}}]{lucia_NLVM}%
  \BibitemOpen
  \bibfield  {author} {\bibinfo {author} {\bibfnamefont {L.~S.}\ \bibnamefont {Ramirez}}, \bibinfo {author} {\bibfnamefont {F.}~\bibnamefont {Vazquez}}, \bibinfo {author} {\bibfnamefont {M.}~\bibnamefont {San~Miguel}},\ and\ \bibinfo {author} {\bibfnamefont {T.}~\bibnamefont {Galla}},\ }\bibfield  {title} {\bibinfo {title} {Ordering dynamics of nonlinear voter models},\ }\href {https://doi.org/10.1103/PhysRevE.109.034307} {\bibfield  {journal} {\bibinfo  {journal} {Phys. Rev. E}\ }\textbf {\bibinfo {volume} {109}},\ \bibinfo {pages} {034307} (\bibinfo {year} {2024})}\BibitemShut {NoStop}%
\bibitem [{\citenamefont {Sznajd-Weron}\ and\ \citenamefont {Sznajd}(2000)}]{Sznajd}%
  \BibitemOpen
  \bibfield  {author} {\bibinfo {author} {\bibfnamefont {K.}~\bibnamefont {Sznajd-Weron}}\ and\ \bibinfo {author} {\bibfnamefont {J.}~\bibnamefont {Sznajd}},\ }\bibfield  {title} {\bibinfo {title} {Opinion evolution in closed community},\ }\href {https://doi.org/10.1142/S0129183100000936} {\bibfield  {journal} {\bibinfo  {journal} {International Journal of Modern Physics C}\ }\textbf {\bibinfo {volume} {11}},\ \bibinfo {pages} {1157} (\bibinfo {year} {2000})}\BibitemShut {NoStop}%
\bibitem [{\citenamefont {Slanina}\ and\ \citenamefont {Lavicka}(2003)}]{Sznajd_MF}%
  \BibitemOpen
  \bibfield  {author} {\bibinfo {author} {\bibfnamefont {F.}~\bibnamefont {Slanina}}\ and\ \bibinfo {author} {\bibfnamefont {H.}~\bibnamefont {Lavicka}},\ }\bibfield  {title} {\bibinfo {title} {Analytical results for the \uppercase{S}znajd model of opinion formation},\ }\href {https://doi.org/10.1140/epjb/e2003-00278-0} {\bibfield  {journal} {\bibinfo  {journal} {The European Physical Journal B - Condensed Matter and Complex Systems}\ }\textbf {\bibinfo {volume} {35}},\ \bibinfo {pages} {279} (\bibinfo {year} {2003})}\BibitemShut {NoStop}%
\bibitem [{\citenamefont {Moran}(1958)}]{moran}%
  \BibitemOpen
  \bibfield  {author} {\bibinfo {author} {\bibfnamefont {P.~A.~P.}\ \bibnamefont {Moran}},\ }\bibfield  {title} {\bibinfo {title} {Random processes in genetics},\ }\href {https://doi.org/10.1017/S0305004100033193} {\bibfield  {journal} {\bibinfo  {journal} {Mathematical Proceedings of the Cambridge Philosophical Society}\ }\textbf {\bibinfo {volume} {54}},\ \bibinfo {pages} {60–71} (\bibinfo {year} {1958})}\BibitemShut {NoStop}%
\bibitem [{\citenamefont {Fichthorn}\ \emph {et~al.}(1989)\citenamefont {Fichthorn}, \citenamefont {Gulari},\ and\ \citenamefont {Ziff}}]{Fichthorn}%
  \BibitemOpen
  \bibfield  {author} {\bibinfo {author} {\bibfnamefont {K.}~\bibnamefont {Fichthorn}}, \bibinfo {author} {\bibfnamefont {E.}~\bibnamefont {Gulari}},\ and\ \bibinfo {author} {\bibfnamefont {R.}~\bibnamefont {Ziff}},\ }\bibfield  {title} {\bibinfo {title} {Noise-induced bistability in a \uppercase{m}onte \uppercase{c}arlo surface-reaction model},\ }\href {https://doi.org/10.1103/PhysRevLett.63.1527} {\bibfield  {journal} {\bibinfo  {journal} {Phys. Rev. Lett.}\ }\textbf {\bibinfo {volume} {63}},\ \bibinfo {pages} {1527} (\bibinfo {year} {1989})}\BibitemShut {NoStop}%
\bibitem [{\citenamefont {Granovsky}\ and\ \citenamefont {Madras}(1995)}]{granovsky}%
  \BibitemOpen
  \bibfield  {author} {\bibinfo {author} {\bibfnamefont {B.~L.}\ \bibnamefont {Granovsky}}\ and\ \bibinfo {author} {\bibfnamefont {N.}~\bibnamefont {Madras}},\ }\bibfield  {title} {\bibinfo {title} {The noisy voter model},\ }\href {https://doi.org/https://doi.org/10.1016/0304-4149(94)00035-R} {\bibfield  {journal} {\bibinfo  {journal} {Stochastic Processes and their Applications}\ }\textbf {\bibinfo {volume} {55}},\ \bibinfo {pages} {23} (\bibinfo {year} {1995})}\BibitemShut {NoStop}%
\bibitem [{\citenamefont {Carro}\ \emph {et~al.}(2016)\citenamefont {Carro}, \citenamefont {Toral},\ and\ \citenamefont {San~Miguel}}]{Carro2016}%
  \BibitemOpen
  \bibfield  {author} {\bibinfo {author} {\bibfnamefont {A.}~\bibnamefont {Carro}}, \bibinfo {author} {\bibfnamefont {R.}~\bibnamefont {Toral}},\ and\ \bibinfo {author} {\bibfnamefont {M.}~\bibnamefont {San~Miguel}},\ }\bibfield  {title} {\bibinfo {title} {The noisy voter model on complex networks},\ }\href {https://doi.org/10.1038/srep24775} {\bibfield  {journal} {\bibinfo  {journal} {Scientific Reports}\ }\textbf {\bibinfo {volume} {6}},\ \bibinfo {pages} {24775} (\bibinfo {year} {2016})}\BibitemShut {NoStop}%
\bibitem [{\citenamefont {Peralta}\ \emph {et~al.}(2018{\natexlab{a}})\citenamefont {Peralta}, \citenamefont {Carro}, \citenamefont {San~Miguel},\ and\ \citenamefont {Toral}}]{Peralta_sto_2018}%
  \BibitemOpen
  \bibfield  {author} {\bibinfo {author} {\bibfnamefont {A.~F.}\ \bibnamefont {Peralta}}, \bibinfo {author} {\bibfnamefont {A.}~\bibnamefont {Carro}}, \bibinfo {author} {\bibfnamefont {M.}~\bibnamefont {San~Miguel}},\ and\ \bibinfo {author} {\bibfnamefont {R.}~\bibnamefont {Toral}},\ }\bibfield  {title} {\bibinfo {title} {Stochastic pair approximation treatment of the noisy voter model},\ }\href {https://doi.org/10.1088/1367-2630/aae7f5} {\bibfield  {journal} {\bibinfo  {journal} {New Journal of Physics}\ }\textbf {\bibinfo {volume} {20}},\ \bibinfo {pages} {103045} (\bibinfo {year} {2018}{\natexlab{a}})}\BibitemShut {NoStop}%
\bibitem [{\citenamefont {Kirman}(1993)}]{Kirman1993}%
  \BibitemOpen
  \bibfield  {author} {\bibinfo {author} {\bibfnamefont {A.}~\bibnamefont {Kirman}},\ }\bibfield  {title} {\bibinfo {title} {Ants, rationality, and recruitment},\ }\href@noop {} {\bibfield  {journal} {\bibinfo  {journal} {The Quarterly Journal of Economics}\ }\textbf {\bibinfo {volume} {108}},\ \bibinfo {pages} {137} (\bibinfo {year} {1993})}\BibitemShut {NoStop}%
\bibitem [{\citenamefont {Llabr\'es}\ \emph {et~al.}(2023)\citenamefont {Llabr\'es}, \citenamefont {San~Miguel},\ and\ \citenamefont {Toral}}]{llabres}%
  \BibitemOpen
  \bibfield  {author} {\bibinfo {author} {\bibfnamefont {J.}~\bibnamefont {Llabr\'es}}, \bibinfo {author} {\bibfnamefont {M.}~\bibnamefont {San~Miguel}},\ and\ \bibinfo {author} {\bibfnamefont {R.}~\bibnamefont {Toral}},\ }\bibfield  {title} {\bibinfo {title} {Partisan voter model: Stochastic description and noise-induced transitions},\ }\href {https://doi.org/10.1103/PhysRevE.108.054106} {\bibfield  {journal} {\bibinfo  {journal} {Phys. Rev. E}\ }\textbf {\bibinfo {volume} {108}},\ \bibinfo {pages} {054106} (\bibinfo {year} {2023})}\BibitemShut {NoStop}%
\bibitem [{\citenamefont {Nyczka}\ \emph {et~al.}(2012{\natexlab{a}})\citenamefont {Nyczka}, \citenamefont {Sznajd-Weron},\ and\ \citenamefont {Cis\l{}o}}]{q-voter_noise}%
  \BibitemOpen
  \bibfield  {author} {\bibinfo {author} {\bibfnamefont {P.}~\bibnamefont {Nyczka}}, \bibinfo {author} {\bibfnamefont {K.}~\bibnamefont {Sznajd-Weron}},\ and\ \bibinfo {author} {\bibfnamefont {J.}~\bibnamefont {Cis\l{}o}},\ }\bibfield  {title} {\bibinfo {title} {Phase transitions in the $q$-voter model with two types of stochastic driving},\ }\href {https://doi.org/10.1103/PhysRevE.86.011105} {\bibfield  {journal} {\bibinfo  {journal} {Phys. Rev. E}\ }\textbf {\bibinfo {volume} {86}},\ \bibinfo {pages} {011105} (\bibinfo {year} {2012}{\natexlab{a}})}\BibitemShut {NoStop}%
\bibitem [{\citenamefont {J\k{e}drzejewski}(2017)}]{Jedrzejewski}%
  \BibitemOpen
  \bibfield  {author} {\bibinfo {author} {\bibfnamefont {A.}~\bibnamefont {J\k{e}drzejewski}},\ }\bibfield  {title} {\bibinfo {title} {Pair approximation for the $q$-voter model with independence on complex networks},\ }\href {https://doi.org/10.1103/PhysRevE.95.012307} {\bibfield  {journal} {\bibinfo  {journal} {Phys. Rev. E}\ }\textbf {\bibinfo {volume} {95}},\ \bibinfo {pages} {012307} (\bibinfo {year} {2017})}\BibitemShut {NoStop}%
\bibitem [{\citenamefont {Artime}\ \emph {et~al.}(2018)\citenamefont {Artime}, \citenamefont {Peralta}, \citenamefont {Toral}, \citenamefont {Ramasco},\ and\ \citenamefont {San~Miguel}}]{Artime_2018}%
  \BibitemOpen
  \bibfield  {author} {\bibinfo {author} {\bibfnamefont {O.}~\bibnamefont {Artime}}, \bibinfo {author} {\bibfnamefont {A.~F.}\ \bibnamefont {Peralta}}, \bibinfo {author} {\bibfnamefont {R.}~\bibnamefont {Toral}}, \bibinfo {author} {\bibfnamefont {J.~J.}\ \bibnamefont {Ramasco}},\ and\ \bibinfo {author} {\bibfnamefont {M.}~\bibnamefont {San~Miguel}},\ }\bibfield  {title} {\bibinfo {title} {Aging-induced continuous phase transition},\ }\href {https://doi.org/10.1103/PhysRevE.98.032104} {\bibfield  {journal} {\bibinfo  {journal} {Phys. Rev. E}\ }\textbf {\bibinfo {volume} {98}},\ \bibinfo {pages} {032104} (\bibinfo {year} {2018})}\BibitemShut {NoStop}%
\bibitem [{\citenamefont {Peralta}\ \emph {et~al.}(2018{\natexlab{b}})\citenamefont {Peralta}, \citenamefont {Carro}, \citenamefont {San~Miguel},\ and\ \citenamefont {Toral}}]{Peralta_2018}%
  \BibitemOpen
  \bibfield  {author} {\bibinfo {author} {\bibfnamefont {A.~F.}\ \bibnamefont {Peralta}}, \bibinfo {author} {\bibfnamefont {A.}~\bibnamefont {Carro}}, \bibinfo {author} {\bibfnamefont {M.}~\bibnamefont {San~Miguel}},\ and\ \bibinfo {author} {\bibfnamefont {R.}~\bibnamefont {Toral}},\ }\bibfield  {title} {\bibinfo {title} {{Analytical and numerical study of the non-linear noisy voter model on complex networks}},\ }\href {https://doi.org/10.1063/1.5030112} {\bibfield  {journal} {\bibinfo  {journal} {Chaos: An Interdisciplinary Journal of Nonlinear Science}\ }\textbf {\bibinfo {volume} {28}},\ \bibinfo {pages} {075516} (\bibinfo {year} {2018}{\natexlab{b}})}\BibitemShut {NoStop}%
\bibitem [{\citenamefont {Artime}\ \emph {et~al.}(2019)\citenamefont {Artime}, \citenamefont {Carro}, \citenamefont {Peralta}, \citenamefont {Ramasco}, \citenamefont {{San Miguel}},\ and\ \citenamefont {Toral}}]{Artime_2019}%
  \BibitemOpen
  \bibfield  {author} {\bibinfo {author} {\bibfnamefont {O.}~\bibnamefont {Artime}}, \bibinfo {author} {\bibfnamefont {A.}~\bibnamefont {Carro}}, \bibinfo {author} {\bibfnamefont {A.~F.}\ \bibnamefont {Peralta}}, \bibinfo {author} {\bibfnamefont {J.~J.}\ \bibnamefont {Ramasco}}, \bibinfo {author} {\bibfnamefont {M.}~\bibnamefont {{San Miguel}}},\ and\ \bibinfo {author} {\bibfnamefont {R.}~\bibnamefont {Toral}},\ }\bibfield  {title} {\bibinfo {title} {Herding and idiosyncratic choices: Nonlinearity and aging-induced transitions in the noisy voter model},\ }\href {https://doi.org/https://doi.org/10.1016/j.crhy.2019.05.003} {\bibfield  {journal} {\bibinfo  {journal} {Comptes Rendus Physique}\ }\textbf {\bibinfo {volume} {20}},\ \bibinfo {pages} {262} (\bibinfo {year} {2019})}\BibitemShut {NoStop}%
\bibitem [{\citenamefont {de~la Lama}\ \emph {et~al.}(2005)\citenamefont {de~la Lama}, \citenamefont {L{\'o}pez},\ and\ \citenamefont {Wio}}]{Sznajd_Lama}%
  \BibitemOpen
  \bibfield  {author} {\bibinfo {author} {\bibfnamefont {M.~S.}\ \bibnamefont {de~la Lama}}, \bibinfo {author} {\bibfnamefont {J.~M.}\ \bibnamefont {L{\'o}pez}},\ and\ \bibinfo {author} {\bibfnamefont {H.~S.}\ \bibnamefont {Wio}},\ }\bibfield  {title} {\bibinfo {title} {Spontaneous emergence of contrarian-like behaviour in an opinion spreading model},\ }\href {https://doi.org/10.1209/epl/i2005-10299-3} {\bibfield  {journal} {\bibinfo  {journal} {Europhysics Letters}\ }\textbf {\bibinfo {volume} {72}},\ \bibinfo {pages} {851} (\bibinfo {year} {2005})}\BibitemShut {NoStop}%
\bibitem [{\citenamefont {Nyczka}\ \emph {et~al.}(2012{\natexlab{b}})\citenamefont {Nyczka}, \citenamefont {Cis{\l}o},\ and\ \citenamefont {Sznajd-Weron}}]{Sznajd_anti}%
  \BibitemOpen
  \bibfield  {author} {\bibinfo {author} {\bibfnamefont {P.}~\bibnamefont {Nyczka}}, \bibinfo {author} {\bibfnamefont {J.}~\bibnamefont {Cis{\l}o}},\ and\ \bibinfo {author} {\bibfnamefont {K.}~\bibnamefont {Sznajd-Weron}},\ }\bibfield  {title} {\bibinfo {title} {Opinion dynamics as a movement in a bistable potential},\ }\href {https://doi.org/https://doi.org/10.1016/j.physa.2011.07.050} {\bibfield  {journal} {\bibinfo  {journal} {Physica A: Statistical Mechanics and its Applications}\ }\textbf {\bibinfo {volume} {391}},\ \bibinfo {pages} {317} (\bibinfo {year} {2012}{\natexlab{b}})}\BibitemShut {NoStop}%
\bibitem [{\citenamefont {Stark}\ \emph {et~al.}(2008)\citenamefont {Stark}, \citenamefont {Tessone},\ and\ \citenamefont {Schweitzer}}]{Stark:2008}%
  \BibitemOpen
  \bibfield  {author} {\bibinfo {author} {\bibfnamefont {H.~U.}\ \bibnamefont {Stark}}, \bibinfo {author} {\bibfnamefont {C.~J.}\ \bibnamefont {Tessone}},\ and\ \bibinfo {author} {\bibfnamefont {F.}~\bibnamefont {Schweitzer}},\ }\bibfield  {title} {\bibinfo {title} {Decelerating microdynamics can accelerate macrodynamics in the voter model},\ }\href {https://doi.org/10.1103/PhysRevLett.101.018701} {\bibfield  {journal} {\bibinfo  {journal} {Physical Review Letters}\ }\textbf {\bibinfo {volume} {101}},\ \bibinfo {pages} {018701} (\bibinfo {year} {2008})}\BibitemShut {NoStop}%
\bibitem [{\citenamefont {Fern\'andez-Gracia}\ \emph {et~al.}(2011)\citenamefont {Fern\'andez-Gracia}, \citenamefont {Egu\'{\i}luz},\ and\ \citenamefont {San~Miguel}}]{aging_juan}%
  \BibitemOpen
  \bibfield  {author} {\bibinfo {author} {\bibfnamefont {J.}~\bibnamefont {Fern\'andez-Gracia}}, \bibinfo {author} {\bibfnamefont {V.~M.}\ \bibnamefont {Egu\'{\i}luz}},\ and\ \bibinfo {author} {\bibfnamefont {M.}~\bibnamefont {San~Miguel}},\ }\bibfield  {title} {\bibinfo {title} {Update rules and interevent time distributions: Slow ordering versus no ordering in the voter model},\ }\href {https://doi.org/10.1103/PhysRevE.84.015103} {\bibfield  {journal} {\bibinfo  {journal} {Phys. Rev. E}\ }\textbf {\bibinfo {volume} {84}},\ \bibinfo {pages} {015103} (\bibinfo {year} {2011})}\BibitemShut {NoStop}%
\bibitem [{\citenamefont {Peralta}\ \emph {et~al.}(2020{\natexlab{a}})\citenamefont {Peralta}, \citenamefont {Khalil},\ and\ \citenamefont {Toral}}]{nonmarkovian_to_markovian}%
  \BibitemOpen
  \bibfield  {author} {\bibinfo {author} {\bibfnamefont {A.~F.}\ \bibnamefont {Peralta}}, \bibinfo {author} {\bibfnamefont {N.}~\bibnamefont {Khalil}},\ and\ \bibinfo {author} {\bibfnamefont {R.}~\bibnamefont {Toral}},\ }\bibfield  {title} {\bibinfo {title} {Reduction from non-\uppercase{m}arkovian to \uppercase{m}arkovian dynamics: the case of aging in the noisy-voter model},\ }\href {https://doi.org/10.1088/1742-5468/ab6847} {\bibfield  {journal} {\bibinfo  {journal} {Journal of Statistical Mechanics: Theory and Experiment}\ }\textbf {\bibinfo {volume} {2020}},\ \bibinfo {pages} {024004} (\bibinfo {year} {2020}{\natexlab{a}})}\BibitemShut {NoStop}%
\bibitem [{\citenamefont {San~Miguel}\ and\ \citenamefont {Toral}(2000)}]{SanMiguelToral:2000}%
  \BibitemOpen
  \bibfield  {author} {\bibinfo {author} {\bibfnamefont {M.}~\bibnamefont {San~Miguel}}\ and\ \bibinfo {author} {\bibfnamefont {R.}~\bibnamefont {Toral}},\ }\bibinfo {title} {Stochastic effects in physical systems},\ in\ \href {https://doi.org/10.1007/978-94-011-4247-2_2} {\emph {\bibinfo {booktitle} {Instabilities and Nonequilibrium Structures VI}}},\ \bibinfo {editor} {edited by\ \bibinfo {editor} {\bibfnamefont {E.}~\bibnamefont {Tirapegui}}, \bibinfo {editor} {\bibfnamefont {J.}~\bibnamefont {Mart{\'i}nez}},\ and\ \bibinfo {editor} {\bibfnamefont {R.}~\bibnamefont {Tiemann}}}\ (\bibinfo  {publisher} {Springer Netherlands},\ \bibinfo {address} {Dordrecht},\ \bibinfo {year} {2000})\ pp.\ \bibinfo {pages} {35--127}\BibitemShut {NoStop}%
\bibitem [{\citenamefont {Abrams}\ and\ \citenamefont {Strogatz}(2003)}]{AS_model}%
  \BibitemOpen
  \bibfield  {author} {\bibinfo {author} {\bibfnamefont {D.~M.}\ \bibnamefont {Abrams}}\ and\ \bibinfo {author} {\bibfnamefont {S.~H.}\ \bibnamefont {Strogatz}},\ }\bibfield  {title} {\bibinfo {title} {Modelling the dynamics of language death},\ }\href {https://doi.org/10.1038/424900a} {\bibfield  {journal} {\bibinfo  {journal} {Nature}\ }\textbf {\bibinfo {volume} {424}},\ \bibinfo {pages} {900} (\bibinfo {year} {2003})}\BibitemShut {NoStop}%
\bibitem [{\citenamefont {Masuda}\ \emph {et~al.}(2010)\citenamefont {Masuda}, \citenamefont {Gibert},\ and\ \citenamefont {Redner}}]{Masuda_2010}%
  \BibitemOpen
  \bibfield  {author} {\bibinfo {author} {\bibfnamefont {N.}~\bibnamefont {Masuda}}, \bibinfo {author} {\bibfnamefont {N.}~\bibnamefont {Gibert}},\ and\ \bibinfo {author} {\bibfnamefont {S.}~\bibnamefont {Redner}},\ }\bibfield  {title} {\bibinfo {title} {Heterogeneous voter models},\ }\href {https://doi.org/10.1103/PhysRevE.82.010103} {\bibfield  {journal} {\bibinfo  {journal} {Phys. Rev. E}\ }\textbf {\bibinfo {volume} {82}},\ \bibinfo {pages} {010103} (\bibinfo {year} {2010})}\BibitemShut {NoStop}%
\bibitem [{\citenamefont {Masuda}\ and\ \citenamefont {Redner}(2011)}]{Masuda_2011}%
  \BibitemOpen
  \bibfield  {author} {\bibinfo {author} {\bibfnamefont {N.}~\bibnamefont {Masuda}}\ and\ \bibinfo {author} {\bibfnamefont {S.}~\bibnamefont {Redner}},\ }\bibfield  {title} {\bibinfo {title} {Can partisan voting lead to truth?},\ }\href {https://doi.org/10.1088/1742-5468/2011/02/L02002} {\bibfield  {journal} {\bibinfo  {journal} {Journal of Statistical Mechanics: Theory and Experiment}\ }\textbf {\bibinfo {volume} {2011}},\ \bibinfo {pages} {L02002} (\bibinfo {year} {2011})}\BibitemShut {NoStop}%
\bibitem [{\citenamefont {Peralta}\ \emph {et~al.}(2020{\natexlab{b}})\citenamefont {Peralta}, \citenamefont {Khalil},\ and\ \citenamefont {Toral}}]{vm_aging}%
  \BibitemOpen
  \bibfield  {author} {\bibinfo {author} {\bibfnamefont {A.~F.}\ \bibnamefont {Peralta}}, \bibinfo {author} {\bibfnamefont {N.}~\bibnamefont {Khalil}},\ and\ \bibinfo {author} {\bibfnamefont {R.}~\bibnamefont {Toral}},\ }\bibfield  {title} {\bibinfo {title} {Ordering dynamics in the voter model with aging},\ }\href {https://doi.org/https://doi.org/10.1016/j.physa.2019.122475} {\bibfield  {journal} {\bibinfo  {journal} {Physica A: Statistical Mechanics and its Applications}\ }\textbf {\bibinfo {volume} {552}},\ \bibinfo {pages} {122475} (\bibinfo {year} {2020}{\natexlab{b}})}\BibitemShut {NoStop}%
\bibitem [{\citenamefont {Llabr{\'e}s}\ \emph {et~al.}(2024)\citenamefont {Llabr{\'e}s}, \citenamefont {Oliver-Bonafoux}, \citenamefont {Anteneodo},\ and\ \citenamefont {Toral}}]{aging_Jaume_Sara}%
  \BibitemOpen
  \bibfield  {author} {\bibinfo {author} {\bibfnamefont {J.}~\bibnamefont {Llabr{\'e}s}}, \bibinfo {author} {\bibfnamefont {S.}~\bibnamefont {Oliver-Bonafoux}}, \bibinfo {author} {\bibfnamefont {C.}~\bibnamefont {Anteneodo}},\ and\ \bibinfo {author} {\bibfnamefont {R.}~\bibnamefont {Toral}},\ }\bibfield  {title} {\bibinfo {title} {Aging in some opinion formation models: A comparative study},\ }\href {https://doi.org/10.3390/physics6020034} {\bibfield  {journal} {\bibinfo  {journal} {Physics}\ }\textbf {\bibinfo {volume} {6}},\ \bibinfo {pages} {515} (\bibinfo {year} {2024})}\BibitemShut {NoStop}%
\bibitem [{Note1()}]{Note1}%
  \BibitemOpen
  \bibinfo {note} {The case $p_\infty =0$ corresponds to a critical functional form for which the system orders, as demonstrated in Ref.~\cite {vm_aging}. However, the canonical model fails to capture this behavior, yielding $A=B=0$. It is expected that higher-order terms in Eq.\protect \eqref {eq:m_general} are required to account for this phenomenon.}\BibitemShut {Stop}%
\bibitem [{\citenamefont {Dall'Asta}\ and\ \citenamefont {Galla}(2008)}]{DallAsta_2008}%
  \BibitemOpen
  \bibfield  {author} {\bibinfo {author} {\bibfnamefont {L.}~\bibnamefont {Dall'Asta}}\ and\ \bibinfo {author} {\bibfnamefont {T.}~\bibnamefont {Galla}},\ }\bibfield  {title} {\bibinfo {title} {Algebraic coarsening in voter models with intermediate states},\ }\href {https://doi.org/10.1088/1751-8113/41/43/435003} {\bibfield  {journal} {\bibinfo  {journal} {Journal of Physics A: Mathematical and Theoretical}\ }\textbf {\bibinfo {volume} {41}},\ \bibinfo {pages} {435003} (\bibinfo {year} {2008})}\BibitemShut {NoStop}%
\bibitem [{\citenamefont {Gastner}\ \emph {et~al.}(2018)\citenamefont {Gastner}, \citenamefont {Oborny},\ and\ \citenamefont {Gulyás}}]{Gastner_2018}%
  \BibitemOpen
  \bibfield  {author} {\bibinfo {author} {\bibfnamefont {M.~T.}\ \bibnamefont {Gastner}}, \bibinfo {author} {\bibfnamefont {B.}~\bibnamefont {Oborny}},\ and\ \bibinfo {author} {\bibfnamefont {M.}~\bibnamefont {Gulyás}},\ }\bibfield  {title} {\bibinfo {title} {Consensus time in a voter model with concealed and publicly expressed opinions},\ }\href {https://doi.org/10.1088/1742-5468/aac14a} {\bibfield  {journal} {\bibinfo  {journal} {Journal of Statistical Mechanics: Theory and Experiment}\ }\textbf {\bibinfo {volume} {2018}},\ \bibinfo {pages} {063401} (\bibinfo {year} {2018})}\BibitemShut {NoStop}%
\bibitem [{\citenamefont {J\k{e}drzejewski}\ \emph {et~al.}(2018)\citenamefont {J\k{e}drzejewski}, \citenamefont {Marcjasz}, \citenamefont {Nail},\ and\ \citenamefont {Sznajd-Weron}}]{Kasia_layer}%
  \BibitemOpen
  \bibfield  {author} {\bibinfo {author} {\bibfnamefont {A.}~\bibnamefont {J\k{e}drzejewski}}, \bibinfo {author} {\bibfnamefont {G.}~\bibnamefont {Marcjasz}}, \bibinfo {author} {\bibfnamefont {P.~R.}\ \bibnamefont {Nail}},\ and\ \bibinfo {author} {\bibfnamefont {K.}~\bibnamefont {Sznajd-Weron}},\ }\bibfield  {title} {\bibinfo {title} {Think then act or act then think?},\ }\href {https://doi.org/10.1371/journal.pone.0206166} {\bibfield  {journal} {\bibinfo  {journal} {PLOS ONE}\ }\textbf {\bibinfo {volume} {13}},\ \bibinfo {pages} {1} (\bibinfo {year} {2018})}\BibitemShut {NoStop}%
\bibitem [{Note2()}]{Note2}%
  \BibitemOpen
  \bibinfo {note} {The ratio $a/(2(1-a))$ was denoted as $\varepsilon $ in Ref.~\cite {Peralta_2018}. However, in this paper this letter denotes the strength of the preference of partisan voter models and it has already appeared in the $q$-voter.}\BibitemShut {Stop}%
\bibitem [{\citenamefont {Llabr{\'e}s}\ \emph {et~al.}(2025)\citenamefont {Llabr{\'e}s}, \citenamefont {Oliver-Bonafoux}, \citenamefont {Anteneodo},\ and\ \citenamefont {Toral}}]{complete_aging}%
  \BibitemOpen
  \bibfield  {author} {\bibinfo {author} {\bibfnamefont {J.}~\bibnamefont {Llabr{\'e}s}}, \bibinfo {author} {\bibfnamefont {S.}~\bibnamefont {Oliver-Bonafoux}}, \bibinfo {author} {\bibfnamefont {C.}~\bibnamefont {Anteneodo}},\ and\ \bibinfo {author} {\bibfnamefont {R.}~\bibnamefont {Toral}},\ }\bibfield  {title} {\bibinfo {title} {Complete aging in the noisy voter model enhances consensus formation},\ }\href {https://doi.org/https://doi.org/10.1016/j.chaos.2025.116153} {\bibfield  {journal} {\bibinfo  {journal} {Chaos, Solitons \& Fractals}\ }\textbf {\bibinfo {volume} {194}},\ \bibinfo {pages} {116153} (\bibinfo {year} {2025})}\BibitemShut {NoStop}%
\bibitem [{\citenamefont {van Kampen}(2007)}]{vanKampen:2007}%
  \BibitemOpen
  \bibfield  {author} {\bibinfo {author} {\bibfnamefont {N.}~\bibnamefont {van Kampen}},\ }\href@noop {} {\emph {\bibinfo {title} {Stochastic Processes in Physics and \\ Chemistry}}},\ \bibinfo {edition} {3rd}\ ed.\ (\bibinfo  {publisher} {North-Holland},\ \bibinfo {address} {Amsterdam},\ \bibinfo {year} {2007})\BibitemShut {NoStop}%
\bibitem [{\citenamefont {Toral}\ and\ \citenamefont {Colet}(2014)}]{Toral2014StochasticNM}%
  \BibitemOpen
  \bibfield  {author} {\bibinfo {author} {\bibfnamefont {R.}~\bibnamefont {Toral}}\ and\ \bibinfo {author} {\bibfnamefont {P.}~\bibnamefont {Colet}},\ }\href@noop {} {\emph {\bibinfo {title} {Stochastic Numerical Methods:\\ An Introduction for Students and Scientists}}}\ (\bibinfo  {publisher} {Wiley-VCH},\ \bibinfo {year} {2014})\BibitemShut {NoStop}%
\bibitem [{\citenamefont {Huang}(1987)}]{Huang87}%
  \BibitemOpen
  \bibfield  {author} {\bibinfo {author} {\bibfnamefont {K.}~\bibnamefont {Huang}},\ }\href@noop {} {\emph {\bibinfo {title} {Statistical Mechanics}}}\ (\bibinfo  {publisher} {John wiley and Sons},\ \bibinfo {year} {1987})\BibitemShut {NoStop}%
\bibitem [{\citenamefont {Cardy}(1988)}]{Cardy:1988}%
  \BibitemOpen
  \bibinfo {editor} {\bibfnamefont {J.~L.}\ \bibnamefont {Cardy}},\ ed.,\ \href {https://doi.org/https://doi.org/10.1016/B978-0-444-87109-1.50001-7} {\emph {\bibinfo {title} {Finite-Size Scaling, Current Physics - \\ Sources and Comments}}},\ Vol.~\bibinfo {volume} {2}\ (\bibinfo  {publisher} {Elsevier},\ \bibinfo {year} {1988})\BibitemShut {NoStop}%
\bibitem [{\citenamefont {Deutsch}(1992)}]{deutsch:92}%
  \BibitemOpen
  \bibfield  {author} {\bibinfo {author} {\bibfnamefont {H.~P.}\ \bibnamefont {Deutsch}},\ }\bibfield  {title} {\bibinfo {title} {Optimized analysis of the critical behavior in polymer mixtures from \uppercase{m}onte \uppercase{c}arlo simulations},\ }\href@noop {} {\bibfield  {journal} {\bibinfo  {journal} {J. Stat. Phys}\ }\textbf {\bibinfo {volume} {67}},\ \bibinfo {pages} {1039} (\bibinfo {year} {1992})}\BibitemShut {NoStop}%
\bibitem [{\citenamefont {Binder}(1981)}]{Binder:cumulant}%
  \BibitemOpen
  \bibfield  {author} {\bibinfo {author} {\bibfnamefont {K.}~\bibnamefont {Binder}},\ }\bibfield  {title} {\bibinfo {title} {{Finite size scaling analysis of Ising model block distribution functions}},\ }\href {https://doi.org/10.1007/BF01293604} {\bibfield  {journal} {\bibinfo  {journal} {Zeitschrift f{\"u}r Physik B Condensed Matter}\ }\textbf {\bibinfo {volume} {43}},\ \bibinfo {pages} {119} (\bibinfo {year} {1981})}\BibitemShut {NoStop}%
\bibitem [{\citenamefont {Yeomans}(1992)}]{yeomans1992statistical}%
  \BibitemOpen
  \bibfield  {author} {\bibinfo {author} {\bibfnamefont {J.}~\bibnamefont {Yeomans}},\ }\href@noop {} {\emph {\bibinfo {title} {Statistical Mechanics of Phase Transitions}}}\ (\bibinfo  {publisher} {Clarendon Press},\ \bibinfo {year} {1992})\BibitemShut {NoStop}%
\bibitem [{\citenamefont {Egu\'{\i}luz}\ \emph {et~al.}(2003)\citenamefont {Egu\'{\i}luz}, \citenamefont {Hern\'andez-Garc\'{\i}a}, \citenamefont {Piro},\ and\ \citenamefont {Klemm}}]{Eguiluz2003}%
  \BibitemOpen
  \bibfield  {author} {\bibinfo {author} {\bibfnamefont {V.~M.}\ \bibnamefont {Egu\'{\i}luz}}, \bibinfo {author} {\bibfnamefont {E.}~\bibnamefont {Hern\'andez-Garc\'{\i}a}}, \bibinfo {author} {\bibfnamefont {O.}~\bibnamefont {Piro}},\ and\ \bibinfo {author} {\bibfnamefont {K.}~\bibnamefont {Klemm}},\ }\bibfield  {title} {\bibinfo {title} {Effective dimensions and percolation in hierarchically structured scale-free networks},\ }\href {https://doi.org/10.1103/PhysRevE.68.055102} {\bibfield  {journal} {\bibinfo  {journal} {Phys. Rev. E}\ }\textbf {\bibinfo {volume} {68}},\ \bibinfo {pages} {055102} (\bibinfo {year} {2003})}\BibitemShut {NoStop}%
\bibitem [{\citenamefont {Vazquez}\ and\ \citenamefont {Egu{\'\i}luz}(2008{\natexlab{b}})}]{VazquezPA}%
  \BibitemOpen
  \bibfield  {author} {\bibinfo {author} {\bibfnamefont {F.}~\bibnamefont {Vazquez}}\ and\ \bibinfo {author} {\bibfnamefont {V.~M.}\ \bibnamefont {Egu{\'\i}luz}},\ }\bibfield  {title} {\bibinfo {title} {Analytical solution of the voter model on uncorrelated networks},\ }\href {https://doi.org/10.1088/1367-2630/10/6/063011} {\bibfield  {journal} {\bibinfo  {journal} {New Journal of Physics}\ }\textbf {\bibinfo {volume} {10}},\ \bibinfo {pages} {063011} (\bibinfo {year} {2008}{\natexlab{b}})}\BibitemShut {NoStop}%
\bibitem [{\citenamefont {Wiseman}\ and\ \citenamefont {Domany}(1995)}]{lack_SA}%
  \BibitemOpen
  \bibfield  {author} {\bibinfo {author} {\bibfnamefont {S.}~\bibnamefont {Wiseman}}\ and\ \bibinfo {author} {\bibfnamefont {E.}~\bibnamefont {Domany}},\ }\bibfield  {title} {\bibinfo {title} {Lack of self-averaging in critical disordered systems},\ }\href {https://doi.org/10.1103/PhysRevE.52.3469} {\bibfield  {journal} {\bibinfo  {journal} {Phys. Rev. E}\ }\textbf {\bibinfo {volume} {52}},\ \bibinfo {pages} {3469} (\bibinfo {year} {1995})}\BibitemShut {NoStop}%
\bibitem [{\citenamefont {Aharony}\ and\ \citenamefont {Harris}(1996)}]{lack_SA_lett}%
  \BibitemOpen
  \bibfield  {author} {\bibinfo {author} {\bibfnamefont {A.}~\bibnamefont {Aharony}}\ and\ \bibinfo {author} {\bibfnamefont {A.~B.}\ \bibnamefont {Harris}},\ }\bibfield  {title} {\bibinfo {title} {Absence of self-averaging and universal fluctuations in random systems near critical points},\ }\href {https://doi.org/10.1103/PhysRevLett.77.3700} {\bibfield  {journal} {\bibinfo  {journal} {Phys. Rev. Lett.}\ }\textbf {\bibinfo {volume} {77}},\ \bibinfo {pages} {3700} (\bibinfo {year} {1996})}\BibitemShut {NoStop}%
\end{thebibliography}%
\end{document}